%% file: casbah_galaxies_astroph.tex
\documentclass[preprint2,times,tighten]{aastex6}

\usepackage{amsmath,amstext}
\usepackage[figure,figure*]{hypcap}
\usepackage{newtxmath} 

\newcommand{\ncasbah}{9}  
\newcommand{\nsdss}{8}
\newcommand{\nlbt}{7}  
\newcommand{\nhectosky}{20} 
\newcommand{\hectorn}{2} 
\newcommand{\medhectsn}{5.3}  
\newcommand{\meddeimossn}{2.9}  

\newcommand{\ngal}{5902}  
\newcommand{\nbad}{356}  
\newcommand{\nstars}{279}  
\newcommand{\zmed}{0.28}  
\newcommand{\Mmed}{10^{10.1}}  

\newcommand{\specdb}{{\sc specDB}} 
\newcommand{\lbcgo}{{\tt LBCgo}}

\newcommand{\mlovi}{\ell_{\rm OVI}(z)}
\newcommand{\lovi}{$\mlovi$}
\newcommand{\mnovi}{\N{OVI}}
\newcommand{\novi}{$\mnovi$}
\newcommand{\ovi}{\ion{O}{6}}
\newcommand{\civ}{\ion{C}{4}}
\newcommand{\mfc}{f_C}
\newcommand{\fc}{$\mfc$}
\newcommand{\mnlim}{\N{OVI}_{\rm lim}}
\newcommand{\nlim}{$\mnlim$}
\newcommand{\mzsys}{z_{\rm sys}}
\newcommand{\zsys}{$\mzsys$}

\newcommand{\ncfield}{7}
\newcommand{\ggrval}{5.48 \pm 0.07 \, \hmpc}
\newcommand{\gggval}{1.33 \pm 0.04}
\newcommand{\gorval}{6.00^{+1.09}_{-0.77} \, \hmpc}
\newcommand{\gogval}{1.25 \pm 0.18}
\newcommand{\bggval}{1.3 \pm 0.1}
\newcommand{\bagval}{1.0 \pm 0.1}
\newcommand{\Mgghalo}{10^{12.1 \pm 0.05}} 
\newcommand{\Maghalo}{10^{11}} 

\newcommand{\mxitgg}{<\xi^{\rm T}_{\rm gg}(\mrperp)>}
\newcommand{\xitgg}{$\mxitgg$}
\newcommand{\mxigg}{\xi_{\rm gg}(r)}
\newcommand{\xigg}{$\mxigg$}
\newcommand{\mxitag}{<\xi^{\rm T}_{\rm ag}(\mrperp)>}
\newcommand{\xitag}{$\mxitag$}
\newcommand{\mxiag}{\xi_{\rm ag}(r)}
\newcommand{\xiag}{$\mxiag$}
\newcommand{\hmpc}{h_{100}^{-1} \, {\rm Mpc}}

\newcommand{\mloz}{\ell_{\rm OVI}(z)}
\newcommand{\loz}{$\mloz$}

\newcommand{\mzem}{z_{\rm em}}
\newcommand{\zem}{$\mzem$}
\newcommand{\msol}{M_\odot}
\newcommand{\mrpar}{R_\parallel}
\newcommand{\rpar}{$\mrpar$}
\newcommand{\mrperp}{R_\perp}
\newcommand{\rperp}{$\mrperp$}
\newcommand{\mrpperp}{R_{\perp,p}}
\newcommand{\rpperp}{$\mrpperp$}
\newcommand{\mrpcom}{R_{\perp,c}}
\newcommand{\rpcom}{$\mrpcom$}

\def\intl{\int\limits}

\def\ltp{\left ( \,}

\def\rtp{\, \right  ) }

\def\smm{\sum\limits}

\newcommand{\cm}[1]{\, {\rm cm^{#1}}}
\newcommand{\mkms}{{\rm \; km\;s^{-1}}}
\newcommand{\kms}{$\mkms$}
\newcommand{\lya}{Ly$\alpha$}
\newcommand{\N}[1]{{N({\rm #1})}}

\shorttitle{CASBaH Galaxies} 

\begin{document}

\title{The COS Absorption Survey of Baryon Harbors: 
The Galaxy Database and Cross-Correlation 
Analysis of \ovi\ Systems\altaffilmark{1}}

\author{
J. Xavier Prochaska\altaffilmark{2,3}, 
Joseph N. Burchett\altaffilmark{2},
Todd M. Tripp\altaffilmark{4},
Jessica K. Werk\altaffilmark{2,5},
Christopher N. A. Willmer\altaffilmark{6},
J.~Christopher Howk\altaffilmark{7},
Scott Lange\altaffilmark{2},
Nicolas Tejos\altaffilmark{8},
Joseph D. Meiring\altaffilmark{4},
Jason Tumlinson\altaffilmark{9,10},
Nicolas Lehner\altaffilmark{7},
Amanda B. Ford\altaffilmark{12},
Romeel Dav\'e\altaffilmark{13}
}

\altaffiltext{1}{Based on observations made with the NASA/ESA Hubble Space Telescope, obtained at the Space Telescope Science Institute, which is operated by the Association of Universities for Research in Astronomy, Inc., under NASA contract NAS 5-26555. These observations are associated with programs 13033 and 11598. Partly based on observations taken at the MMT Observatory, a joint facility operated by the Smithsonian Institution and the University of Arizona.}
\altaffiltext{2}{University of California, Santa Cruz; 1156 High St., Santa Cruz, CA 95064, USA; xavier@ucolick.org}
\altaffiltext{3}{Kavli Institute for the Physics and Mathematics of the Universe (Kavli IPMU)
The University of Tokyo; 5-1-5 Kashiwanoha, Kashiwa, 277-8583, Japan}
\altaffiltext{4}{Department of Astronomy, University of Massachusetts, 710 North Pleasant Street, Amherst, MA 01003-9305}
\altaffiltext{5}{University of Washington, Department of Astronomy, Seattle, WA 98195}
\altaffiltext{6}{Steward Observatory, University of Arizona, 933 North Cherry Avenue, Tucson, AZ 85721}
\altaffiltext{7}{Department of Physics, University of Notre Dame, 225 Nieuwland Science Hall, Notre Dame, IN 46556}
\altaffiltext{8}{Instituto de F\'isica, Pontificia Universidad Cat\'olica de Valpara\'iso, Casilla 4059, Valpara\'iso, Chile}
\altaffiltext{9}{Space Telescope Science Institute, Baltimore, MD, 21218}
\altaffiltext{10}{Department of Physics and Astronomy, Johns Hopkins University, Baltimore, MD, 21218}
\altaffiltext{11}{Institute for Computational Cosmology and Centre for Extragalactic Astronomy, Department of Physics, Durham University, South Road, Durham, DH1 3LE, UK}
\altaffiltext{12}{Google, 1600 Amphitheatre Parkway, Mountain View, CA 94043}
\altaffiltext{13}{Institute for Astronomy, Royal Observatory, University of Edinburgh, EH9 3HJ, UK}

\begin{abstract}
We describe the survey for galaxies in the fields surrounding \ncasbah\ sightlines
to far-UV bright, $z \sim 1$ quasars that define the 
COS Absorption Survey of Baryon Harbors (CASBaH) program.
The photometry and spectroscopy that comprise the dataset come from a mixture of public
surveys (SDSS, DECaLS) and our dedicated efforts on private facilities 
(Keck, MMT, LBT).  We report the redshifts and stellar masses for 
\ngal\ galaxies within $\approx 10$ comoving-Mpc (cMpc) of the sightlines with a 
median of $\bar z=\zmed$ and $\bar M_* \approx \Mmed \msol$.
This dataset, publicly available as the CASBaH {\sc specDB}, forms the basis
of several recent and ongoing CASBaH analyses.
Here, we perform a clustering analysis of the galaxy sample with itself
(auto-correlation) and against the set of 
\ion{O}{6} absorption systems
(cross-correlation) discovered in the CASBaH quasar spectra
with column densities $\N{O^{+5}} \ge 10^{13.5} \, \rm cm^{-2}$.
For each, we describe the measured
clustering signal with a power-law correlation function $\xi(r) = (r/r_0)^{-\gamma}$
and find that $(r_0,\gamma) = (\ggrval,\gggval)$ for the auto-correlation
and $(\gorval, \gogval)$ for galaxy-\ion{O}{6} cross-correlation.
We further estimate a bias factor of $b_{gg} = \bggval$ from 
the galaxy-galaxy auto-correlation indicating the galaxies are hosted by
halos with mass $M_{\rm halo} \approx \Mgghalo \msol$.
Finally, we estimate an \ion{O}{6}-galaxy bias factor
$b_{\rm OVI} = \bagval$ from the cross-correlation
which is consistent with 
\ion{O}{6} absorbers being hosted by
dark matter halos with typical mass
$M_{\rm halo} \approx \Maghalo \msol$.
Future works with upcoming datasets (e.g.,\ CGM$^2$) will improve upon these
results and will assess whether any of the detected \ion{O}{6} arises
in the intergalactic medium.

\end{abstract}


\section{Introduction}
\label{sec:intro}

The cosmic web is the filamentary network of dark matter and baryons
predicted by cosmological simulations to permeate our universe
\citep[[]{mco+96,lukic+15}.  It forms under the competing influences of
gravitational collapse and cosmic expansion, modulated by
hydrodynamic heating and cooling during collapse, and also
ionization balance with the extragalactic radiation field.  
While the web holds as a ubiquitous
prediction of dark matter cosmology, 
tests of this paradigm are relatively scarce.

Using luminous galaxies as tracers, wide-field surveys 
have revealed patterns of
large-scale structure that resemble theoretical prediction.  And,
within the limited statistical measures afforded by these data, the
distributions match model predictions \citep[e.g.,][]{davis+85,bond+96}.
As the statistical sample increases and pushes to higher
redshift, topology diagnostics afford tests of the
cosmic web morphology \citep[e.g.,][]{cautun+13,tempel+14}.  These experiments, however,
are inherently limited by the sparseness of galaxies and the surveys'
inherent biases \citep[e.g.,][]{smith+03}.

Absent a means to directly image the diffuse emission predicted
from the cosmic web \citep{GW96,cantalupo14},
one relies on the inverse approach of detecting the
web's threads in absorption.  
Absorption lines in the spectra of luminous
background sources (typically quasars) yield a one-dimensional description
of the matter distribution across cosmic time.  Quantitative 
comparison of cosmological predictions with the resultant \ion{H}{1} \lya\ forest 
have lent further support to this model \citep[e.g.,][]{mco+96,cwb+03}.
The agreement is sufficiently compelling that
modern efforts have inverted the practice, adopting the cosmic
web paradigm
to constrain parameters of the cosmology and other properties of the
universe \citep[e.g.,][]{sik+13,palanque+13}.

Within these same quasar spectra, one also identifies absorption
from transitions of heavy elements (e.g., \civ, \ovi) that record
prior enrichment by galaxies.
The high incidence of this metal absorption, especially at high-$z$
where the data quality is exquisite and the rest-frame ultraviolet (UV) transitions shift
into the optical bandpass and become observable from ground based telescopes,
requires this enrichment to extend
far beyond the galaxies' interstellar medium (ISM) and possibly 
beyond their
local environs (aka the circumgalactic medium or CGM) and
into the IGM \citep[e.g.,][]{simcoe04,booth+12}. 
The distribution of enrichment
on these scales is a complex interplay between the timing of the metal
production, the galaxies involved, the processes that eject/transport
the matter from star-forming regions, and the underlying 
potential well of the galaxy and its environment.
Unfortunately, the degree of complexity is sufficient that even
a precise accounting of the incidence and degree of heavy element
absorption along multiple sightlines is insufficient to fully resolve
the underlying astrophysics \citep[e.g.,][]{ford+16}.

Of greater potential power for analyzing the cosmic web and
its enrichment is to combine absorption-line studies with surveys
of the galaxies surrounding the absorption-line sightlines.
At $z \sim 0$, where galaxies are more easily
observed, several studies have examined UV spectroscopy of the 
\lya\ forest to provide constraints on the present-day cosmic web
\citep{mwd+93,bowen+02,pss02,cpw+05,araciletal06,pwc+11,tejos+12,wakker09}.
These have established the connection between intergalactic filaments
and \lya\ absorption \citep{wakker+15}
and also glimpses of the voids which are 
expected to fill the volume \citep{tejos+12}.
Studies focused on the association of heavy elements to the
cosmic web are more rare \citep{stockeetal06,araciletal06,cm09,pwc+11}
and these are stymied by smaller samples.
\cite{pwc+11} argued that the 
majority (and possibly all) of the metal-line
detections in quasar spectra
arise within a few hundred kpc of galaxies, casting some doubt
for any enrichment in the IGM.  These results, however, were
tempered by the limitations of sample variance and the signal-to-noise and detection sensitivity of the UV absorption spectra.

Extending absorber-galaxy analysis to $z>0$ is challenged
by several evolving factors.  For $z \sim 0-1$, the paucity of
UV-bright quasars limits the number and quality of absorption-line
spectra available.  Furthermore, present wide-field surveys (SDSS, 2dF)
are generally complete only at $z<0.1$.  Primary exceptions are those targeting
large samples of 
luminous red galaxies \citep[LRGs;][]{eisenstein+01} and, more recently,
emission line galaxies \citep[ELGs;][]{eBOSS}.
While these $z \sim 0.5$ galaxy
surveys have enabled important works on a subset of 
absorption lines \citep[primarily MgII;][]{zhu+14,lm18},
detailed exploration of the cosmic web has required 
dedicated follow-up surveys of the rare fields hosting UV-luminous
quasars \citep{ovi_paper4,tejos+14,johnson+15,keeney+18}.  

The most comprehensive work at $z \sim 0.5$
has been carried out by a group in Durham with results published by \cite{tejos+14} 
and \cite{finn+16}. Their first paper \citep{tejos+14} focused
on \ion{H}{1} absorption and its clustering to galaxies on scales
of $\sim 10 \hmpc$.  Their analysis confirmed previous assertions
\citep{mwd+93,cpw+05,tripp+98}
that the \ion{H}{1} \lya\ forest is roughly divided into a low-density
population tracing the cosmic web and a higher-density 
component associated with dark matter halos.
Their second paper \citep{finn+16} measured the cross-correlation
of \ovi\ with galaxies which they interpreted as evidence for
\ovi\ distributed away from galaxies but following the same underlying 
mass distribution on $\sim$\,Mpc scales. 

%
%
With the installation of the Cosmic Origins Spectrograph (COS) on {\it HST} in 2009, we were 
motivated to pursue a new survey dedicated to investigations
of the cosmic web at $z \sim 0-1.5$.  With this goal in mind, the
COS Absorption Survey of Baryon Harbors (CASBaH, HST Programs 11741 \& 13846, PI Tripp)
was initiated to obtain high S/N
spectra of $\sim 10$ quasars at $z \gtrsim 1$ and
to assess \ion{H}{1} and heavy element absorption.  
A full description of the CASBaH design and data handling procedures, as well as the first release of the absorption-line database, are provided by \citet{casbah}. In brief, this survey observed nine QSOs (see Table~\ref{tab:fields}) with the high-resolution COS G130M, G160M, G185M, and G225M gratings as well as the Space Telescope Imaging Spectrograph (STIS) E230M echelle mode.\footnote{For information on COS and STIS, see \citet{cos} and \citet{STIS} respectively.}
This set of observations was designed to provide complete spectral coverage from observed wavelength $\lambda _{\rm ob}$ = 1152 \AA\ to $(1+z_{\rm QSO})\times$1215.67 \AA , i.e., the spectra cover the entire Ly$\alpha$ forest for each QSO with good spectral resolution (FWHM $\approx 10 - 20$ km s$^{-1}$).  In the far-UV range ($\lambda _{\rm ob}$ = 1152$-$1800 \AA ), the survey was designed to detect weak metal lines such as the \ion{Ne}{8} doublet and affiliated species \citep[e.g.,][]{tmp+11,mtw+13}, and accordingly the exposure times were set to provide signal-to-noise (S/N) ratios of $\approx 15-50$ per resolution element. 
The near-UV spectra were obtained to detect strong \ion{H}{1} lines that are crucial for proper line/system identification and to extend the coverage of strong \ion{O}{6} lines to higher redshifts; for these purposes, the near-UV data did not require high S/N and typically have S/N $\approx 5-20$ per resel. The achieved S/N levels of the far-UV spectra (the COS G130M and G160M data) afford unparalleled insight into the diffuse
gas of the cosmic web and access to extreme UV lines
(e.g., \ion{Ne}{8}) that have been only rarely observed.

A crucial component of the CASBaH
program is a dedicated, deep
survey of galaxies around the quasar sightlines.
This paper describes the CASBaH galaxy redshift survey and provides the current 
CASBaH galaxy database.  The presentation of this database culminates many years of 
observing to gather $\sim 10,000$ spectra in 7 quasar fields.
Given the tremendous legacy value of the CASBaH absorption-line
database, there are certain to be additional surveys of
galaxies in these fields \citep[e.g., QSAGE;][]{qsages}.
This manuscript additionally offers a first analysis of the nature
of \ovi\ absorption in the cosmic web.  
Other upcoming works from CASBaH include detailed analyses of the gas ionization state and 
absorption kinematic structure that leverage the very high S/N in the FUV and NUV of the CASBaH spectra.

This paper is outlined as follows.
Section~\ref{sec:sample} describes the galaxy selection criteria
and the related photometry.
Section~\ref{sec:spectra} presents the spectroscopy and redshift
measurements and Section~\ref{sec:derived} lists estimates for
several derived quantities (e.g.\ stellar mass).
Lastly, Section~\ref{sec:clustering} presents a clustering
analysis of these galaxies with themselves and against the population
of \ion{O}{6} absorbers along the CASBaH sightlines.
Throughout the analysis we adopt the Planck15 cosmology,
as encoded in the {\sc astropy}\footnote{\url{http://www.astropy.org/}} package.

\section{Sample Selection}
\label{sec:sample}

\subsection{Overview}

The basis of our CASBaH galaxy survey are the fields
surrounding the \ncasbah\ quasars observed for the 
project with {\it HST} \citep{casbah}.  These quasars are presented
in Table~\ref{tab:fields}, where we also list the ancillary
data available in the public domain (as of May 16, 2018)
and those collected by our team\footnote{Numerous other
public imaging survey datasets (e.g.,\ {\it WISE}) also cover these
fields and are not listed but are employed in our galaxy characterization.  
These did not inform the sample selection.}.
The quasar coordinates were taken from the Simbad
database, and we adopt the QSO emission redshift
measured from SDSS spectra by
\cite{hw11}.

\input{tab_fields.tex}

Given that the scientific goals of the CASBaH project include
the analysis of gas from $z \sim 0$ to $z \sim 1$
\citep{casbah}, we pursued
galaxies to faint magnitude limits, i.e., much fainter than
typical of public spectroscopic datasets (although 
any such data is included).
In general, our approach
was two-pronged:
(i) we obtained spectra to the SDSS imaging limit over a wide
field-of-view (FOV)
using the Hectospec spectrometer \citep{hectospec}
on the MMT telescope, and
(ii) we obtained deep LBT/LBC imaging and faint object
spectra with the DEIMOS spectrometer \citep{fpk+03}
on the Keck-II telescope over a narrower FOV.
Each of these activities had basic requirements, e.g.,\
SDSS imaging for MMT/Hectospec, visibility from 
Keck, etc.  In addition, the data collected
flowed from the vagaries of time assignment committees
and weather.  These factors resulted in more heterogeneous
sampling of each field than may be desired.

\subsection{Photometry}

\subsubsection{SDSS}

For the \nsdss\ fields within the SDSS imaging footprint, we
retrieved the photometric measurements from their archive
with the {\sc astroquery}\footnote{\url{http://github.com/astropy/astroquery}} package.  
Specifically, we retrieved
all photometric sources in the SDSS-DR12 catalog within 2\,deg
of each field, requesting Petrosian magnitudes and errors.
We then cross-matched these to the spectroscopic catalog and
cut on $z > 0.001666$ ($v > 500 \mkms$) to trim  stars.\footnote{Galaxies at $v < 500$ km s$^{-1}$ are difficult to use at any rate because the H~I Ly$\alpha$ line is lost within the Milky Way damped Ly$\alpha$ profile and the geocoronal Ly$\alpha$ emission, which affects a substantial region in COS spectra.}
These data form the primary public dataset integrated within
our database.

For the fields observed with Hectospec, we further queried
the SDSS photometric catalogs to generate a set of targets.
Again, we use Petrosian magnitudes and errors.  
A full description of the Hectospec targeting is given in
$\S$~\ref{sec:hecto_targ}.

\input{tab_lbt_obs.tex}

\subsubsection{LBT/LBC}

For \nlbt\ fields, we obtained deep multi-band ($UBVI$ or $griz$) images
with the LBC on LBT under a variety of conditions (PIs: Howk, Ford).
Table~\ref{tab:lbt-obs} summarizes the observations.
The two LBC cameras described by \cite{lbt-lbc} sit at the 
prime foci of the twin 8.4-m mirrors of the LBT. The blue ($U,B$) and red ($V,I$) 
LBCs each provide a 23\arcmin\ field of view using a four-chip mosaic. We used dithered observations (typically a 9-step dither pattern for these data) to fill in the inter-chip spacings, and we used twilight sky flats to perform flat-field corrections. 
Total exposure times are typically 3000 sec for the $U,I$ band images and 420 sec for the $B,V$ bands. For a subset of the images, one of the CCDs in the mosaic was unavailable (CCD \#3). In those cases we filled in the missing area with additional dithers, which provided additional exposure time for other areas of the field, so these exposure times should be taken as representative only.

The LBC data were reduced with a development version of the Python-based \lbcgo\ data reduction pipeline \citep{howk2019}.\footnote{https://github.com/jchowk/LBCgo} \lbcgo\ performs basic image processing steps, such as removing the overscan strip, deriving and applying flat fields, etc., following standard practice. After basic image processing, \lbcgo\ uses several {\tt Astromatic.net}\footnote{http://www.astromatic.net/} codes to project the images onto a common WCS frame and coadd them, following an approach described first by \cite{sand2009}. On a chip-by-chip basis (by default) \lbcgo\ uses Source Extractor \citep{bertin1996} to find sources detected in each chip. It then uses {\tt SCAMP} \citep{bertin2002} to derive the astrometric solution for each chip based after matching detected sources with the GAIA-DR1 catalog \citep{gaia-collaboration2016a, gaia-collaboration2016b}. The astrometric solution is critical given the distortions over the full 23\arcmin\ LBC field. The individual chips are then resampled, background subtracted, and coadded using {\tt SWARP} \citep{bertin2002}. The astrometric solution provided by {\tt SCAMP} has a typical reported rms $\sim0\farcs05$, with values typically ranging from $\sim0\farcs03 - 0\farcs10$ per exposure.

\begin{figure}
\begin{center}
\includegraphics[width=3.5in]{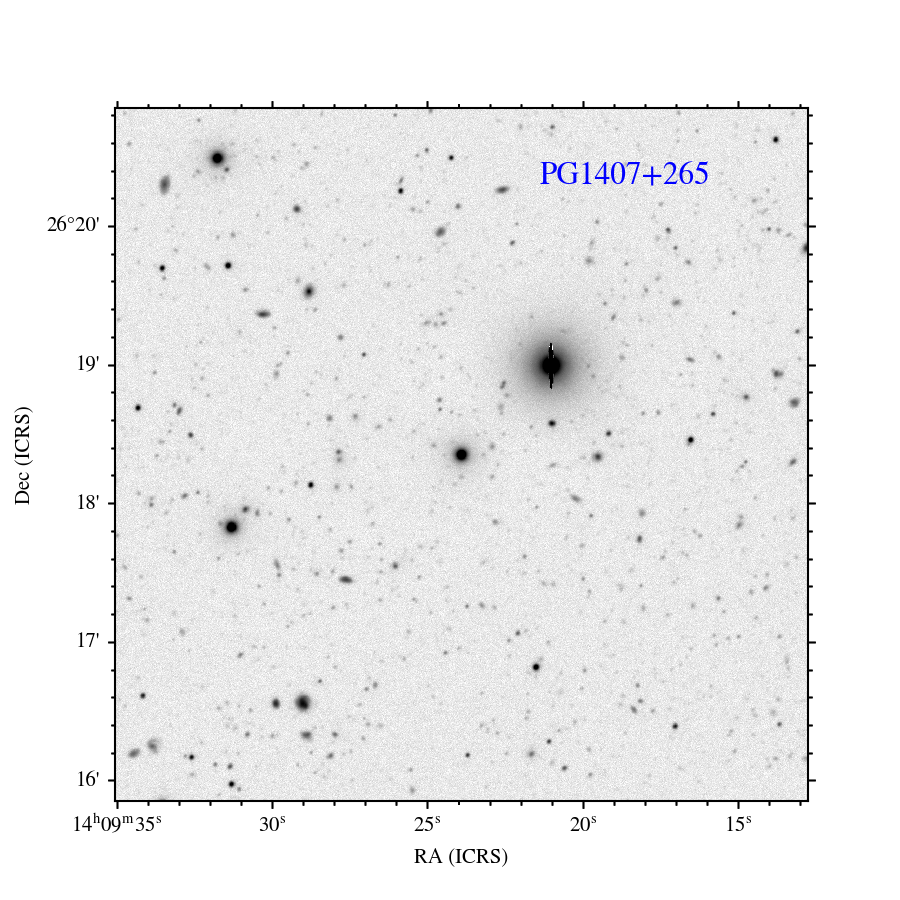}
\caption{
$V$-band image of the field surrounding PG1407+265 (centered)
obtained with the LBC on the LBT.  
[See the Journal version for the full image.]
}
\label{fig:LBC_image}
\end{center}
\end{figure}

We then adopted published zero points
for the instrument\footnote{http://abell.as.arizona.edu/$\sim$lbtsci/Instruments/\\LBC/lbc\_description.html\#zeropoints.html} 
and corrected for airmass but not Galactic extinction:
U (SDT\_USpec)=27.33,
B (Bessel)=27.93,
V (Bessel)=27.94, 
I (Bessel)=27.59,
g (Sloan)=28.31,
r (Sloan)=27.75.
With our typical total exposure times, we achieved
the sensitivities listed in Table~\ref{tab:lbc_depth}.
Figure~\ref{fig:LBC_image} shows the
$V$-band image of the field surrounding
PG1407+265, which is typical of our full dataset. Additional examples of the LBT imaging are presented in \citet{rlh+11}, 
\citet{tmp+11}, \citet{mtw+13}, and \citet{Burchett:2013qy}; 
these examples are more zoomed-in and thus show the depth and morphological information provided by the LBT imaging in greater detail.


On each of the reduced images, we ran the SExtractor
software package to generate a catalog of sources.
We adopted a standard parameter suite, including the following extra parameters: $CLASS\_STAR$, $A\_IMAGE$, $B\_IMAGE$, $THETA\_IMAGE$
and $MU\_THRESHOLD$. 
The image parameters are included in the database, although
we caution that the uncertainties are large for faint
and/or compact sources.
For source detection, we adopted three pixels as the minimum number of pixels above a detection threshold of 2.5$\sigma$. Each filtered image was processed separately and sources were cross-matched in custom software based on the astrometry.  

\input{tab_lbc_depth.tex}


\begin{figure*}
\begin{center}
\includegraphics[width=7.0in]{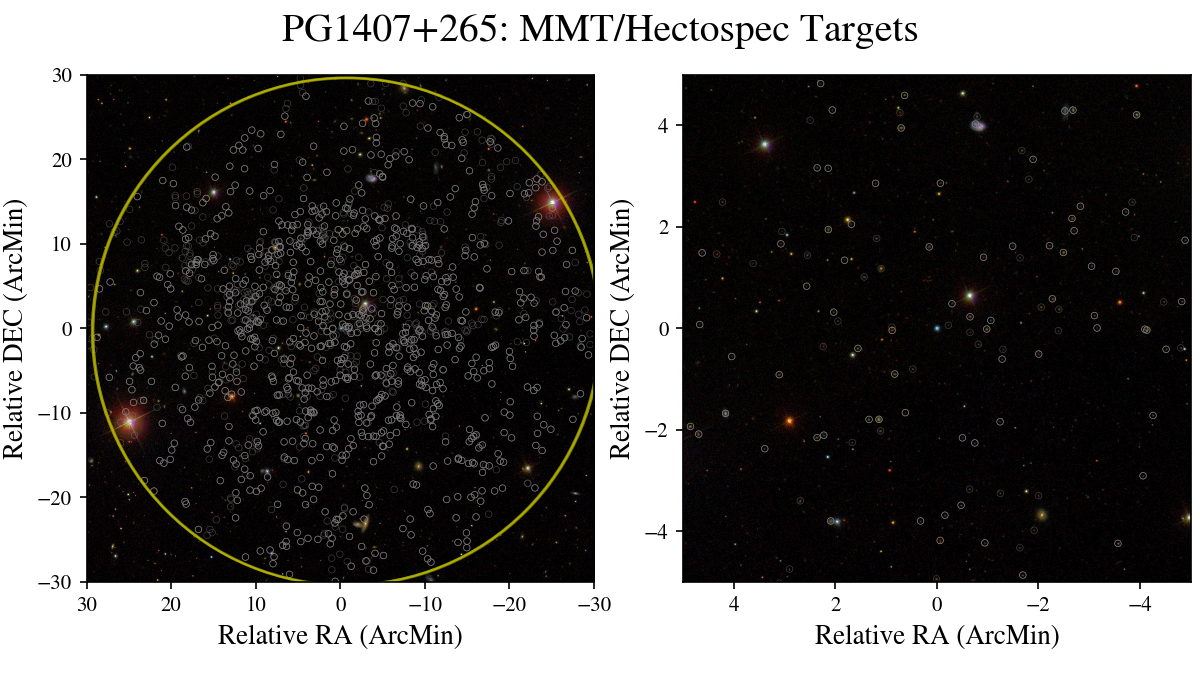}
\caption{
Illustration of the targeting strategy for Hectospec
observations of the PG1407+265 field.
The left panel shows the entire Hectospec field-of-view,
centered on PG1407+265. 
Note the `wedding cake' strategy employed, i.e.,\ a denser
set of targets closer to the quasar.
The right-hand panel shows a zoom-in to illustrate the
depth of targeting.
[See the Journal version for the full image.]
}
\label{fig:hecto_target}
\end{center}
\end{figure*}

\subsection{Targeting}
\label{sec:target}

\subsubsection{Hectospec}
\label{sec:hecto_targ}

From the SDSS imaging data described above, we generated target lists to observe
with the MMT/Hectospec spectrograph employing a `wedding-cake'
strategy that sampled the inner angular offsets from the quasar
to fainter  magnitudes.  Specifically, we targeted galaxies with $r<22$\,mag
for $\theta < 5'$, $r<21$\,mag for $\theta = [5',10']$,
and $r<20$\,mag to $\theta = 30'$. This was done because at lower redshifts, it is desirable to cover a larger FOV to probe similar impact parameter ranges as the higher redshift data, but the survey does not need to go as deep as the higher$-z$ observations to reach similar galaxy luminosities.

\begin{figure}
\begin{center}
\includegraphics[width=3.5in]{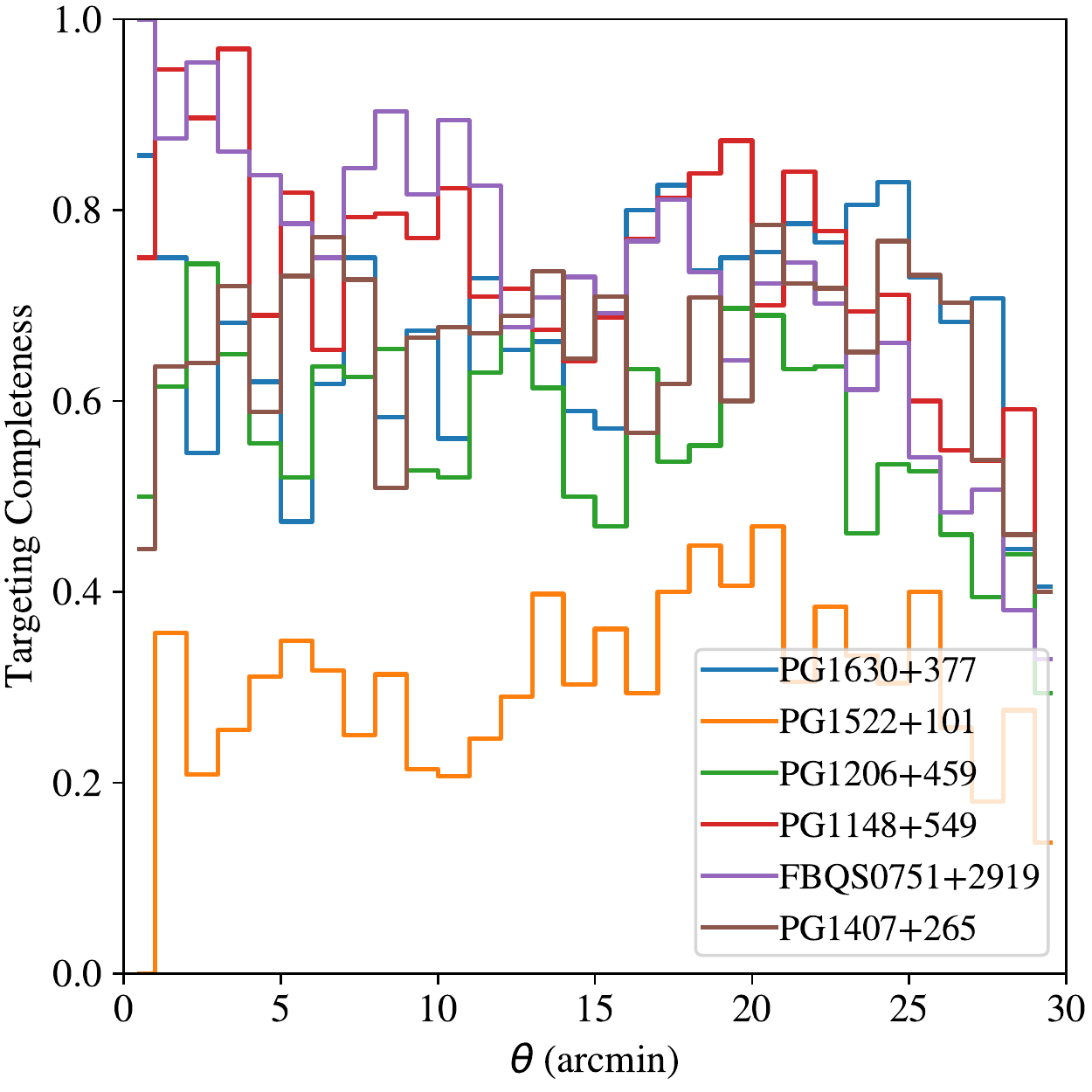}
\caption{
Completeness fraction for the targeting of galaxies
(i.e.,\ fraction of fibers on desired galaxies for Hectospec
only) 
selected from SDSS imaging according to our 
targeting criteria.
The value is independent of whether a precise redshift
was measured from the resultant spectrum.
Note that the sparsest field (PG1522+101) was observed with
only 2 configurations.
}
\label{fig:hecto_complete}
\end{center}
\end{figure}

Figure~\ref{fig:hecto_target} shows an example
of target selection
taken from the PG1407+265 field and 
Figure~\ref{fig:hecto_complete} presents the
completeness for all of the fields with the 
wedding-cake criteria above.  The targeting completeness
reported here is the percentage of targets with a 
fiber placed upon them during our observing runs.
Details on the spectroscopy and redshift determinations
are provided in the following section.


\subsubsection{DEIMOS}

\begin{figure}
\begin{center} 
\includegraphics[width=3.5in]{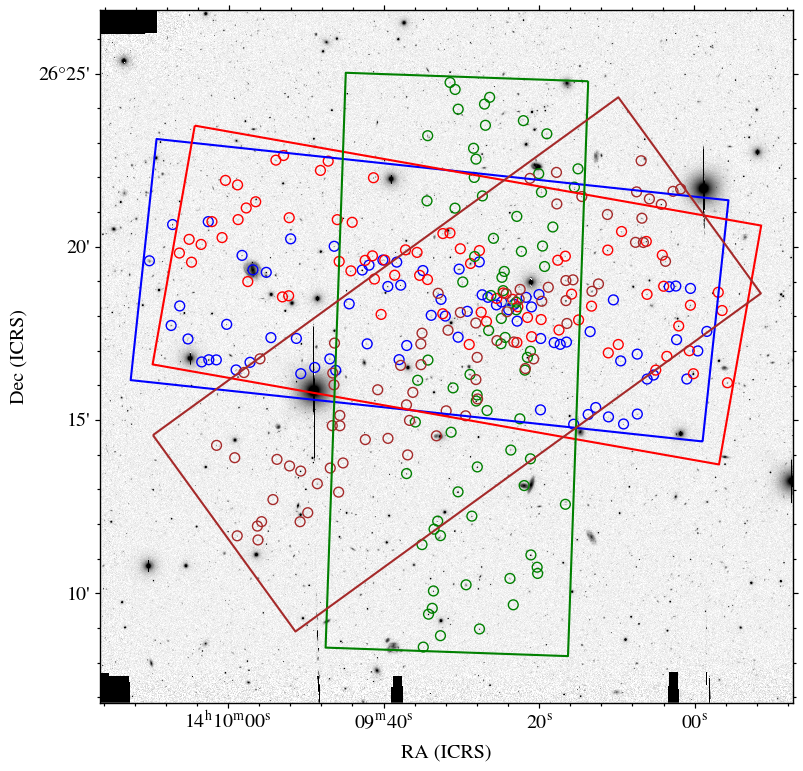}
\caption{
Slitmask FOVs (rectangles) and associated targets (circles) for
the PG1407+265 field observed with Keck/DEIMOS.
This quasar is surrounded by the densest distribution
of targets among those in our survey.
[See the Journal version for the full image.]
}
\label{fig:deimos_masks}
\end{center}
\end{figure}

With Keck/DEIMOS, we pursued fainter galaxies that are more effectively surveyed with this telescope/instrument combination, which has a 
smaller field-of-view but also a larger primary mirror.  For these observations,
we again gave higher priority to sources close to the quasar
(in angular offset) but had additional, observing-related criteria
that further affected our slit-mask designs.  
Figure~\ref{fig:deimos_masks} illustrates the targeting
for the PG1407+265 field, where we adopted the following criteria:
(1) $\theta < 15'$, 
(2) $14 < V < 24.5$,
and (3) SExtractor star-galaxy classifier
$S/G < 0.9$, except for sources within $20''$ 
of the quasar where no S/G criterion was applied.
Other fields had small differences from these criteria,
as summarized in Table~\ref{tab:deimos_criteria}.

\begin{figure}
\begin{center} 
\includegraphics[width=3.5in]{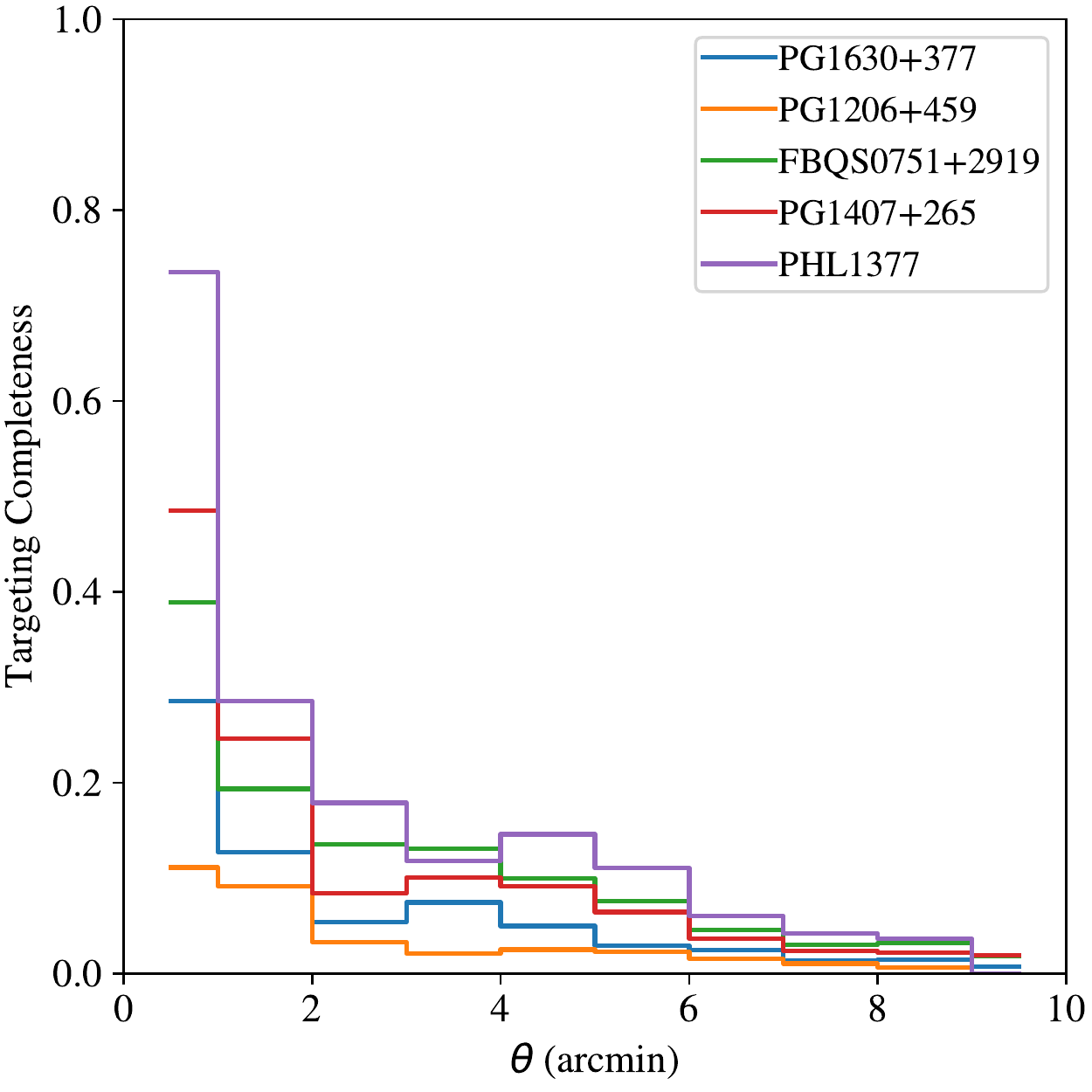}
\caption{
Completeness of targeting for the fields observed with
Keck/DEIMOS defined as the fraction of targets observed.
The value is independent of whether a precise redshift
was measured from the resultant spectrum.
}
\label{fig:deimos_complete}
\end{center}
\end{figure}

In general, we avoided targeting sources previously observed
by SDSS or our own Hectospec program.  Similar to 
Figure~\ref{fig:hecto_complete}, Figure~\ref{fig:deimos_complete}
shows the completeness for the DEIMOS fields.
In contrast to the Hectospec survey, the DEIMOS survey has sparser
coverage and higher incompleteness which resulted mainly from
poor weather.

\input{tab_deimos_criteria}



\input{tab_hectolog}

\input{tab_deimos_spectraobslog}

\section{Spectroscopy}
\label{sec:spectra}

\subsection{Observations}

\subsubsection{Hectospec}

For all of the Hectospec data collected in our CASBaH survey,
we employed an identical setup of 
300 1.5$''$ fibers and
the G270 grating, yielding
 $R \approx 1,000$ with wavelength coverage
$\lambda _{\rm ob} \approx 3,700 - 9,200$\AA.  
The observations used three or more 
exposures with times ranging from 900s to 1800s each,
for a total exposure of 3600s or 5400s as 
detailed in Table~\ref{tab:hectolog}.
Each fiber configuration included $\sim$ \nhectosky\ fibers placed on 
`blank' sky.  

All of these spectra were reduced by the HSREDv2\footnote{ \url{http://www.mmto.org/node/536}} data
reduction pipeline to wavelength calibrate, extract, sky subtract, and
flux the fiber data.  The $1\sigma$ error array assumes Gaussian
statistics and a \hectorn\ electron read noise term.
Each exposure was reduced separately, and the final 1-D spectra were co-added in wavelength space weighted by the inverse variance of the individual exposures.
The pipeline, in our case, produced a wavelength solution
calibrated in air and unfluxed.  Therefore, we converted
to vacuum with the Ciddor equation described at NIST\footnote{\url{https://emtoolbox.nist.gov/Wavelength/Documentation.asp}}.
The spectra were fluxed using a sensitivity function derived
from Feige~34 observed on another program.
We caution that the absolute fluxes do not include corrections
for fiber losses, airmass, telluric absorption or variable observing conditions.

\begin{figure}
\begin{center} 
\includegraphics[width=3.5in]{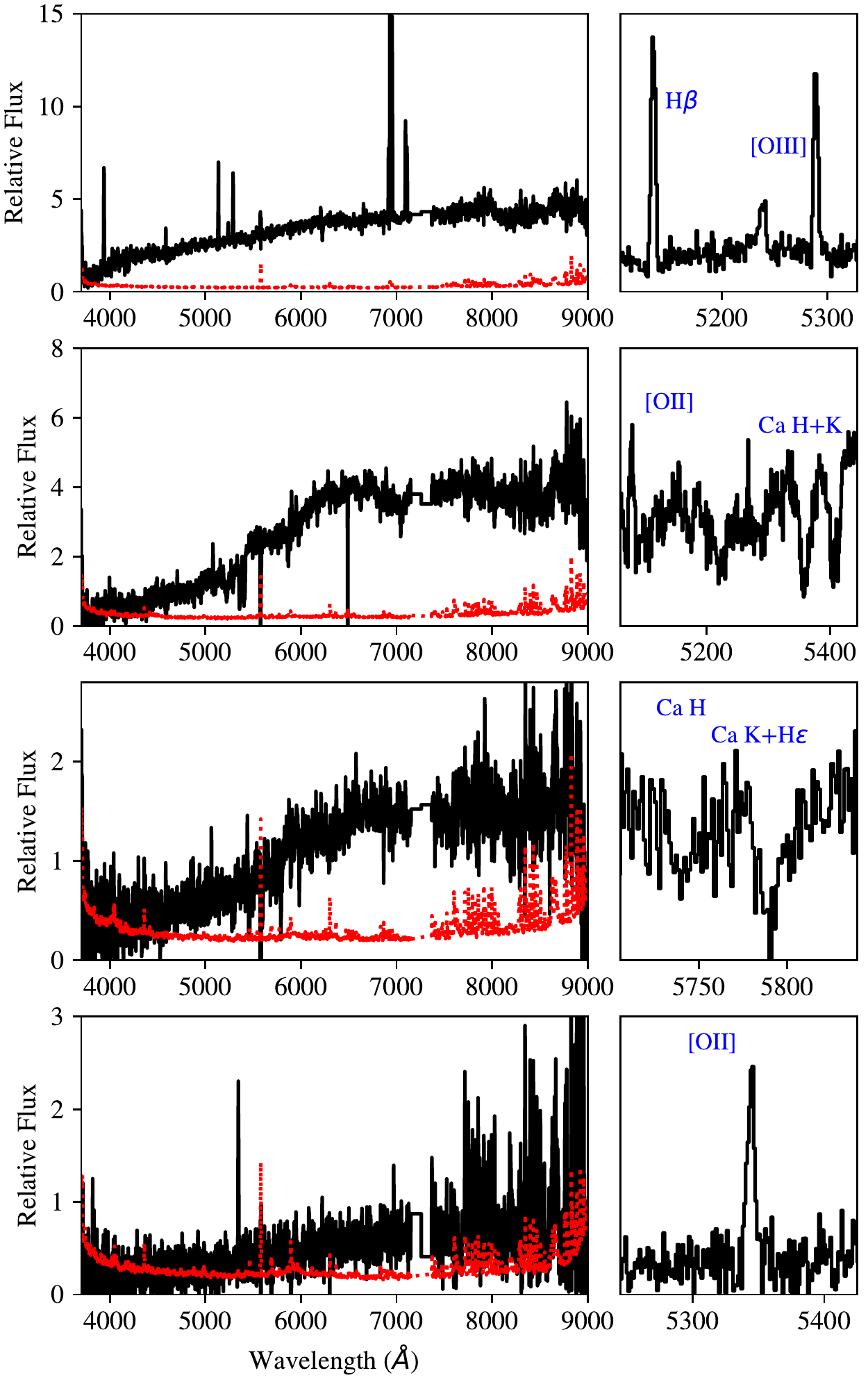}
\caption{
Representative spectra from the Hectospec observations.
Cut-outs on the right show examples of key spectral features used in the
redshift determination.
}
\label{fig:hecto_spectra}
\end{center}
\end{figure}

Figure~\ref{fig:hecto_spectra} shows several examples
of spectra for sources spanning the dynamic range of
observed magnitudes ($r \approx 18.1-21.9$\,mag).
The median S/N of all spectra is approximately
\medhectsn\ per 1.2\AA\ pixel at $\lambda _{\rm ob} \approx 5,000$\AA.

\subsubsection{DEIMOS}

For the DEIMOS observations of the CASBaH target fields (Table \ref{tab:deimos_obs}), 
we designed slitmasks with the {\sc DSIMULATOR} software taking into account 
atmospheric dispersion with an attempt to optimize targets
and observing time.  We employed the G600~grating, which yields
a spectral resolution $R \approx 1600$ for our $1''$ slits,
a dispersion of $\approx 0.5$\AA\ per pixel,
and an approximate wavelength coverage of
$\lambda _{\rm ob} \approx 5,000-10,000$\AA.
The spectral images include arc and quartz lamp calibration frames.
All of these data were reduced with the {\sc spec2d} data
reduction pipeline
developed by M. Cooper for the DEEP survey \citep{Newman2013}.
The pipeline produces optimally extracted, wavelength-calibrated 
spectra  (in air and converted later to vacuum).
Multiple exposures taken with a given mask on the same night
are combined in 2D by the DRP.
Masks exposed on separate nights were extracted
separately and the individually,
extracted 1D-spectra were 
coadded with a custom algorithm.

From observations of two spectrophotometric
standards, Feige 110 and 34, on three separate nights, November 14, 2012, 
December 13, 2012 (Feige 110), and May 8, 2013 (Feige 34) we generated a combined sensitivity function that was applied to the entire CASBaH DEIMOS spectroscopic dataset. Again, we made no correction for slit losses,
airmass, or variable observing conditions.
Furthermore, vignetting at the edges of the detector
leads to fluxing error at the longest and shortest
wavelengths. 

\begin{figure}
\begin{center} 
\includegraphics[width=3.5in]{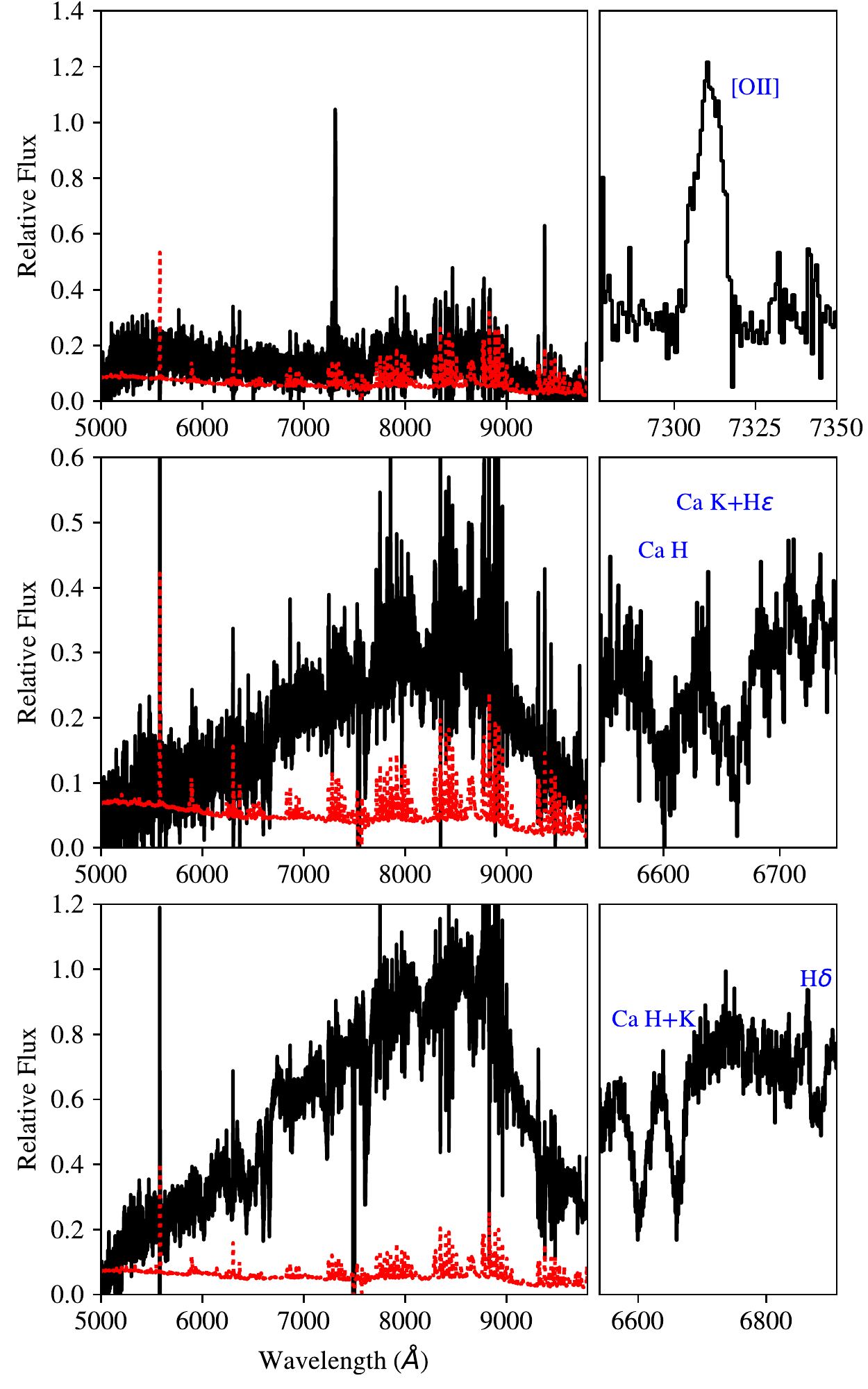}
\caption{
Gallery of example spectra from the Keck/DEIMOS observations,
ordered from top to bottom by increasing flux.
Cut-outs on the right show examples of key spectral features used in the
redshift determination.
}
\label{fig:deimos_spectra}
\end{center}
\end{figure}

Representative spectra spanning the dynamic range of sources
observed with DEIMOS are illustrated in Figure~\ref{fig:deimos_spectra}.
The median S/N of the complete dataset 
at $\lambda _{\rm ob} \approx 6,500$\AA\ is $\approx \meddeimossn$
per 0.6\AA\ pixel.

The standard mode of the {\sc spec2d} extraction algorithm
searches for additional sources in each slit (in part to 
assist sky subtraction) and extracts them.
We have used the mask information 
and our astrometry to assign
each an RA and DEC coordinate.  Where possible, we have also
matched them to our photometric catalog.
All of these `serendipitous' spectra are included in the database
and were analyzed in a similar manner as the
primary targets.

\subsection{Redshift Analysis}

Redshift analysis proceeded in two stages. 
The first stage employed a template-fitting
algorithm custom to the spectrograph (see
following sections).  These
results were vetted by one or more
co-authors and a quality flag $Z_Q$
was assigned to each source with
$Z_Q = 0,1,2,3,4$ as follows:
(0) Data too poor for any assessment;
(1) Data poor and redshift highly uncertain;
(2) Data quality good but no redshift determined;
(3) Data quality good and redshift is highly likely
but not confirmed by multiple lines (or one line is
of low S/N);
(4) Highly certain and confirmed by multiple spectral
features \citep[see also][]{Newman2013}. 
We then analyzed each spectrum with
the {\sc RedRock} software 
package\footnote{https://github.com/desihub/redrock}
v0.7, which is under development by the Dark
Energy Spectroscopic Instrument project.
This code also compares a set of galaxy and
star templates to the spectra to generate a set
of best-fitting models and corresponding
redshift estimates.  We then inspected every
{\sc RedRock} solution offset by more than $50 \mkms$ from
the original estimate and resolved the conflict
based on the observed spectral features (if any).
Only sources with $Z_Q \ge 3$ are considered
reliable.
We now detail specific aspects of the analysis
and results for each instrument.

\subsubsection{Hectospec}

\begin{figure}
\begin{center} 
\includegraphics[width=3.5in]{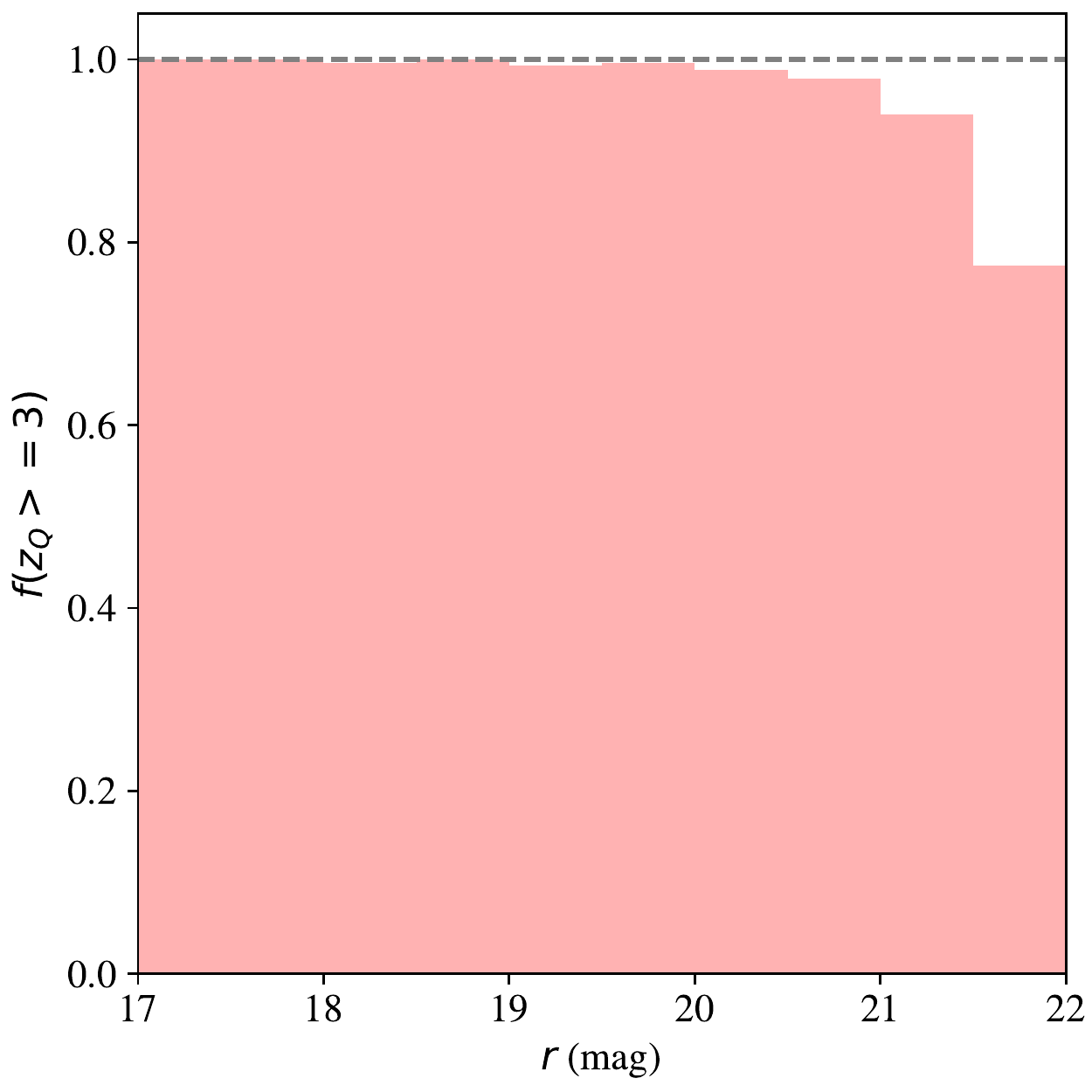}
\caption{
Redshift completeness fraction for sources observed with
Hectospec as a function of $r$ magnitude.  
}
\label{fig:hecto_zQ}
\end{center}
\end{figure}

The first algorithm adopted for Hectospec redshift calculation is 
a modified version of {\sc zfind} developed by the 
SDSS project, which
cross-correlates a series of templates (stars, galaxies and quasars) in 
measurement space (i.e., no Fourier transform is applied).
These are then ranked in $\chi^2$ and the lowest value is selected by the redshift code.
Figure~\ref{fig:hecto_zQ}
summarizes the fraction of sources
with $Z_Q \ge 3$ as a function of observed magnitude.
For $r < 21$\,mag, the redshift success rate exceeded 95\%\
and, as expected, 
declined with fainter sources.  Nevertheless, the 
success rate remains high to the magnitude-limit of the survey.

Spectra that showed redshift discrepancies of the order of 50\kms\ 
between the RedRock and Hectospec solutions were visually inspected, and the solution that showed the best matching locations of certain prominent galaxy lines was selected. 
We found that the majority of cases favored the RedRock determination, though in cases where the S/N was low, the Hectospec/{\sc zfind} determination frequently showed the higher confidence redshift. 

The reported redshift uncertainty from these $\chi^2$-minimization
codes is frequently several times $10^{-5}$, i.e., $\sigma_v \approx 10 \mkms$.
We consider such small errors to be overly optimistic and suggest
one assume a minimum of $30 \mkms$ uncertainty due to systematic
error (e.g.,\ wavelength calibration).

\subsubsection{DEIMOS}

For the DEIMOS spectra, we derived redshifts with two separate
algorithms with two different sets of co-authors:
(i)  a modified version of the SDSS {\sc zfind} algorithm
by JW and SL and
(ii) the {\sc RedRock} software package v0.7
developed by the DESI experiment. 
For nearly 500 sources,
the two packages reported redshifts that matched
to within $50 \mkms$.  For these we simply adopted
the {\sc RedRock} estimate and assigned $Z_Q=3$.
For all other cases (approximately 600~sources,
the majority of which had no previously determined
redshift),
we vetted each of these manually.  After visually 
inspecting the spectra, we assigned the solution
with coincident, prominent spectral features.
In a few percent of the cases, we assigned a redshift
of our own estimation.  This included most of the galaxies
at $z>1$ because we had limited our automatic search to less
than this redshift.

\begin{figure}
\begin{center} 
\includegraphics[width=3.5in]{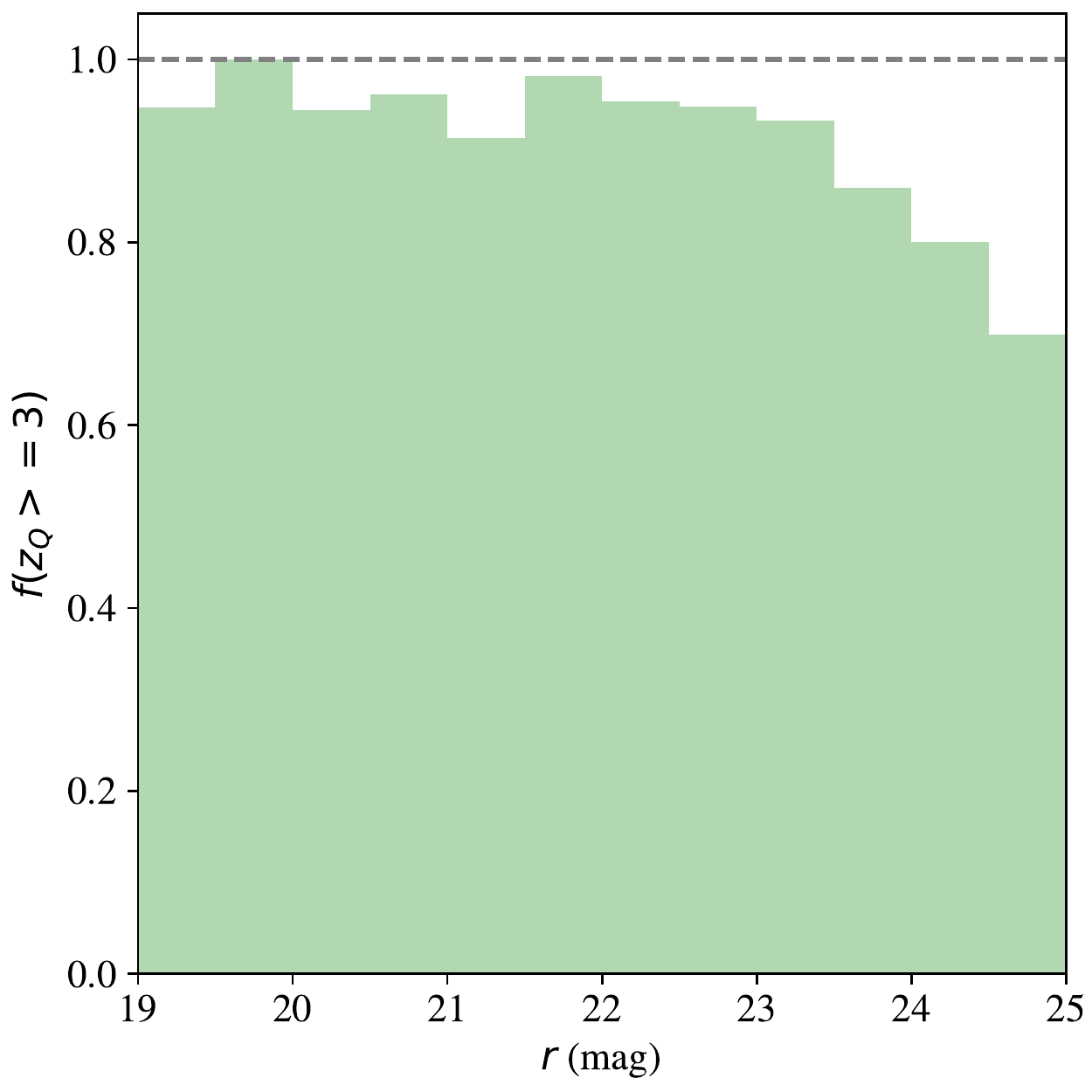}
\caption{
Redshift success rate for sources observed with
DEIMOS as a function of $r$ magnitude.  
}
\label{fig:deimos_zQ}
\end{center}
\end{figure}

Figure~\ref{fig:deimos_zQ} shows the redshift 
measurement success rates from our DEIMOS analysis where a secure redshift
is defined with $Z_Q \ge 3$.  This analysis is limited 
to the primary targets, i.e.,\ we ignore 
serendipitous sources that entered the DEIMOS slits.  
The success rate is 
approximately 95\%\ to 22~mag, declines to $\approx 80\%$
at 24~mag, and drops rapidly from there.

\subsection{Galaxy Spectroscopy Summary}

Integrating the redshift measurements from Hectospec, DEIMOS,
and the SDSS database, we performed internal comparisons 
between the $\sim 175$ sources common to two or more of the
sub-surveys.  Ignoring catostrophic failures (described below),
the measured RMS values between Hectosec/SDSS and DEIMOS/Hectospec
are $\approx 35 \mkms$ and $\approx 36 \mkms$ respectively.
Therefore, we advise one adopt a minimum redshift uncertainty of 
$35 \mkms$ for galaxies drawn from the CASBaH database.
We find no variation with emission/absorption-line
properties among this common set of galaxies.
This exercise also revealed 13~cases where the 
recovered redshifts were substantially offset
($\delta v \gg 100 \mkms$).
In nearly all of the cases, at least
one of the two spectra was of poor data quality.
All but one of the non-SDSS spectra already showed
$Z_Q < 3$ and we have now downgraded the few from
SDSS accordingly.  In two cases, there were multiple
sources with separation $<1''$ in the slit/fiber; 
we have corrected the catalog accordingly.

\begin{figure}
\begin{center} 
\includegraphics[width=3.5in]{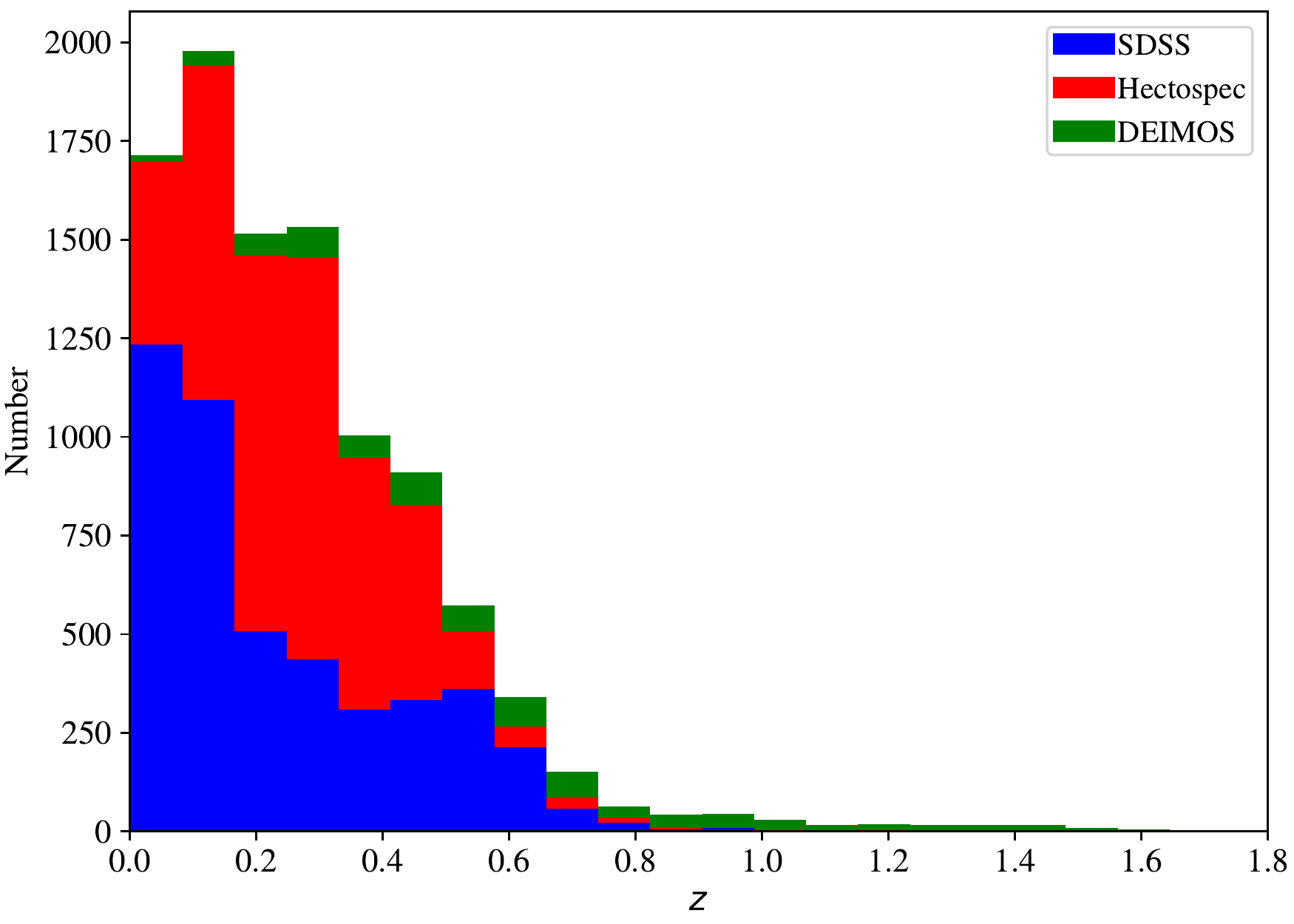}
\caption{
Histogram of the galaxy redshifts from the three main
sources for the CASBaH survey: SDSS, Hectospec,
and DEIMOS.  The histograms are stacked, not overlapping,
i.e.,\ the DEIMOS histogram is placed above Hectospec.
}
\label{fig:z_hist}
\end{center}
\end{figure}

The CASBaH spectroscopic redshift survey is summarized in Table~\ref{tab:z}. Altogether in the \ncasbah\ fields, we have
\ngal\ galaxies with high quality redshifts ($Z_Q \ge 3$)
and $z_{\rm em} > 0.00166$.  Their redshift distribution
is shown in Figure~\ref{fig:z_hist} for the three primary datasets
of our program\footnote{The full database includes several
sources observed with Keck/LRIS, several with MMT/BCS, 
and a small set of galaxies
from \cite{johnson+15} that they kindly made public.}.
Clearly, our Hectospec survey provides the majority of
spectroscopic redshifts.  These lie predominantly at 
$z_{\rm em} = [0,0.5]$.  The DEIMOS dataset
contributes primarily at $z_{\rm em}>0.5$, as designed.
We also report \nbad\ sources with $Z_Q < 3$ and no
secure redshift measured as well as \nstars\ spectra from DEIMOS
and Hectospec of stars.

\begin{figure*}
\begin{center} 
\includegraphics[width=6.5in]{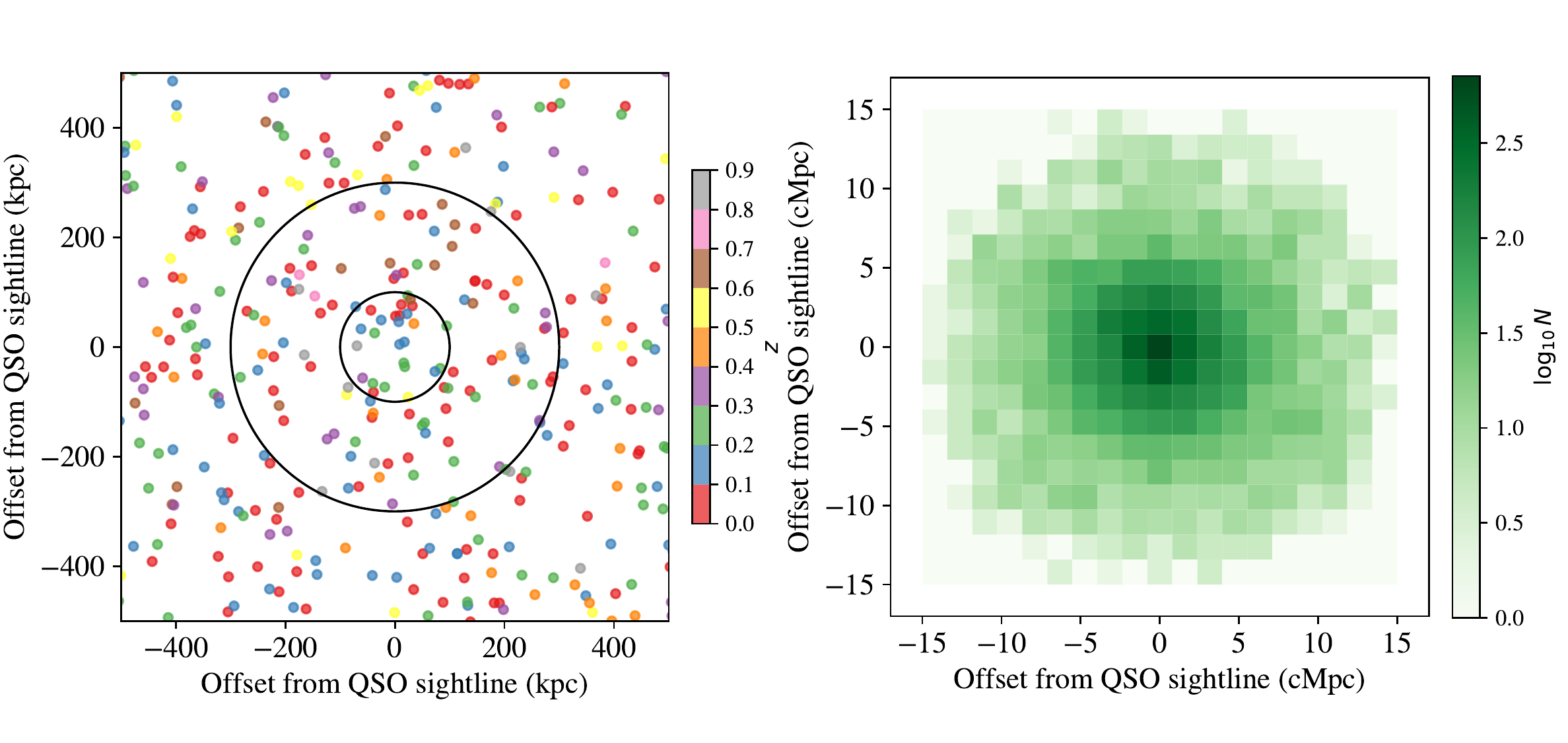}
\caption{
Distributions of galaxies around the quasar sightlines of
CASBaH on small-scales (left), which probe the circumgalactic
medium of the galaxies and on large scales (right), which
trace the distribution of the intergalactic medium.  Annuli in
the left panel indicate impact parameters of 100 kpc and 250 kpc.
}
\label{fig:bull}
\end{center}
\end{figure*}

With our adopted cosmology, we may translate the galaxy
redshifts and their angular offsets from their corresponding
quasar sightline to estimate the impact parameter (physical
$R_{\rm phys}$ and comoving $R_{\rm com}$).
The distributions on small ($R_{\rm phys} < 450$\,kpc)
and large ($R_{\rm com} \sim 10$\,cMpc) scales are summarized
in Figure~\ref{fig:bull}.
These distributions show the statistical power of 
CASBaH for studies of the CGM and IGM, respectively.

\input{tab_sub_redshifts}

\section{Derived Quantities}
\label{sec:derived}

The previous sections described measurements made
(nearly) directly from the imaging and spectroscopic
data.  This section describes a few quantities derived
from the combined measurements, e.g.,\ two or more filters
and/or a combination of photometry and spectroscopy.
We generally adopt standard techniques, assumptions,
and software in the analysis and warn that the
uncertainties (especially systematic error) can be 
large.

\subsection{SED Fitting}

To estimate the stellar masses and SFRs of all galaxies in our database, we have employed our spectroscopic redshifts, the photometric measurements from our own LBT/LBC observations, and photometry from various publicly available survey catalogs spanning the optical and near-infrared.  We fit the spectral energy distributions (SEDs) of each galaxy with stellar population model spectra, models for dust attenuation, and emission from nebular lines using the CIGALE software package \citep{Noll:2009aa}.  While myriad high-quality SED fitting codes are available, we chose CIGALE for several reasons, 
including its ability to easily handle our heterogeneous dataset given the variety of sources from which our data are derived (e.g.,\ certain fields were covered by a given survey while others were not, and within a field, not all objects were detected in a single survey). 
Furthermore, CIGALE natively supports several choices for 
stellar-population models, dust models, etc., across a wide range of parameters.  We discuss our choices for these parameters below so that the reader may recreate or improve upon our estimates.

We assembled our photometric dataset for SED fitting by cross-matching all galaxies in our spectroscopic database having redshift quality scores $Z_Q \ge 3$ 
with several public survey catalogs:  SDSS DR12 \citep[][$ugriz$ coverage of all fields except PHL1377]{Alam:2015kq}, the Canada-France-Hawaii Telescope Legacy Survey \citep[CFHTLS;][$ugriz$ coverage of the PHL1377 field]{Hudelot:2012aa}, the Wide-field Infrared Survey Explorer all sky survey \citep[][3.4, 4.6, 12, and 22 $\mu$m bands for all fields]{Cutri:2013aa}, BASS \citep{Zou:2017aa,Zou:2017ab}, UKIDSS \citep{Warren:2007aa, Lawrence:2007aa}, KiDS \citep{de-Jong:2017aa}, DECALS \citep{Dey:2018aa}, MzLS \citep{Dey:2018aa}, and Pan-STARRS \citep{Flewelling:2016aa,Chambers:2016aa}.

We then corrected all magnitudes for Galactic reddening using the \citep{Schlafly:2011aa} extinction values provided by the Nasa Extragalactic Database\footnote{The NASA/IPAC Extragalactic Database (NED) is operated by the Jet Propulsion Laboratory, California Institute of Technology, under contract with the National Aeronautics and Space Administration.} service accessed through the \textsc{astroquery} framework.  Reddening values are returned for a limited number of filter bandpasses (UBVRIugrizJHKL'), and we dereddened our photometry data using those most closely matching the central wavelengths of filters used in our compiled dataset.  CIGALE integrates fitted stellar population models over any filter response curve provided; we employed the SVO Filter Profile Service \citep{svoFilterService} as well as individual survey websites to obtain filter curves and zeropoints corresponding to each band.

As mentioned above, CIGALE includes a variety of models to include in the fitting.  For stellar populations, we used the \citet{Bruzual:2003aa} models, assuming a \citet{Chabrier:2003pd} initial mass function, with metallicities ranging from $<1/100$th solar to 2.5x solar.  The star formation histories included in our models (via the `sfhdelayed' module) spanned 0.25-12 Gyr for the oldest population with e-folding times 0.1-8 Gyr.  In addition to the stars, we included nebular emission with the default values and reprocessed dust emission using the \cite{Dale:2014aa} model with slopes $\alpha = 1-2.5$ and 0\% AGN fraction.  

The final component of our modeling is the dust attenuation curve, for which we adopted a model based on that of \citet{Calzetti:2000lr} but also 
includes a `bump' in the UV.  The \citet{Calzetti:2000lr} form, although originally derived for starburst galaxies, was later validated by \cite{Battisti:2016aa} for local star forming galaxies more generally, and \citet{Battisti:2017aa} found evidence for a bump feature in inclined galaxies.  We adopt the modification by \citet{Buat:2011aa} and implemented in CIGALE by \citet{Buat:2011aa} and \citet{Lo-Faro:2017aa} with 35.6 nm and 217.5 nm the bump width and wavelength, respectively.  The bump amplitude setting is 1.3, and the slope is modified by $-0.38$.  
We reduce the color excess of old population relative to the young stars by a factor of 0.44 and allow the $E(B-V)$ 
for the young population to vary between 0.12 and 1.98.

\subsection{Stellar Mass, SFR, and Rest-frame Color}
\begin{figure}
\begin{center} 
\includegraphics[width=3.5in]{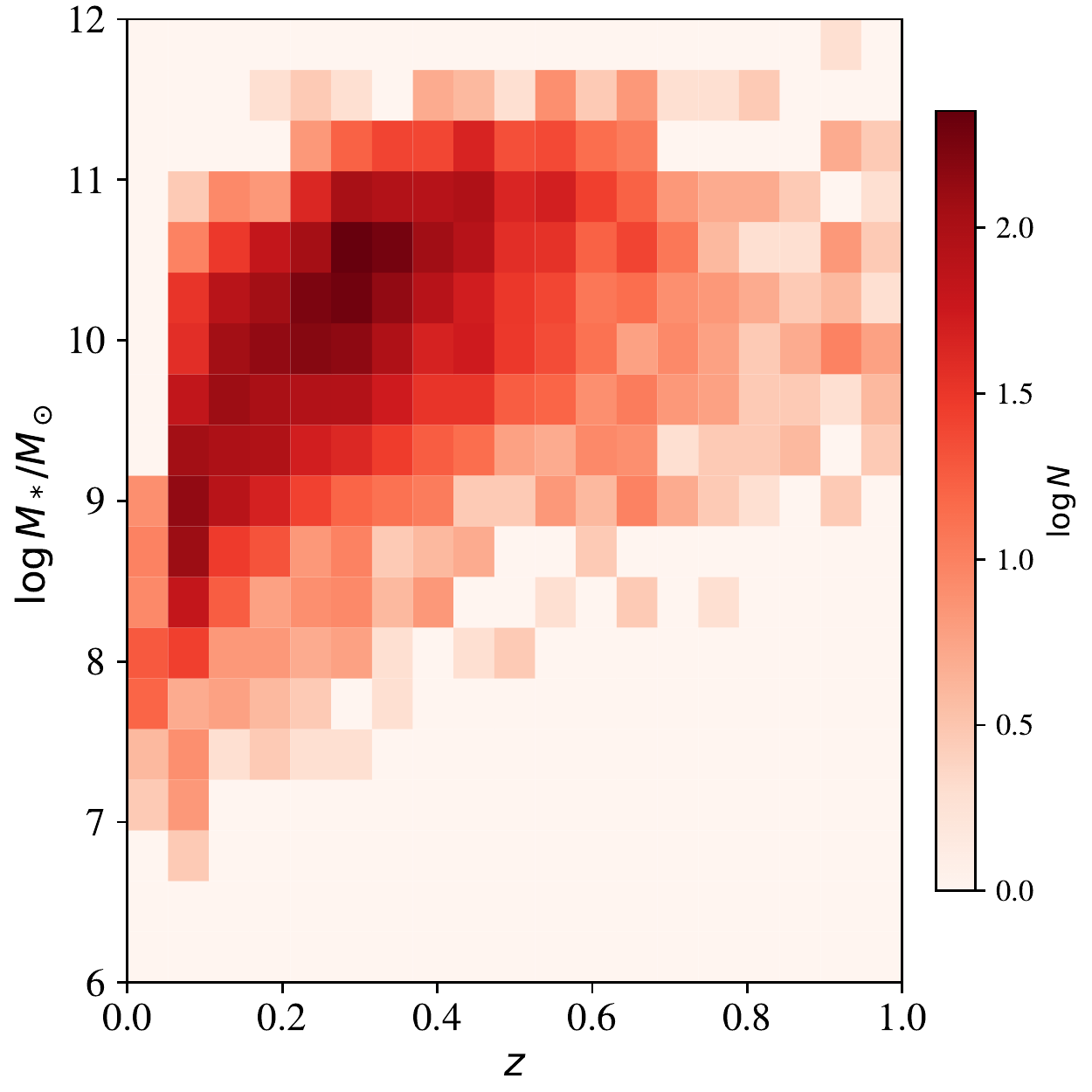}
\caption{
Distribution of stellar mass vs.\ redshift for the spectroscopically
confirmed galaxies comprising the CASBaH sample.  The 
median of the sample occurs
at $\bar z=\zmed$ and $\bar M_* \approx \Mmed \msol$.
}
\label{fig:stellar_mass}
\end{center}
\end{figure}

\begin{figure}
\begin{center} 
\includegraphics[width=3.5in]{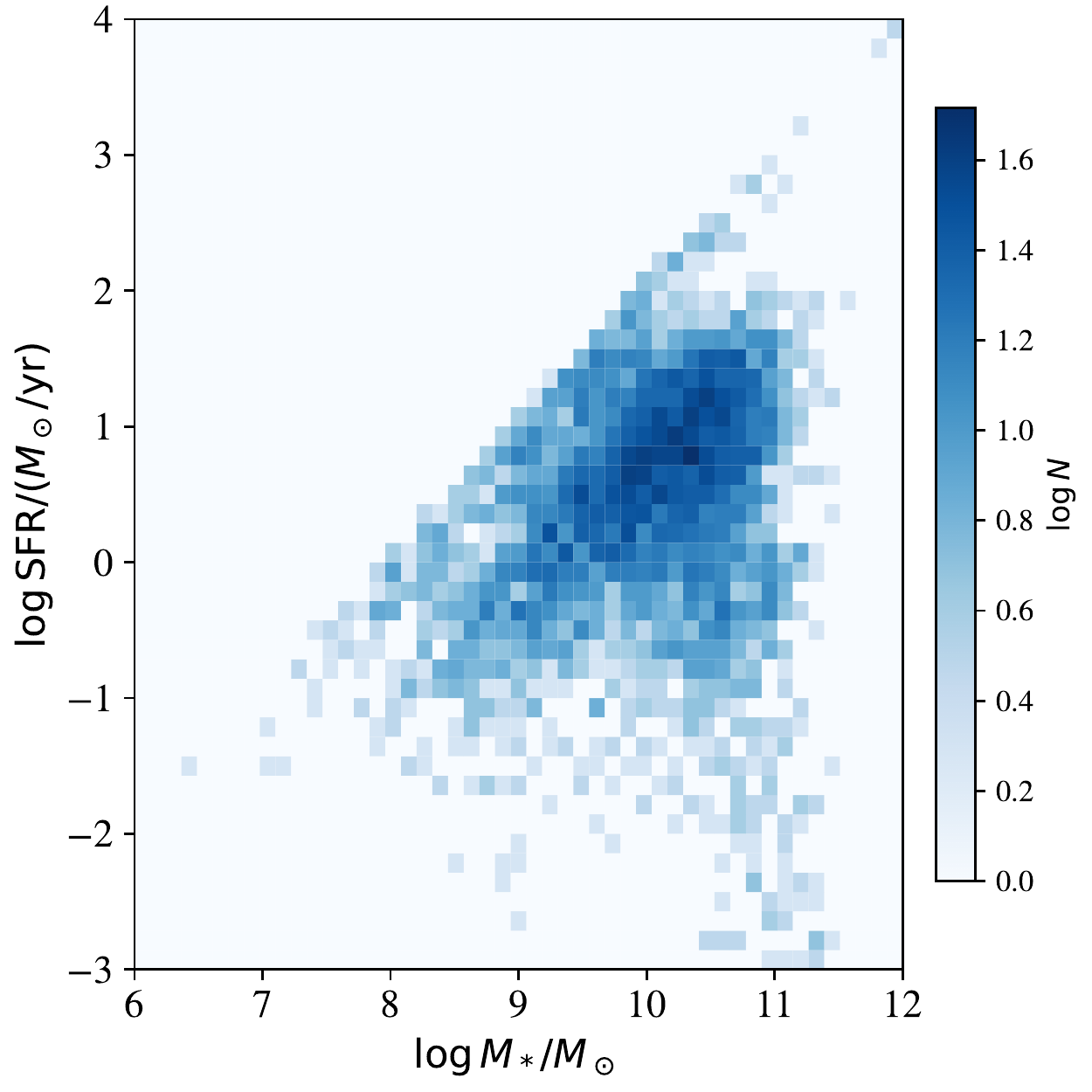}
\caption{
SFR vs.\ stellar mass from SED fitting for the CASBaH galaxy sample, where
one well recognizes the locus of galaxies 
known as the star-forming main-sequence.  The hard diagonal
edge at the highest SFR for each stellar mass bin occurs due to failures/outliers
fitted with models at the boundaries of parameter space.
}
\label{fig:SFR}
\end{center}
\end{figure}

\begin{figure}
\begin{center} 
\includegraphics[width=3.5in]{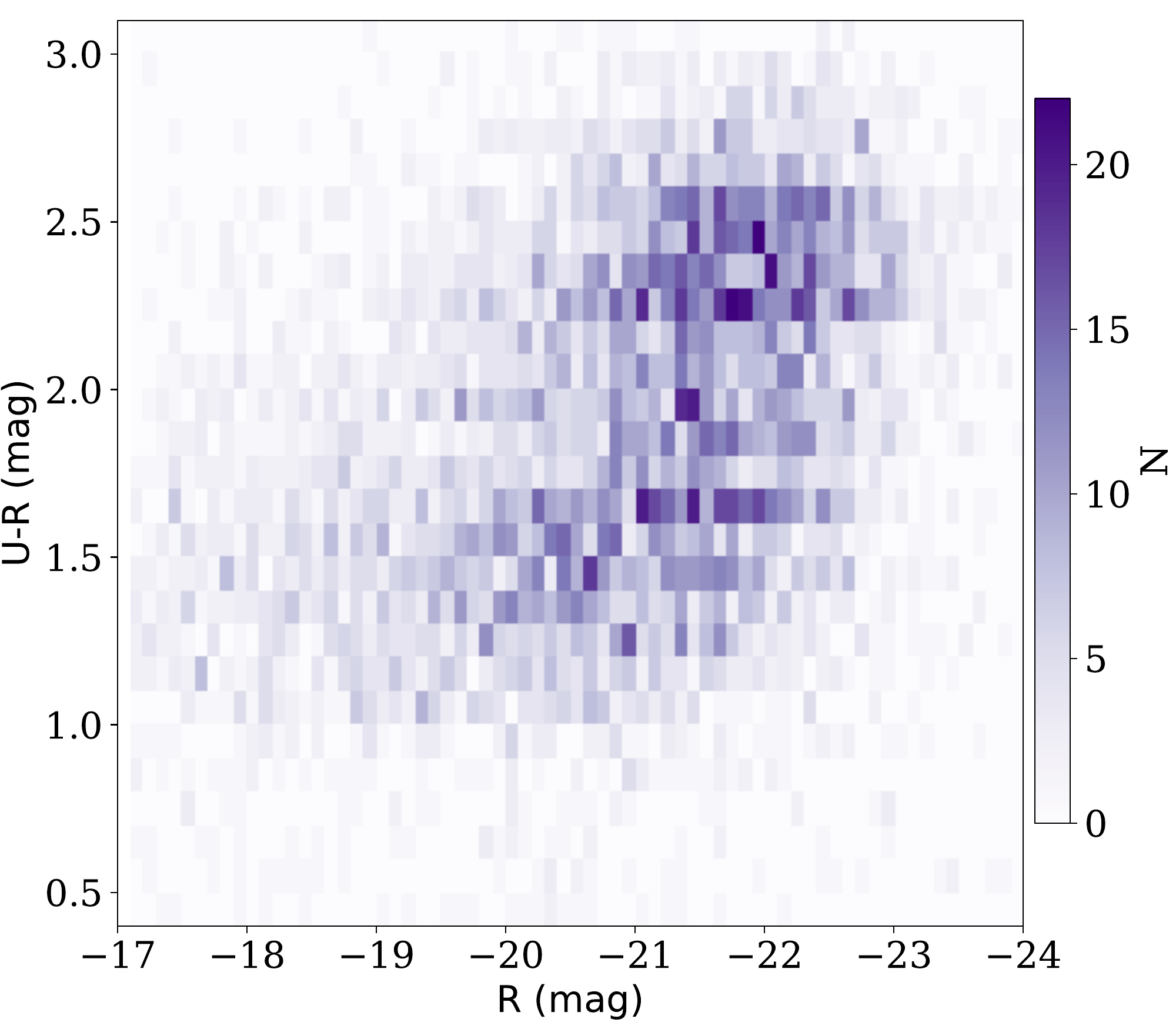}
\caption{
Color-magnitude diagram (rest-frame) for 
galaxies comprising the CASBaH sample.
Evident is the well-known bimodal populations 
of star-forming and `red-and-dead' galaxies.
}
\label{fig:color_mag}
\end{center}
\end{figure}

Once each galaxy's SED has been fitted, a number of physical parameters as well as the
rest-frame absolute magnitudes are then
extracted from the resulting stellar population models.
Figures~\ref{fig:stellar_mass}-\ref{fig:color_mag}
illustrate the measurements gleaned from SED fitting
for the complete CASBaH galaxy database.  As shown in Figures \ref{fig:stellar_mass}
and \ref{fig:SFR}, 
the locus of measured stellar masses spans
from $M_* \approx 10^8 - 10^{11} \msol$ with a median 
$\bar M_* \approx \Mmed \msol$.
at $\bar z=\zmed$.
These values are driven by our Hectospec sample.
The SFRs in Figure~\ref{fig:SFR} show that the majority
of our galaxies are star-forming.  We have compared our 
results to measurements for other $z<1$ surveys
\citep[e.g., PRIMUS;][]{moustakas+13} and find similar results.
Lastly, the color-magnitude diagram (Figure \ref{fig:color_mag} 
reveals the bimodal populations
of star-forming and red-and-dead galaxies.  In summation,
the basic measured properties of galaxies discovered in our survey 
reproduce/follow the typical trends
and distributions characteristic of  any other low-$z$ sample.

\section{Identification and Measurement of \ion{O}{6} Absorbers}

In the following section, we will study the study the clustering of \ion{O}{6} absorbers with galaxies using the CASBaH database.  The full description of the CASBaH ultraviolet QSO spectroscopy and data handling is presented in Tripp et al. (2019, in prep.); in this section, we summarize aspects of the \ion{O}{6} identifications and measurements that are important for the absorber sample definition and \ion{O}{6}-galaxy clustering analysis.  



We identified the \ion{O}{6} absorption lines using the multipass line-identification procedure described by \citet{tripp08}.  In the first pass through the data, we simply searched for lines with the signature of the \ion{O}{6} doublet, i.e., lines with the relative separation and relative strengths of the \ion{O}{6} 1031.926 and 1037.617 \AA\ transitions \citep[see, e.g.,][]{Verner94}.  In this pass, we did not require detection of any corresponding \ion{H}{1} or metal lines; we only searched for the \ion{O}{6} doublet by itself.  This first pass identified the majority of the \ion{O}{6} absorbers, but in some cases, evidence of blending with lines from other redshifts was clearly evident.  This is not surprising given the moderately high density of lines in the CASBaH QSO spectra \citep[see Fig. 1 in][]{Tripp13}.  In addition, in some cases, one line of the doublet is so severely blended with a strong feature from a different redshift that both lines of the doublet could not be directly recognized in our first pass through the data.  To overcome these blending issues, we iteratively made subsequent passes through the spectra in which we added information about absorption systems established by the presence of \ion{H}{1} lines (often with many lines of the Lyman series) and various metals.  CASBaH absorbers often show many metal and \ion{H}{1} lines with distinctive component structure \citep[e.g.,][]{tmp+11,ribaudo11,lht+13,mtw+13}, and we used the detailed correspondence of candidate \ion{O}{6} lines with other metals at the same redshift to identify additional \ion{O}{6} absorbers when one or both of the \ion{O}{6} lines was affected by blending \citep[for examples with \ion{Ne}{8} lines, see][]{burchett19a}.   Since the CASBaH spectra fully cover the \ion{H}{1} Ly$\alpha$ region from $\lambda _{\rm ob}$ = 1216 to $(1 + z_{\rm QSO})\times 1216$, we are able to identify nearly all of the lines and systems in the spectra (not just the \ion{O}{6} systems), including the lines that are blended with the \ion{O}{6} doublets. Often the interloping lines that are blended with an \ion{O}{6} absorber can be modeled and removed based on lines recorded elsewhere in the CASBaH spectra, and then comparison of the deblended data further corroborates the identification.

In this paper, we are only interested in the foreground/intervening absorption systems (far from the background QSOs) and their relationships with foreground galaxies.  To avoid contaminating our intervening-absorber sample with ``proximate'' ($z_{\rm abs} \approx z_{\rm QSO}$) absorbers that often comprise material ejected by the QSO that is close to the QSO's central engine \citep{misawa+07,ganguly+13}, we exclude \ion{O}{6} absorbers within 5000 km s$^{-1}$ of the QSO redshift.\footnote{In addition, in a few cases we exclude systems at somewhat higher ejection velocities when they exhibit obvious characteristics of ``mini broad absorption line (BAL)'' systems such as partial covering, smooth absorption profiles that are much broader than intervening absorption profiles, and strong absorption by exotic and highly ionized species such as \ion{Na}{9} and \ion{Mg}{10} \citep[see examples in][]{muzahid+13}. However, inclusion or exclusion of these mini-BAL systems has no impact on the analysis in this paper because they are all at higher redshifts than our \ion{O}{6}-galaxy clustering sample.}  

All of the intervening \ion{O}{6} absorbers that we identify exhibit \ion{H}{1} absorption at very similar velocities and often show other metal lines (at least \ion{C}{3} or \ion{O}{4}, and often other metals); we do not find any unambiguous \ion{O}{6} systems without any corresponding \ion{H}{1}.  This is consistent with previous studies of low-redshift \ion{O}{6} absorbers, which have shown that the ``\ion{H}{1}-free'' \ion{O}{6} systems are only found in proximate cases with $z_{\rm abs} \approx z_{\rm QSO}$ \citep{tripp08}.  We note that some intervening \ion{O}{6} systems have interesting {\it individual} \ion{O}{6} components that have corresponding \ion{H}{1} that is very weak \citep[or absent altogether, see, e.g.,][]{tripp08,savage10}, but these cases are not \ion{H}{1}-free; on the contrary, these absorbers have very strong corresponding \ion{H}{1} absorption in some of the components of the overall system.  For example, \citet{savage10} have analyzed the \ion{O}{6} system at z = 0.167 in the spectrum of PKS0405-123.  This system includes an \ion{O}{6} component at $v = -278$ km s$^{-1}$ that is not detected in \ion{H}{1}, but the same system also shows \ion{O}{6} components at $v \approx -125, -50,$ and 0 km s$^{-1}$, and these lower-velocity components have strong corresponding \ion{H}{1} absorption \citep[see Fig.3 in][]{savage10}.  Thus, this example has an \ion{H}{1}-free component, but it is not an \ion{H}{1}-free system; similar intervening systems are found in the CASBaH database.  Only the proximate absorption {\it systems} are entirely free of \ion{H}{1} in all of the components \citep[some examples of such proximate systems are shown in Fig.7 and Fig. 20 of][]{tripp08}.

\begin{figure*}
\includegraphics[width=17cm,angle=0]{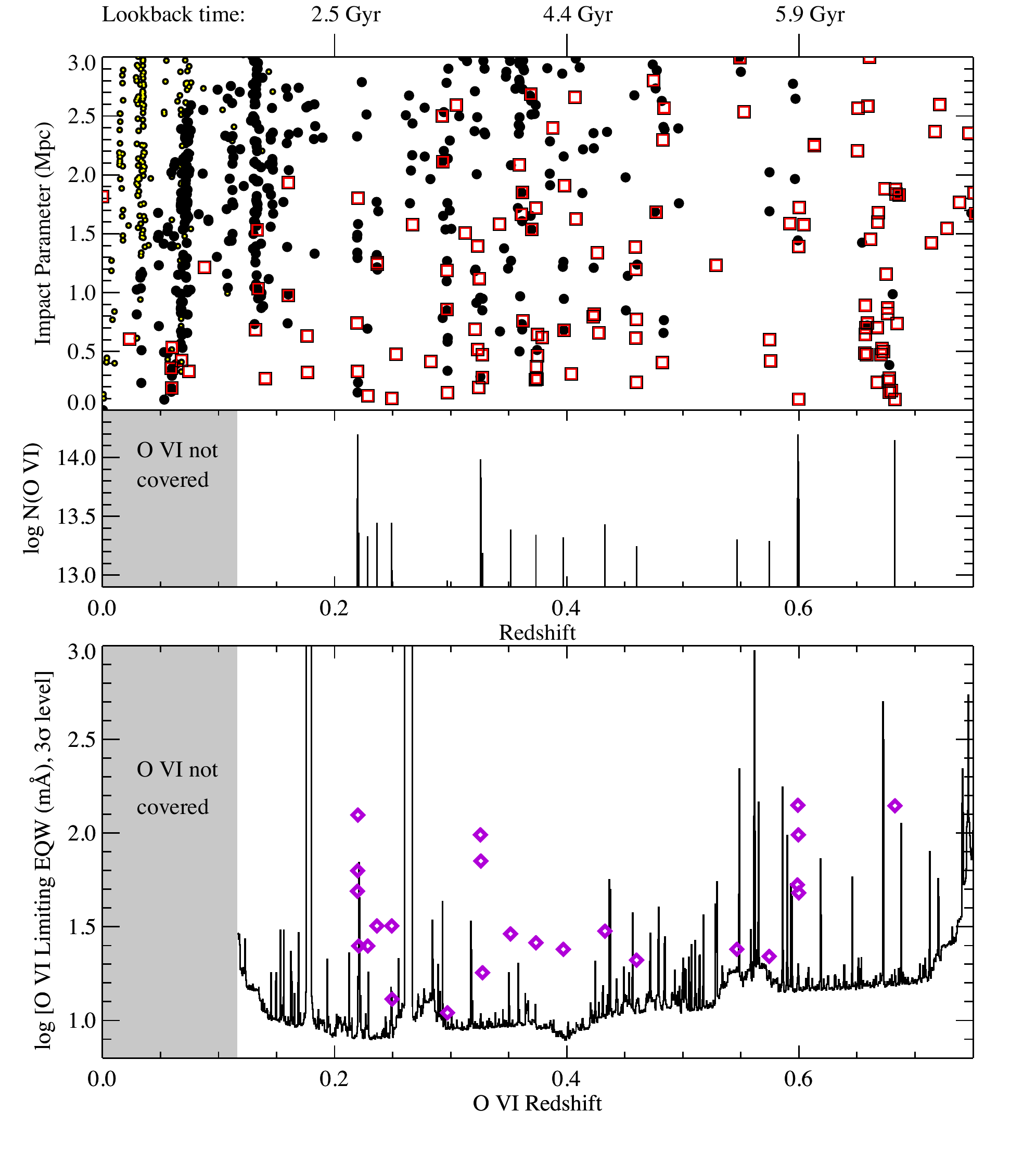}
\caption{\textsl{Top:} Galaxies in the CASBaH database for PG1407+265 as a function of redshift, within impact parameters of 3 Mpc or less at $z_{\rm gal} < 0.75$. Galaxies with redshifts from SDSS are indicated by small circles with yellow centers, galaxy redshifts measured with Hectospec are represented with larger filled-black circles, and galaxies measured with DEIMOS are shown with red squares. For reference, lookback times are indicated on the top axis. \textsl{Middle:} logarithmic column densities and redshifts of the \ion{O}{6} absorption lines detected in the UV spectra of PG1407+265.  Each vertical line marks the redshift of an \ion{O}{6} component, and the height of the line indicates log $N$(\ion{O}{6}). \textsl{Bottom:} The 3$\sigma$ limiting equivalent width for detection of \ion{O}{6} in the PG1407+265 data vs. redshift.  The equivalent widths of detected \ion{O}{6} components are plotted with purple diamonds.\label{fig_o6_montage}}
\end{figure*}

To measure the absorption-line properties, here we primarily rely on fitting multicomponent Voigt profiles to the data using the software developed by \citet{burchett+15}.  These models constrain the redshift, line width, and column density of each \ion{O}{6} component, and we aggregate the components into ``systems'' as described below.  To assess the significance of the candidate lines, we use the Voigt-profile parameters to calculate the equivalent width, which we then compare to the limiting equivalent width \citep[calculated using the method of][]{tripp08} at that wavelength.  In this paper we have focused on well-detected lines ($>4\sigma$ significance) that are not involved in very complicated blends that preclude robust profile fitting. In this conservative sample, we find a paucity of lines with log $N$(\ion{O}{6}) $<$ 13.5, and we also impose a lower limit on the \ion{O}{6} column density of absorbers that are included in the clustering analysis (see below).

Figure~\ref{fig_o6_montage} provides a snapshot of the overall database (galaxy and \ion{O}{6} absorber information) using the PG1407+265 sight line and field as an example.  In the top panel of this figure, we show the full galaxy database in a cylinder with radius = 3 Mpc centered on the QSO.  The middle panel illustrates the column densities of the \ion{O}{6} lines are their locations with respect to the nearby galaxies and large-scale structures that can be seen in the upper panel, and the lower panel shows the 3$\sigma$ limiting (rest-frame) equivalent width as a function of \ion{O}{6} redshift with the equivalent widths of the detected \ion{O}{6} lines overplotted.  Comparing the \ion{O}{6} and galaxy impact parameters, we see that \ion{O}{6} lines are typically detected when there is a galaxy close to the line of sight, but there is substantial scatter in the \ion{O}{6} column at a given impact-parameter value.  This is consistent with previous studies \citep[e.g.,][]{stockeetal06,ttw+11,johnson+15}.  We will present analyses of the connections of individual \ion{O}{6}-galaxy papers in a separate paper (Burchett et al., in prep.).   We also see from Figure~\ref{fig_o6_montage} that the CASBaH database provides extensive information on galaxy structures on large scales, which is the focus on the galaxy-absorber analyses in this paper.

\section{OVI-Galaxy Clustering}
\label{sec:clustering}

With the galaxy spectral database of CASBaH constructed,
we proceed to measure the clustering of CASBaH
galaxies with themselves (auto-correlation) 
and with \ovi\
absorption systems (cross-correlation)
in the $z<1$ Universe.
Our scientific motivations are two-fold:
(i) to further characterize the galaxy sample
of the CASBaH survey;
and
(ii) to provide new estimates on the masses of galaxies
associated with \ovi\ absorption.
Regarding the latter goal, we emphasize that the
results derived will only apply for \ovi\ systems
with properties similar to those drawn from CASBaH.
Furthermore, any estimate on mass follows from 
the ansatz that the majority of these \ovi\ systems
arise within dark matter halos.
We also emphasize that incompleteness in 
the galaxy and \ovi\ samples is accounted for
by the estimator and procedures adopted in the
clustering analysis.
Last, we restrict the analysis to comoving separations
in excess of $1 \hmpc$ to isolate the so-called
two-halo term that results from large-scale clustering.

\subsection{Setup}
\label{sec:setup}

This sub-section describes cuts on our absorber
and galaxy databases
used to generate a well-defined set of systems
and galaxies for the analysis.
To measure \ovi-galaxy clustering, 
we define a discrete
set of \ovi\ systems along the sightlines, 
with each system 
characterized by a systemic redshift \zsys\ and 
column density \novi.  
To achieve this, our approach in this CASBaH paper is to 
synthesize components \citep[provided by][]{casbah}
into
absorption systems. 
Using the MeanShift clustering algorithm from 
the scikit-learn Python package \citep{scikit-learn}, 
we grouped components clustered in velocity space into absorption systems, setting the redshift of each system to the center of its component cluster. 
The MeanShift algorithm accepts a ‘bandwidth’ argument, which defines an approximate width within which to group components. We chose a bandwidth value 
of 600 \kms\ for component clustering, finding that this value was large enough to produce systems with components spread over a 
large range of velocity ($\sim 1000$ \kms),
such as the post-starburst outflow analyzed by \cite{tmp+11}, 
but small enough to separate systems with components clustered about 
discrete center points with large separations (also
on the order of 1000\kms). 
For systems composed of multiple components, we 
sum their individual column densities
to yield a total \novi\ for each system.

\begin{figure}
\begin{center} 
\includegraphics[width=3.5in]{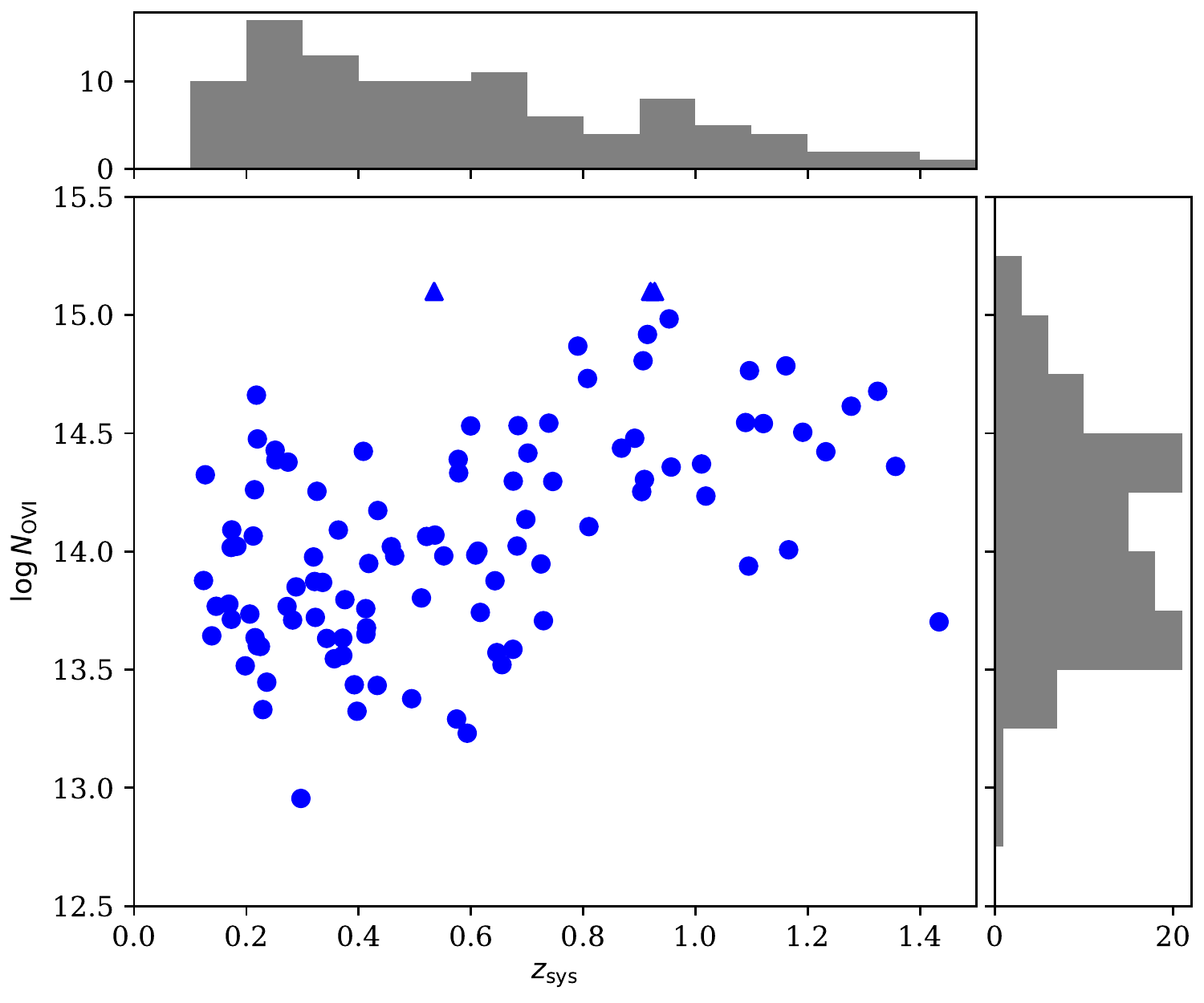}
\caption{
Scatter plot and histograms of the column densities
versus redshift for the \ovi\ systems detected
in the CASBaH survey. 
Typical uncertainties in $\N{OVI}$ are $\approx 0.03$\,dex.
The measurements with $\N{OVI} > 10^{15} \cm{-2}$
are saturated, are reported as lower limits, and 
are designated with triangles. 
The lack of of weaker \ovi\ absorbers at $z>0.75$ is driven by the shift of the \ovi\ doublet from the FUV to the NUV bandpass where the spectra have lower signal-to-noise levels. 
}
\label{fig:NOVI_vs_z}
\end{center}
\end{figure}

These values are shown in Figure~\ref{fig:NOVI_vs_z}
as a function of system redshift.
At $z < 0.75$, the sample scatters from 
$\N{OVI} \approx 10^{13.3} - 10^{14.5}$ with no discernible
redshift evolution (the cut-off at $z \approx 0.12$
is due simply to lowest observed wavelength of our FUV spectra, $\lambda _{\rm ob}$ = 1152 \AA ).
At higher redshifts, the
$\N{OVI}$ values are systematically higher \citep[see discussion in][]{casbah}.
These issues motivate sample criterion 1:  
the clustering analysis is
restricted to $0.12 < z < 0.75$. 
This criterion is also motivated by the small
number of galaxies that we have observed at higher redshift
(e.g.,\ Figure~\ref{fig:z_hist}).

Further examining the \ovi\ column density histogram, one notes a marked
decline in the number of systems with $\log \N{OVI} < 13.5$;
e.g.,\  only one system exhibits $\log \N{OVI} < 13.2$.
We emphasize that this dropoff is not entirely driven by sensitivity.
Indeed, a substantial fraction of the CASBaH FUV spectra have a 
$2\sigma$ limit on \novi\ that is lower than $10^{13.5} \cm{-2}$
(for \ovi\ $\lambda$1031
over an integration window of $60 \mkms$).
Therefore, the observed distribution implies
a physical turn-over in the $\N{OVI}$ frequency distribution
as also reported by \cite{danforth16};
this will be explored further 
in a future manuscript. 
For the clustering study,  we are motivated
to set a minimum column density threshold
\nlim\ for including \ovi\
systems in the analysis.
This provides a well-defined sample 
and we can set \novi\ to be sufficiently high
to include nearly all of the COS-FUV data. 
We find that  $\mnlim = 10^{13.5} \cm{-2}$ satisfies this goal
and also includes the majority of \ovi\ systems detected;
this motivates sample criterion 2: We restrict the \ovi~systems to those
with $\mnovi \ge \mnlim$.  
The primary exceptions to an approximately uniform sensitivity
in the COS-FUV spectra
are spectral regions absorbed
by other, unrelated systems (e.g., Lyman series absorption
by a system at higher redshift).  
For each galaxy at $0.12 <z<0.75$, we have assessed the 
spectra in a $\pm 30 \mkms$
window centered at the \ovi\ $\lambda$1031 wavelength and find
that $\approx 16\%$ have a significant blend that prohibits
sensitivity to \nlim.
This motivates sample criterion 3:  
We ignore galaxies whose redshift places them 
in a blend that prohibits measuring down to \nlim, despite
the overall high S/N of the spectra.

\begin{figure}
\begin{center} 
\includegraphics[width=3.5in]{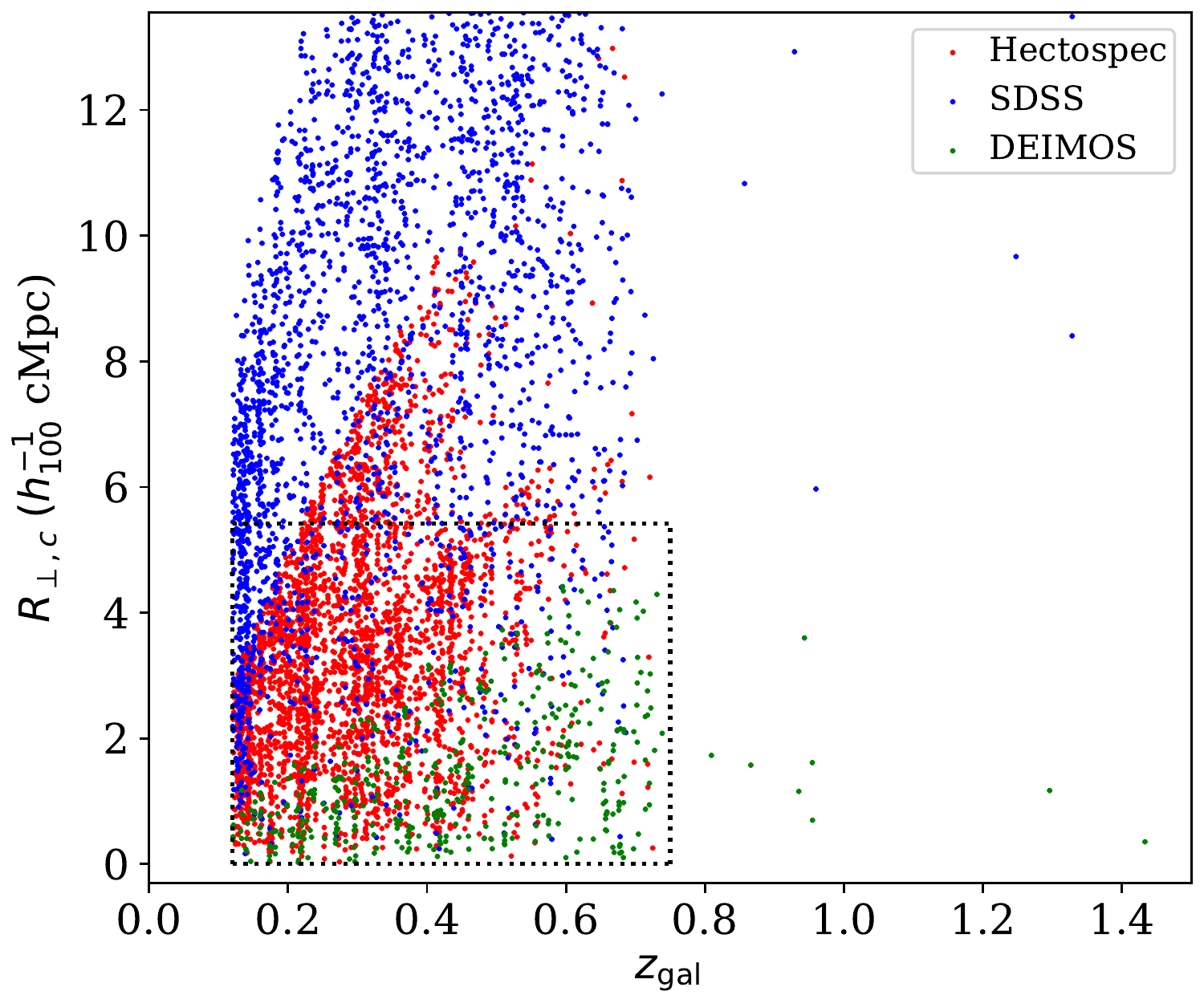}
\caption{
Scatter plot of the comoving impact parameter \rpcom\ 
versus galaxy redshift for the $\approx 6,000$ galaxies 
that have redshifts corresponding to CASBaH absorption
data that are sufficiently
sensitive to detect an \ovi\ system 
with $\N{OVI} \ge 10^{13.5} \cm{-2}$.
The dotted box illustrates the \rpcom,$z_{\rm gal}$ parameter
space used for our OVI-galaxy cross-correlation analysis.
}
\label{fig:OVI_gal}
\end{center}
\end{figure}

Figure~\ref{fig:OVI_gal} illustrates the set of galaxies
satisfying the three sample criteria,
plotted at their comoving distance from the quasar sightline.
To enforce the \nlim\ criterion,
we estimated the uncertainty in column density 
by integrating the apparent optical
depth at \ovi\ $\lambda$1031 in a window of $\pm 30\mkms$ centered on each galaxy redshift. 
We then flagged those with $2 \sigma(\N{OVI}) < 10^{13.5} \cm{-2}$, 
without a blend, and with $z_{\rm gal} > 0.12$.
Altogether we have $\approx 6,000$ galaxies in the CASBaH survey 
satisfying these criteria.
It is apparent from Figure~\ref{fig:OVI_gal}
that very few galaxies with $z_{\rm gal} > 0.75$ 
have sufficient S/N at \ovi\ to satisfy the sensitivity limit
\citep[see also][]{casbah}.
One also notes that, at fixed redshift, 
the CASBaH survey has a roughly constant galaxy
sampling to $\mrpcom \approx 5.4\, \rm h_{100}^{-1} cMpc$ 
which declines monotonically with increasing $z_{\rm gal}$.
Beyond $\mrpcom \approx 5.4 \, \hmpc$, one identifies striping
related to the wedding cake design of the Hectospec observations
($\S$~\ref{sec:hecto_targ}).
This leads to sample criterion 4:  
restrict the cross-correlation analysis
to $\mrpcom < 5.4\, \hmpc$.

Lastly we restrict the analysis to the \ncfield~fields observed with
Hectospec, DEIMOS, or both (i.e.,\ PG1338+416 and LBQS1435$-$0134 are not used in this paper).
The fields with only SDSS coverage have too few galaxies (i.e., 
too large shot noise) for a meaningful analysis.

\subsection{Galaxy-Galaxy Auto-correlation}
\label{sec:gg}

Interpretation of the results from the \ovi-galaxy cross-correlation
analysis will depend on the nature of the galaxies that comprise
the CASBaH survey.  We have assessed several intrinsic properties
in the previous section (e.g., Figure~\ref{fig:stellar_mass});
here we perform an auto-correlation analysis to further assess
the halo mass of the population. 

\begin{figure}
\begin{center} 
\includegraphics[width=3.5in]{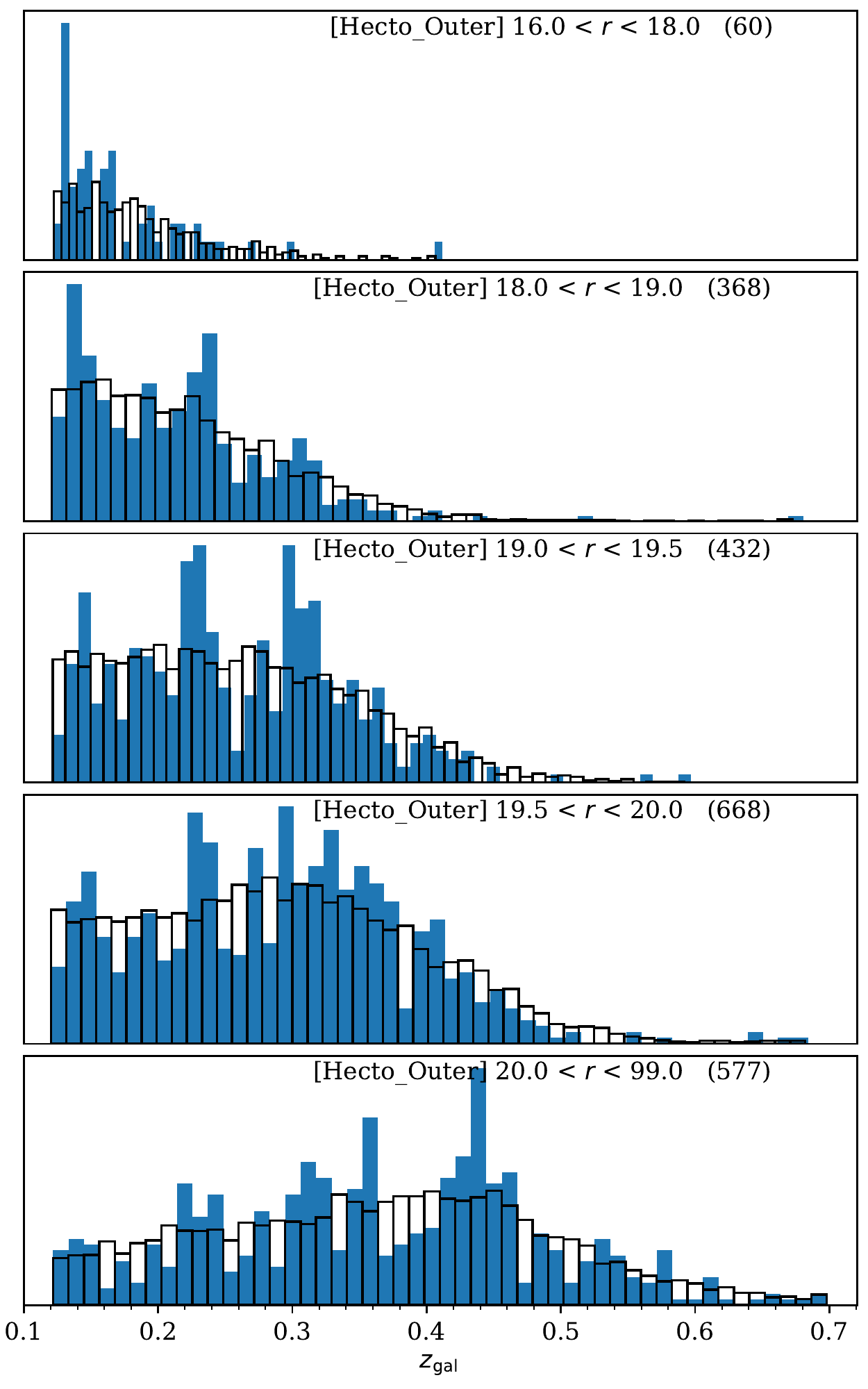}
\caption{
Blue, filled histograms show the redshift distribution
for galaxies observed in the outer layer of the Hectospec
dataset ($\theta > 10'$), split by $r$-band magnitude.
The number in parenthesis lists the total in each interval.
Overlaid in the open, black histogram is a normalized,
random distribution drawn from a Cubic-spline fit to a
Gaussian-smoothed version of the real histogram. 
This smooths out the small-scale clustering of the
galaxies while maintaining the overall distribution.
}
\label{fig:z_sens}
\end{center}
\end{figure}

Our methodology follows closely that of \cite{tejos+14}
who studied the clustering of \lya\ absorption with
galaxies\footnote{All of the code is available in the PYIGM
repository on GitHub (\url{https://github.com/pyigm/pyigm})}.  Their approach compares the incidence of 
galaxy-galaxy (or galaxy-absorber) pairs at a given
comoving separation with the incidence of `random' 
pairs derived from properties of the survey design.
Specifically, they adopt the 
Landy-Szalay formalism \citep[L-S;][]{landy93}.
Of particular importance to the analysis
is matching the redshift
and impact parameter distributions of the random
and real samples as a function of apparent magnitude.   
Regarding the former, 
Figure~\ref{fig:z_sens} compares the observed 
redshift distributions for galaxies discovered with
Hectospec (outer layer, $\theta > 10'$; 
see~$\S$~\ref{sec:hecto_targ})
as a function of $r$-band magnitude against
a random distribution drawn from a
Cubic-spline representation fit to a Gaussian-smoothed 
histogram of the real distributions. 
We refer to these
cubic-splines as sensitivity functions because
they depend on the magnitude limit of the galaxies
targeted and the quality of the spectroscopy. 
Importantly,  the sensitivity functions
are designed to smooth out redshift `spikes' in the real 
observations while maintaining the general 
distribution.  Similar sensitivity functions were
derived from each sub-set of the spectroscopic survey
(i.e.,\ SDSS, DEIMOS, other Hectospec layers), also
in cuts of galaxy magnitude.
The agreement between data and randoms is shown in
Figure~\ref{fig:z_sens} for the
Hectospec (outer layer) sub-set.
The sensitivity functions were derived by combining data
from all of the fields with the exception of the 
PG1630+377 field.  We found it necessary to generate
a custom sensitivity function for the Hectospec-Outer
subset due to a large overdensity at $z\approx 0.4$
in that field.

\begin{figure}
\begin{center}
\includegraphics[width=3.5in]{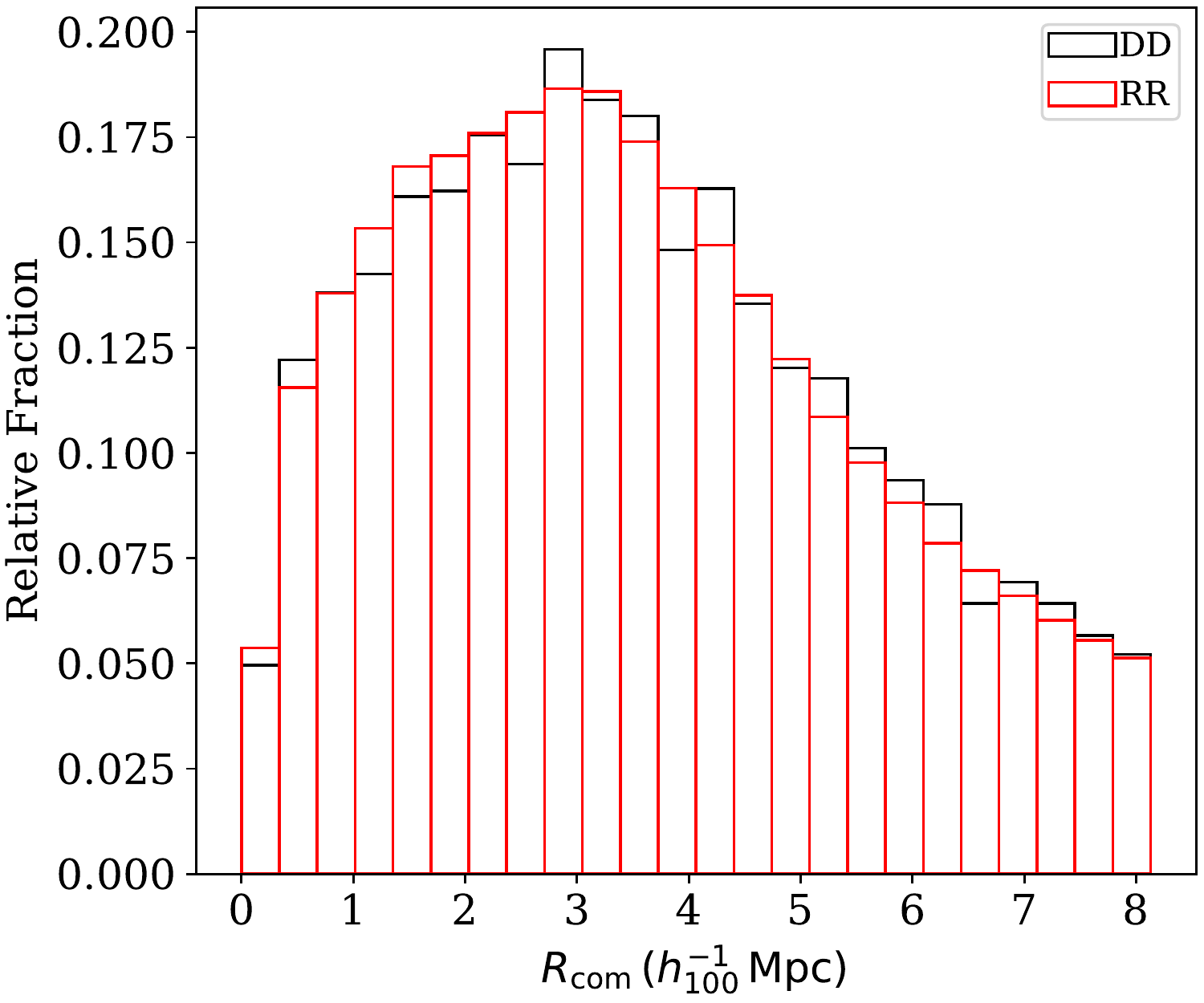}
\caption{
Normalized histograms comparing the distribution of projected
separations from the quasar sightline for the real galaxy
sample (DD; black) and the constructed set of random galaxies (RR; red).
The close agreement is critical to an unbiased estimate of  \xigg.
}
\label{fig:Rcom_real_rand}
\end{center}
\end{figure}

\begin{figure}
\begin{center} 
\includegraphics[width=3.5in]{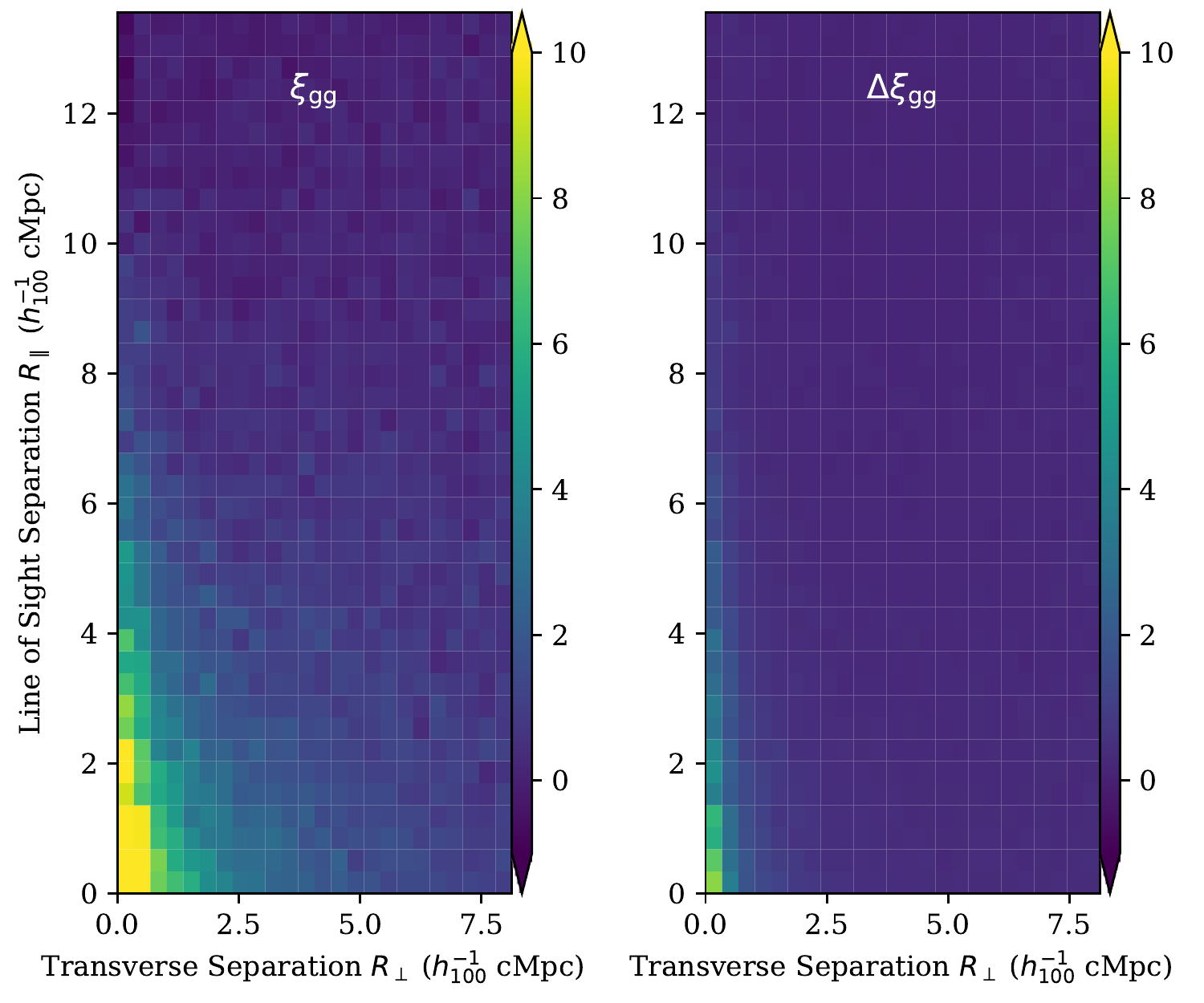}
\caption{
(left) Binned evaluations 
of the galaxy-galaxy
auto-correlation function \xigg\ measured from $z<0.75$ galaxies
in the CASBaH survey  ($0.359 \, \hmpc$ grid). 
(right) Estimates of the uncertainty in
\xigg\ derived from the variance of the Landy-Szalay
estimator (Equation~\ref{eq:ls_var}).
}
\label{fig:gal_gal}
\end{center}
\end{figure}

For each real galaxy, a set of $n_{\rm rand} = 100$ galaxies
were placed at its RA/DEC with
redshifts drawn randomly from the appropriate sensitivity 
function.   
Figure~\ref{fig:Rcom_real_rand} compares the distribution
of comoving separations of the real and random galaxies
for the sightlines.  The close correspondence is 
vital to the analysis.
For each field, we then evaluated the number of
data-data ($D_g D_g$)
data-random ($D_g R_g$)
and random-random ($R_g R_g$) 
galaxy-galaxy
pairs in bins of $0.339 \hmpc$ in the radial (\rpar, line-of-sight)
and tangential (\rperp, plane-of-sky)
directions.  We then sum all of the fields and use the 
L-S estimator: 

\begin{equation}
\mxigg^{LS} = \frac{D_g D_g/n_{gg}^{DD} - 2 D_g R_g / n_{gg}^{DR}}{R_g R_g/n_{gg}^{RR}} + 1  
\end{equation}
to evaluate \xigg\ 
with $n_{gg}^{DD}$,  $n_{gg}^{DR}$, and $n_{gg}^{RR}$, 
the normalization factors.
Figure~\ref{fig:gal_gal} shows the binned evaluation.

Uncertainties in \xigg\ have been estimated from the analytic approximation of the variance presented by \citet[][]{landy93} as (in our notation),

\begin{equation}
\sigma^2_{\xi_{gg}}(r)^{LS} \approx \frac{(1+\xi_{gg}^{LS})^2}{n_{gg}^{DD} (R_g R_g/n_{gg}^{RR})} \approx \frac{(1+\xi_{gg}^{LS})^3}{D_g D_g}  \;\; .
\label{eq:ls_var}
\end{equation}
As typical of such analysis, one observes an 
asymmetry due to peculiar motions in converting redshift into
distance along the line-of-sight (i.e.,\ redshift distortions).  
Examining the measurements in the transverse separation \rperp,
one also identifies significant \xigg\ signal at large separations
which we now quantify.

\begin{figure}
\begin{center} 
\includegraphics[width=3.5in]{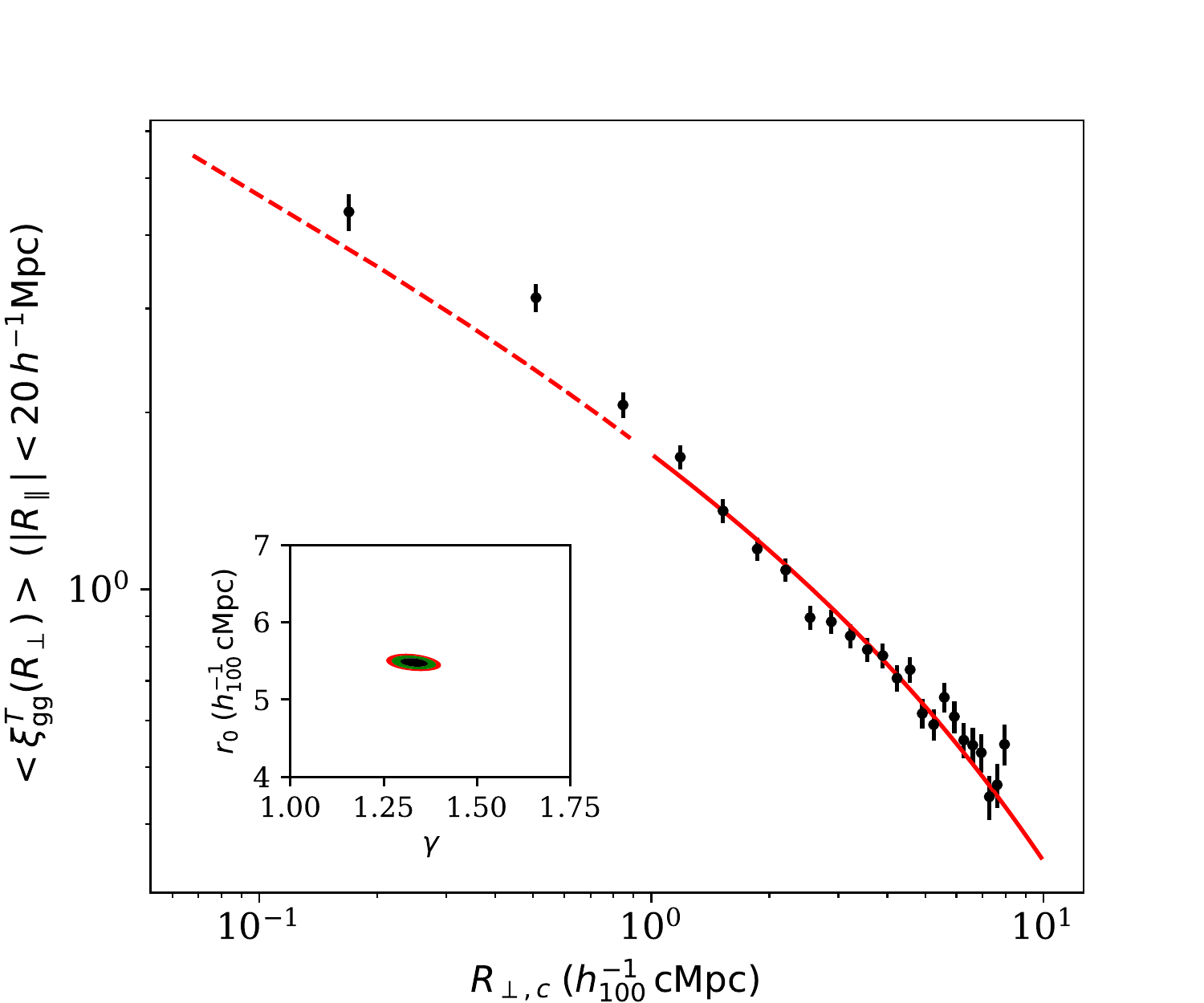}
\caption{
Evaluations of the galaxy-galaxy correlation function 
averaged along the line-of-sight to $R_\parallel = 13.55 \, \hmpc$,
\xitgg, in bins of transverse separation.  
The solid red curve shows the evaluation of \xitgg\ for
the best-fit 3D power-law for \xigg\ 
to the data over $\mrpcom = [1-10] \, \hmpc$.
The dashed curve is an extrapolation of this model to 
$\mrpcom < 1 \, \hmpc$ where one notes the data significantly
exceed the evaluation.  This offset is attributed to galaxy-galaxy
clustering within dark matter halos, i.e.,\ the one-halo term.
The inset shows the confidence contours from a maximum likelihood
analysis which yields  $r_0 = \ggrval \, \hmpc$
and $\gamma = \gggval$ at 68\%\ c.l.
}
\label{fig:xi_gg_T}
\end{center}
\end{figure}

To reduce the effects of redshift distortion and
also to parameterize the \xigg\ measurements, we have evaluated the mean
transverse correlation function by averaging \xigg\ along the
line-of-sight to $\mrpar = 13.55 \, \hmpc$,

\begin{equation}
\mxitgg = \frac{1}{N} \smm_i^N \xi_{\rm gg}^i (r=\mrperp)  \;\;\; ,
\end{equation}
where the sum is over the 40 bins of $0.339\,\hmpc$ in the line-of-sight 
dimension.
This quantity is presented in Figure~\ref{fig:xi_gg_T},
with uncertainties derived from 
the L-S estimator (see equation~\ref{eq:ls_var}) for the collapsed pair-counts.  
Overlaid on the data is the best-fit power-law model for
the 3D correlation function 
$\mxigg = (r/r_0)^{-\gamma}$
estimated from standard
maximum likelihood techniques assuming a Gaussian deviate.
The best-fit values are $r_0 = \ggrval \, \hmpc$
and $\gamma = \gggval$ with uncertainties referring to 
68\%\ confidence intervals.
Specifically, we averaged \xigg\ over $\mrpar = [0,13.55] \hmpc$ 
at the center of each \xitgg\ bin and constructed
the resultant likelihood function by varying $\gamma$
and $r_0$.  The likelihood evaluation is limited to the transverse
bins in the interval $\mrpcom = [1-10] \, \hmpc$
to isolate the so-called two-halo term of large-scale
galaxy-galaxy clustering.
The best-fit correlation length is typical of star-forming
galaxies at $z \sim 0.3$
\citep{coil+17}, which is consistent with the properties
of our sample \citep[Figure~\ref{fig:stellar_mass},][]{casbah_ovi}
We also note that the reported
uncertainties are likely underestimated 
because we have not included all sources of variance, 
e.g.,\ field-to-field
variations \citep{tejos+14}.

The dashed-line in Figure~\ref{fig:xi_gg_T}
is an extrapolation of the \xitgg\ evaluation to
$\mrpcom < 1 \, \hmpc$.  At these separations,
the measurements well exceed the model which is 
generally interpreted as galaxy-galaxy
clustering within individual halos, aka. the one-halo term.  
We have also examined \xitgg\ for sub-samples of the full
galaxy dataset.  Splitting the sample into 
two redshift bins at $z_{\rm gal} = 0.45$, we estimate
\xitgg\ values that are approximately 2 times higher for
the higher redshift galaxies.  This follows from the
fact that they are intrinsically more luminous and have
higher average stellar mass.
Furthermore, the high-$z$ set includes many LRGs from
SDSS which have a very high clustering amplitude 
\citep[e.g.,][]{nuza13}.

\subsection{\ovi-galaxy Clustering}
\label{sec:ag}

Inherent to an absorber-galaxy cross-correlation analysis is the 
notion that absorption systems are measured to occur more (or less)
frequently in the proximity of a galaxy than at random.
For absorption systems, one can assess the random incidence by
surveying many sightlines to estimate the average number per redshift
interval\footnote{It is also common to evaluate $\ell(X)dX$ with
$X$ defined to give a constant incidence if the physical
cross-section and comoving number density of the population is constant
in time \citep{bp69}.}, 
$\mloz dz$.  
Blind surveys for
\ovi\ systems along low-$z$ sightlines have yielded 
direct estimates of \loz.  \cite{tripp08} report 
$\mloz = 15.6_{-2.4}^{+2.9}$ at $z = [0.1,0.5]$ 
from a sample of 51 systems along 16 sightlines
for an equivalent width limit of 30\,m\AA\ 
(i.e., about $ \log \mnovi \simeq 13.3$).  
\citet{danforth16} have extended the
analysis to 82 sightlines at $z_{\rm QSO}<0.85$ and we estimate
$\mloz \approx 17$ for
$\mnovi > \mnlim = 10^{13.5} \cm{-2}$ from their reported
statistics (their Table 5; redshift path $\Delta z_{\rm O\,VI} \approx 14.5$).
One of the future goals of the CASBaH survey is to measure
\loz\ from our sightlines to $z\sim1$ \citep{casbah}.   
As a first estimate, we 
report 59~systems with $\mnovi > \mnlim$ over 
\ncfield~sightlines giving a redshift path of 
$\Delta z \approx 7 (0.75-0.12) \simeq 4.4$.
Therefore, we estimate $\mloz \approx 13.5$, consistent
with the previous literature (this preliminary estimate is lower because our 
redshift path is overestimated as it does not take into 
account parts of the spectra that can be blocked to the \ovi\ absorption).
In the following, we adopt $\mloz = 13.5$ at $z=0.2$
and assume that $\ell(X)$ is constant throughout our analysis
window.

Before assessing the \ovi-galaxy cross-correlation function
\xiag, we begin with an estimate of the
covering fraction \fc\ of \ovi\ gas around
$z<1$ galaxies.  
To associate \ovi\ with an individual galaxy, one must adopt a redshift
(or velocity) window.  Previous work on the CGM has found that
the majority of associated gas occurs within a few hundred \kms\ 
of the galaxy redshift \citep[e.g.,][]{pwc+11,werk+13}.  This also holds
for CASBaH \citep{burchett19a}.  In the following, we adopt a window
of $\delta v = \pm 400 \mkms$. 
We further note that this window is small enough
that a chance association with \ovi\ is relatively low.
Taking \lovi\ from above,  this implies
an average of $\mathcal{N}_{\rm OVI} = \ell(z) \delta z \approx 0.03$ systems
for $\delta z = (\delta v / c)/(1+z)$. 

\begin{figure}
\begin{center} 
\includegraphics[width=3.5in]{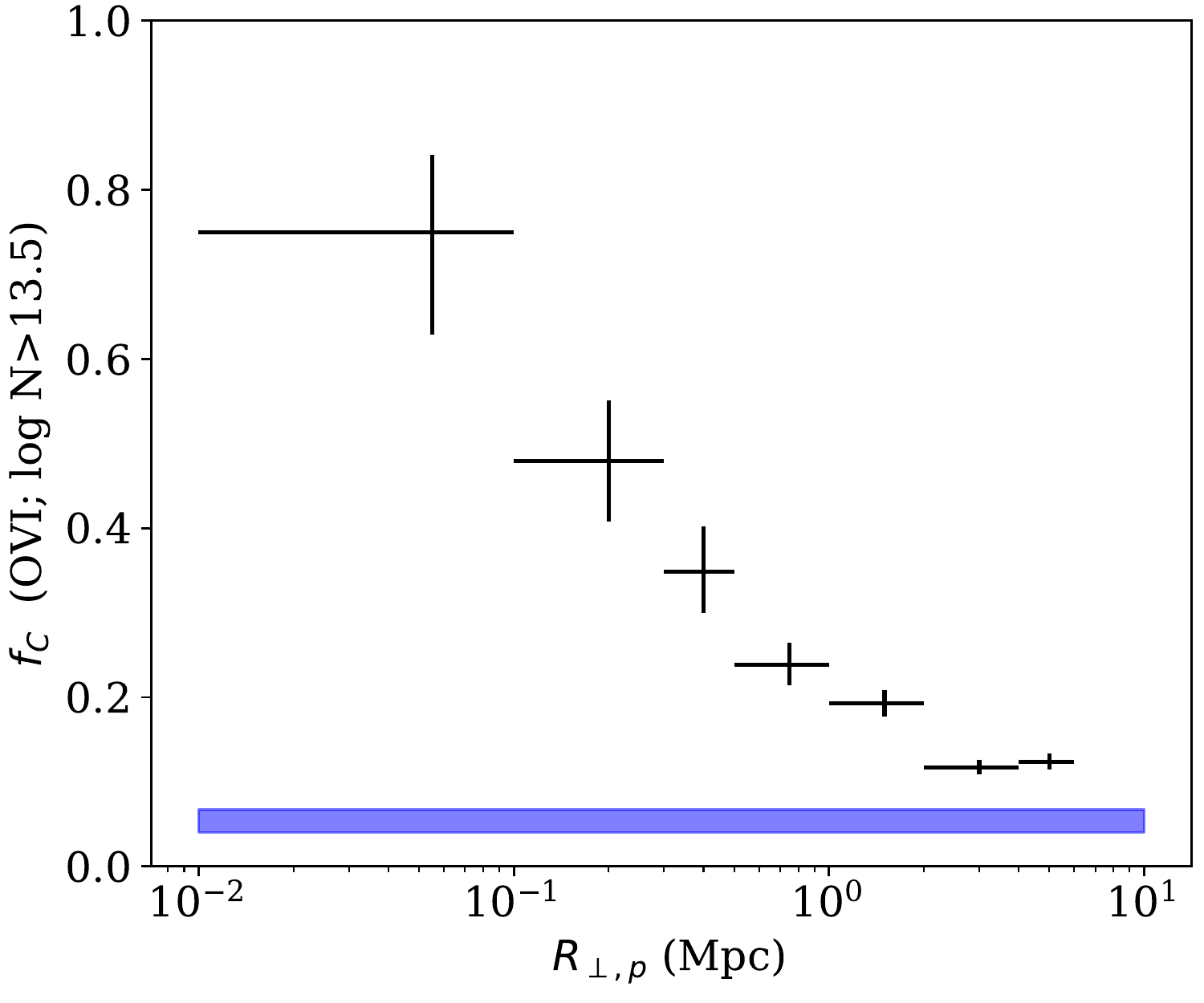}
\caption{
Covering fraction of \ovi\ gas versus
physical impact parameter for galaxies in the CASBaH
survey to a sensitivity limit of $\mnlim = 10^{13.5} \cm{-2}$
with a redshift coincidence within $\pm 400 \mkms$.  
There is a high incidence on small scales
that may be attributed to gas within galactic halos (i..e the CGM).
At larger impact parameters ($>1$\,Mpc), one still recovers 
$f_C$ in excess of random expectation (blue-band), indicating
significant \ovi-galaxy clustering.
}
\label{fig:fc_OVI}
\end{center}
\end{figure}

In a series of arbitrary bins of 
{\it physical} impact parameter \rpperp, we have 
assessed the fraction of CASBaH
galaxies with one or more \ovi\ systems\footnote{
As noted in $\S$~\ref{sec:setup}, we ignore galaxies 
with significant blends at the expected location of \ovi\
in the COS-FUV spectra.}
occurring within $\pm 400 \mkms$.
Any number of galaxies may be associated with a given \ovi\ system.
Figure~\ref{fig:fc_OVI} shows the incidence (or
covering fraction, $f_C$) with uncertainties derived from 
binomial counting statistics.
At small impact parameters ($\mrpperp < 100\, \rm kpc$),
the incidence is very high: $f_C \approx 75\%$.
This excess is generally attributed to gas within galaxy
halos, i.e.,\ the CGM 
\citep[see][for analysis of the \ovi\ CGM in CASBaH]{casbah_ovi}.
The covering fraction declines monotonically with \rpperp, as expected, 
but remains $\approx 2\times$ higher than random
expectation at the largest offsets probed by CASBaH ($\approx 8\, \rm pMpc$).
This implies significant OVI-galaxy clustering on these scales,
which we now assess.

For the cross-correlation analysis of \ovi-galaxy clustering,
we adopt two approaches.  The first follows the analysis developed\footnote{
Note that we have also corrected their approximation for the 
probability estimate to be truly Poisson.} 
by \cite{QPQ2} to evaluate the clustering of optically thick gas
around luminous, $z \sim 2$ quasars \citep[see also,][]{qpq6}. 
This analysis uses a maximum likelihood approach
to estimate the 3D cross-correlation function \xiag\ with an
assumed functional form of $\mxiag = (r/r_0)^{-\gamma}$.
The likelihood function is given by 
$\mathcal{L} = (\Pi_i P^{\rm hit}_i) (\Pi_j P^{\rm miss}_j)$ 
with $P^{\rm hit}$ and $P^{\rm miss}$ the probability of observing one
(or more) \ovi\ systems or none, respectively.
A galaxy is considered a `hit' if one or more \ovi~systems
occurs within $\pm 400 \mkms$ and a miss otherwise,
and the likelihood is evaluated from the full dataset
satisfying the sample criteria ($\S$~\ref{sec:setup}).
The probability of zero absorbers within a velocity window 
of $\delta v = 400 \mkms$ is given by Poisson statistics,

\begin{equation}
P^{\rm miss} = \exp \ltp -[1+\chi_\perp] \, \ell_{\rm OVI}(z_{\rm gal}) \delta z \rtp
\;\;\; ,
\end{equation}
where $\mlovi dz$ is the mean incidence and $\chi_\perp$
expresses the boost from clustering
i.e.,\ $1 + \chi_\perp$, with 

\begin{equation}
\chi_\perp \approx \frac{a H(z)}{2 \delta v}
\intl_{\rm -\delta v/[aH(z)]}^{\rm \delta v/[aH(z)]} d\mrpar \, 
\xi_{\rm ag} \left ( \sqrt{\mrperp^2 + \mrpar^2} \right )  
\;\;\; .
\label{eqn:chi}
\end{equation}
It follows trivially that $P^{\rm hit} = 1 - P^{\rm miss}$.  

\begin{figure}
\begin{center} 
\includegraphics[width=3.5in]{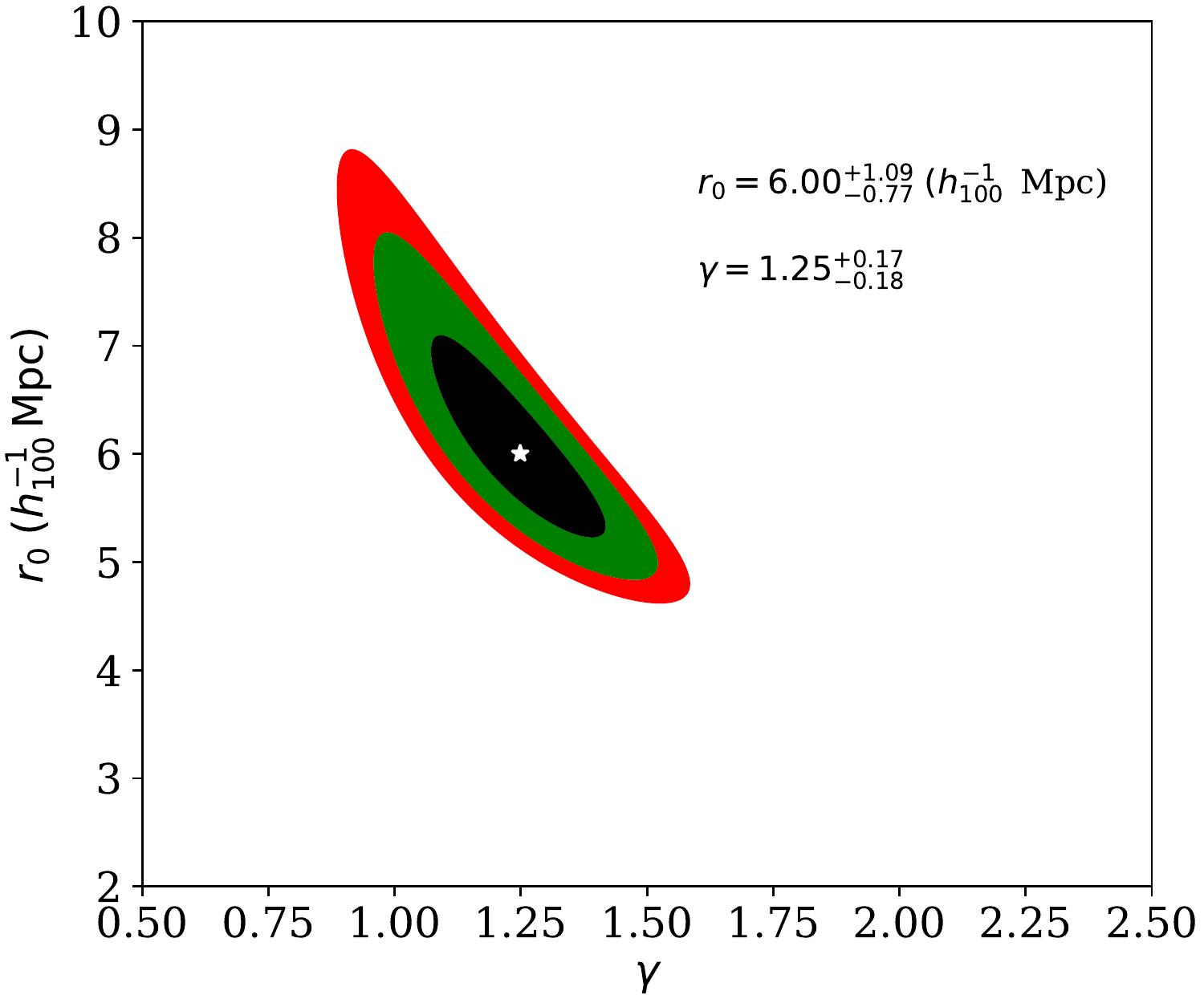}
\caption{
Constraints on $r_0$ and $\gamma$ from the \ovi-galaxy
cross-correlation analysis, resticted to galaxies with 
$\mrpcom = [1,8] \, \hmpc, z=[0.12,0.75]$ and
those without a significant line-blend at \ovi.
}
\label{fig:r0_gamma}
\end{center}
\end{figure}

\begin{figure}
\begin{center} 
\includegraphics[width=3.5in]{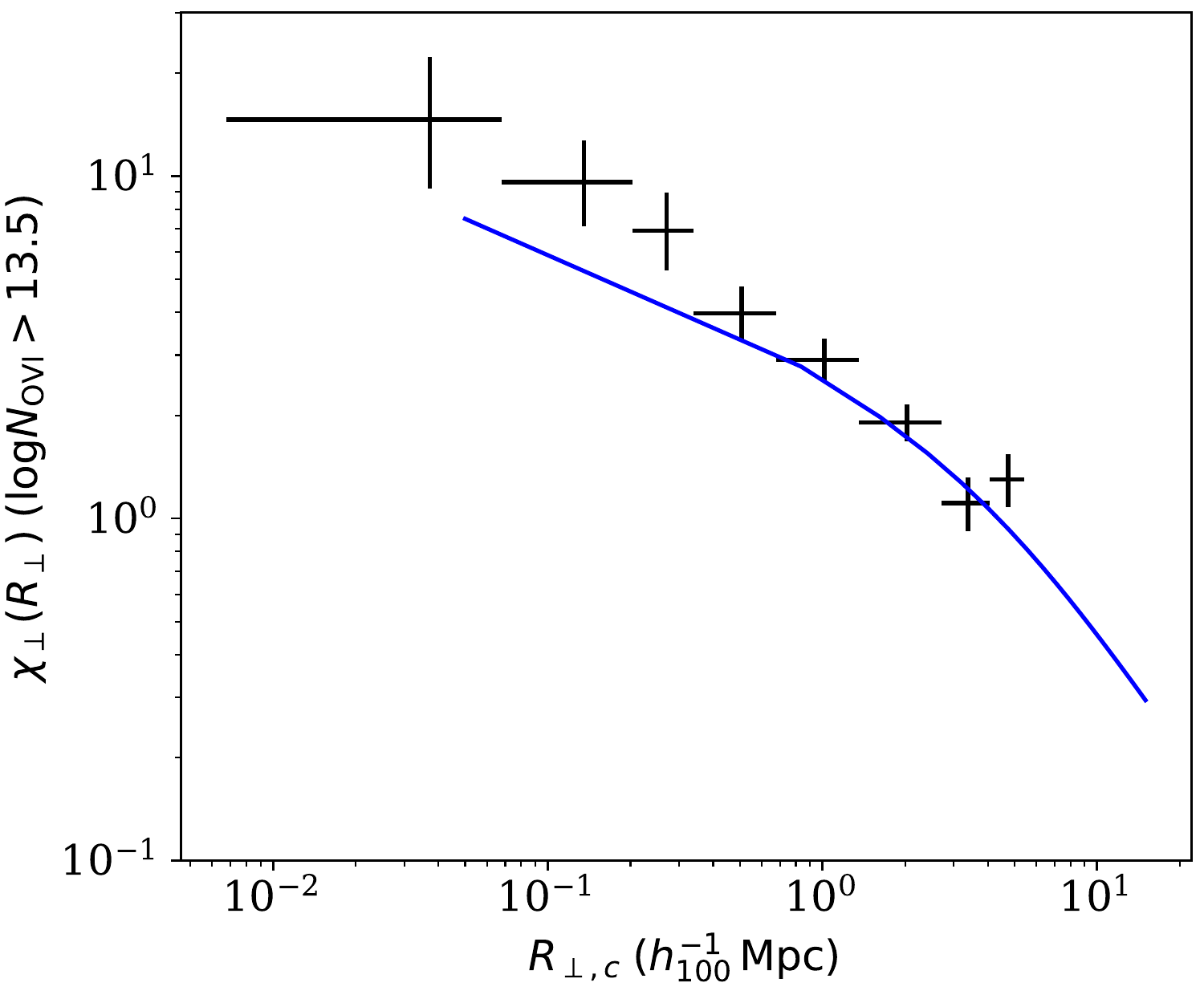}
\caption{
Binned evaluations of $\chi_\perp$ derived from the ratio of 
galaxies with associated \ovi\ absorption relative to random
expectation.  
Overplotted is the evaluation of Equation~\ref{eqn:chi}
using the best-fit parameters derived from a maximum
likelihood analysis (Figure~\ref{fig:r0_gamma}).
}
\label{fig:chi_perp}
\end{center}
\end{figure}

We constructed a grid of $\mathcal{L}$ by varying $r_0$ and
$\gamma$ over a range of values and then found the maximum. 
Figure~\ref{fig:r0_gamma} presents the constraints on $\gamma$
and $r_0$ for the subset of CASBaH galaxies analyzed:
galaxies with 
$\mrpcom = [1,8] \, \hmpc, z=[0.12,0.75]$ and also having no
substantial blend at the expected location of \ovi.
We have estimated the uncertainty by integrating $\mathcal{L}$
down to several confidence limits.
The best-fit model is shown on
a binned evaluation of $\chi_\perp$ in Figure~\ref{fig:chi_perp}.
This model provides a good description of the observations
at large values and, as with the galaxy-galaxy clustering,
we identify a putative one-halo term at $R < 0.5 \hmpc$, seen as an
excess $\chi_\perp$ over the best fit to larger \rperp. We note that our results contrast with the \ion{O}{6}-galaxy clustering study by \citet{finn+16}, who find that the \ion{O}{6}-galaxy signal is lower than the galaxy-galaxy autocorrelation on all scales, 
while here we find similar correlation lengths between them. Still, given the somewhat shallower slope for the \ion{O}{6}-galaxy ($\gamma\approx 1.25$) compared to the galaxy-galaxy one ($\gamma\approx 1.33$), we find that $\xi_{\rm ag}/\xi_{\rm gg}$ should be $\lesssim 1$ on scales $\lesssim 1.3 \hmpc$ (see below).
As deeper and more extensive galaxy surveys are completed in the future, it will be valuable to revisit this topic.

We have generated a separate estimate of \xiag\ by analyzing the 
absorber-galaxy pair counts
using the formalism applied to \xigg\ in $\S$~\ref{sec:gg}.
In addition to constructing a set of random galaxies,
we must also introduce random absorbers    
for the absorber-galaxy analysis. 
These were placed along the sightlines with a uniform
redshift distribution in the interval $z=[0.12, 0.75]$, avoiding
strong Galactic ISM absorption (e.g.,\ \ion{Si}{2}~1526) which would
preclude the detection of \ion{O}{6}.  

\begin{figure}
\begin{center} 
\includegraphics[width=3.5in]{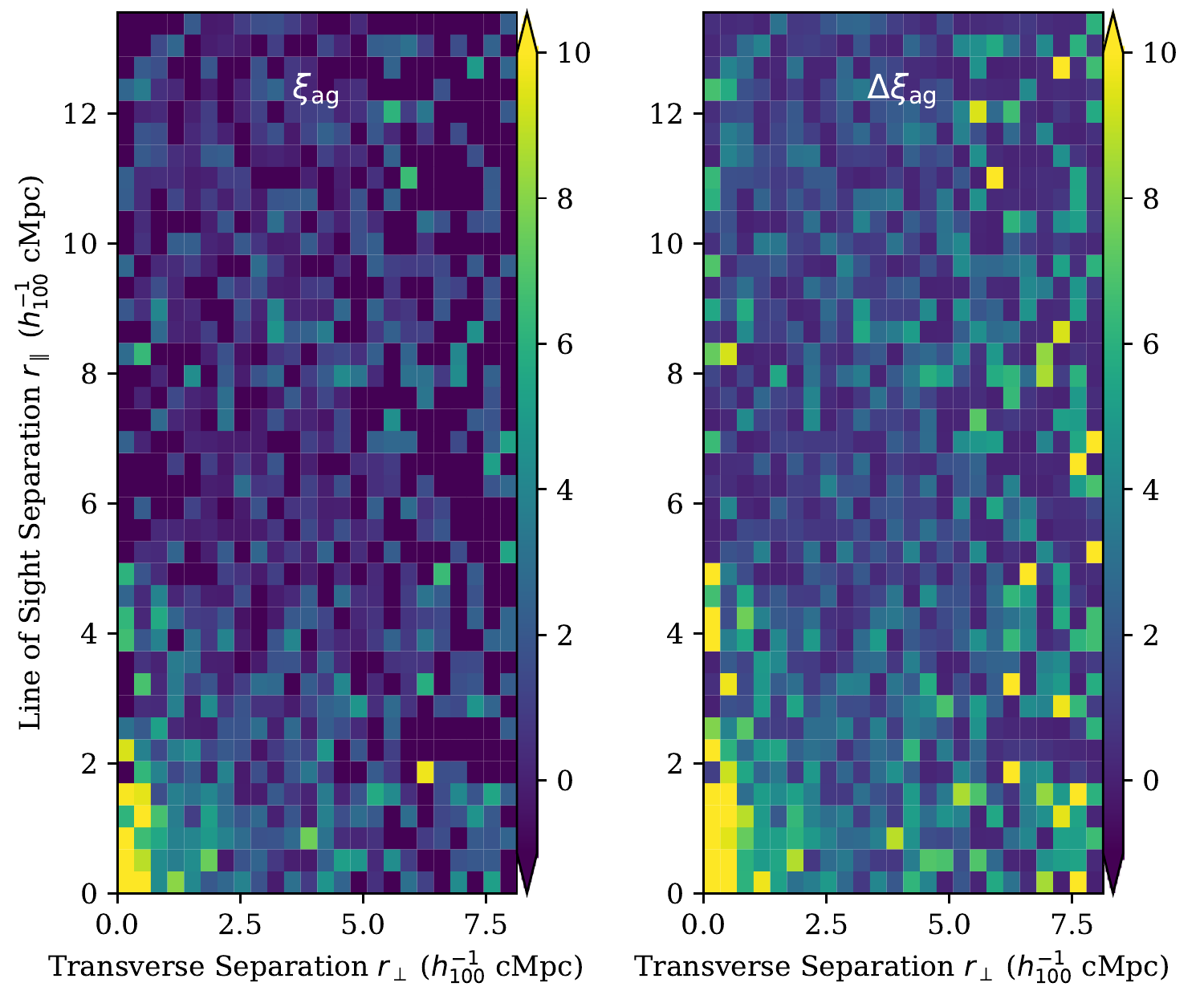}
\caption{
Binned evaluations of \xiag\ and its uncertainty
as a function of transverse and line-of-sight separation.
Similar to the galaxy-galaxy auto-correlation, we observe the effects
of redshift distortions and also detect a significant signal to
large separations in $R_\perp$.
}
\label{fig:xiag}
\end{center}
\end{figure}

\begin{figure}
\begin{center} 
\includegraphics[width=3.5in]{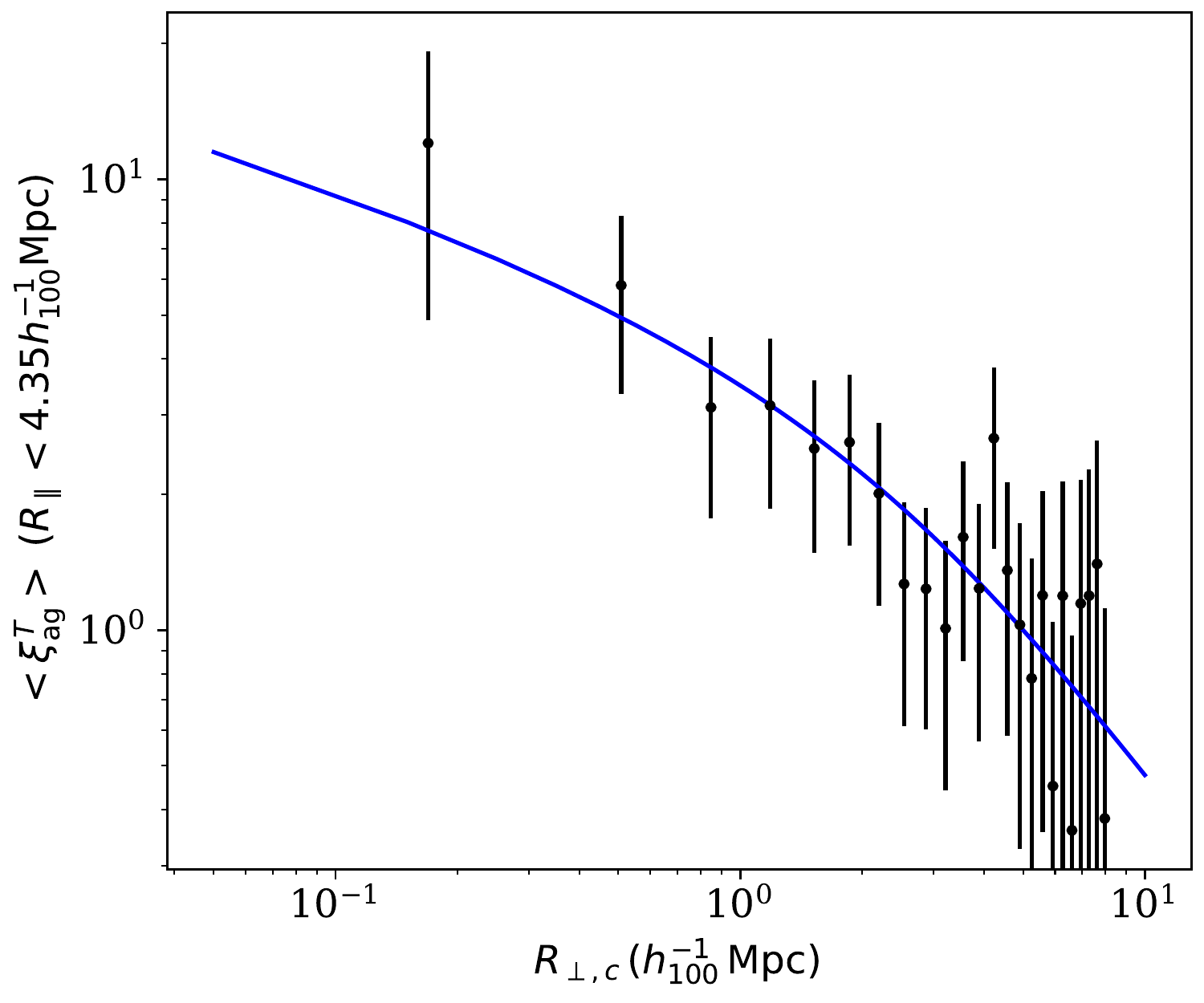}
\caption{
Evaluation of the transverse \ovi-galaxy cross-correlation
function averaged to $\mrpar = 4.4 \, \hmpc$ (corresponding
to $\approx 400 \mkms$) from the pair analysis.  Measurements
and uncertainties were derived from the L-S estimator.
Overplotted on these values is an evaluation of
\xitag\ using the best-fit model for \xiag\ from
the $\chi_\perp$ analysis.
}
\label{fig:xiag_T}
\end{center}
\end{figure}

Figure~\ref{fig:xiag} shows the binned evaluations of 
\xiag\ and its uncertainty.  
Similar to the galaxy-galaxy auto-correlation, we observe the effects
of redshift distortions and also detect a significant signal to
large separations in $R_\perp$.

To compare with the $\chi_\perp$ analysis from above, we 
calculate \xitag\ by averaging to $\mrpar = 4.4\, \hmpc$ 
which corresponds to approximately 400\kms\ at $z = 0.3$.
These estimates are shown in Figure~\ref{fig:xiag_T}
and we also overplot the estimate for \xitag\ based on the
best-fit model from Figure~\ref{fig:r0_gamma}.
There is good overall 
agreement between the two techniques, although the
pair-counting analysis does yield an $\approx 20\%$ lower
amplitude at most scales.
%
In the following, we use the pair analysis \xiag\ to
compare with the auto-correlation function \xigg\
but use the power-law fit to $\chi_\perp$ 
for any further discussion of $r_0, \gamma$.

\subsection{Discussion}


We now synthesize the results of the previous sub-sections
to derive new insight on the physical association of \ovi\ 
absorption to galaxies.  We first remind the reader that the analysis
was restricted to \ovi\ systems with $\N{OVI} \ge 10^{13.5} \cm{-2}$
and redshift $0.12 < z < 0.75$.  Furthermore, the galaxy sample
is dominated by the Hectospec observations and these have
$z \sim 0.2 - 0.4$ and stellar mass of a few $10^{10} M_\odot$
(Figure~\ref{fig:stellar_mass}).

In the regime of linear bias, we may relate the ratio of the correlation
functions to their bias factors,

\begin{equation}
b_{\rm OVI} = b_{gg} \frac{\xi_{ag}}{\xi_{gg}} \;\; ,
\label{eqn:bias}
\end{equation}
and further relate the galaxy-galaxy bias $b_{gg}$ to dark matter
clustering ($b_{gg}^2 = \xi_{gg}/\xi_{DM}$)
to infer the `mean' halo mass hosting \ovi\ gas.
On the latter point, we have made evaluations of $\xi_{DM}$
from the code\footnote{http://cosmicpy.github.io/index.html}
of \cite{smith+03} to estimate the
galaxy-galaxy bias function at
$z=0.3$: $b_{gg} = \bggval$.
Following the HOD analysis of \cite{zzw+11}
for the SDSS main survey ($z \sim 0.1$), we relate
$b_{gg}$ to a characteristic halo mass $M_h \approx \Mgghalo M_\odot$.

\begin{figure}
\begin{center} 
\includegraphics[width=3.5in]{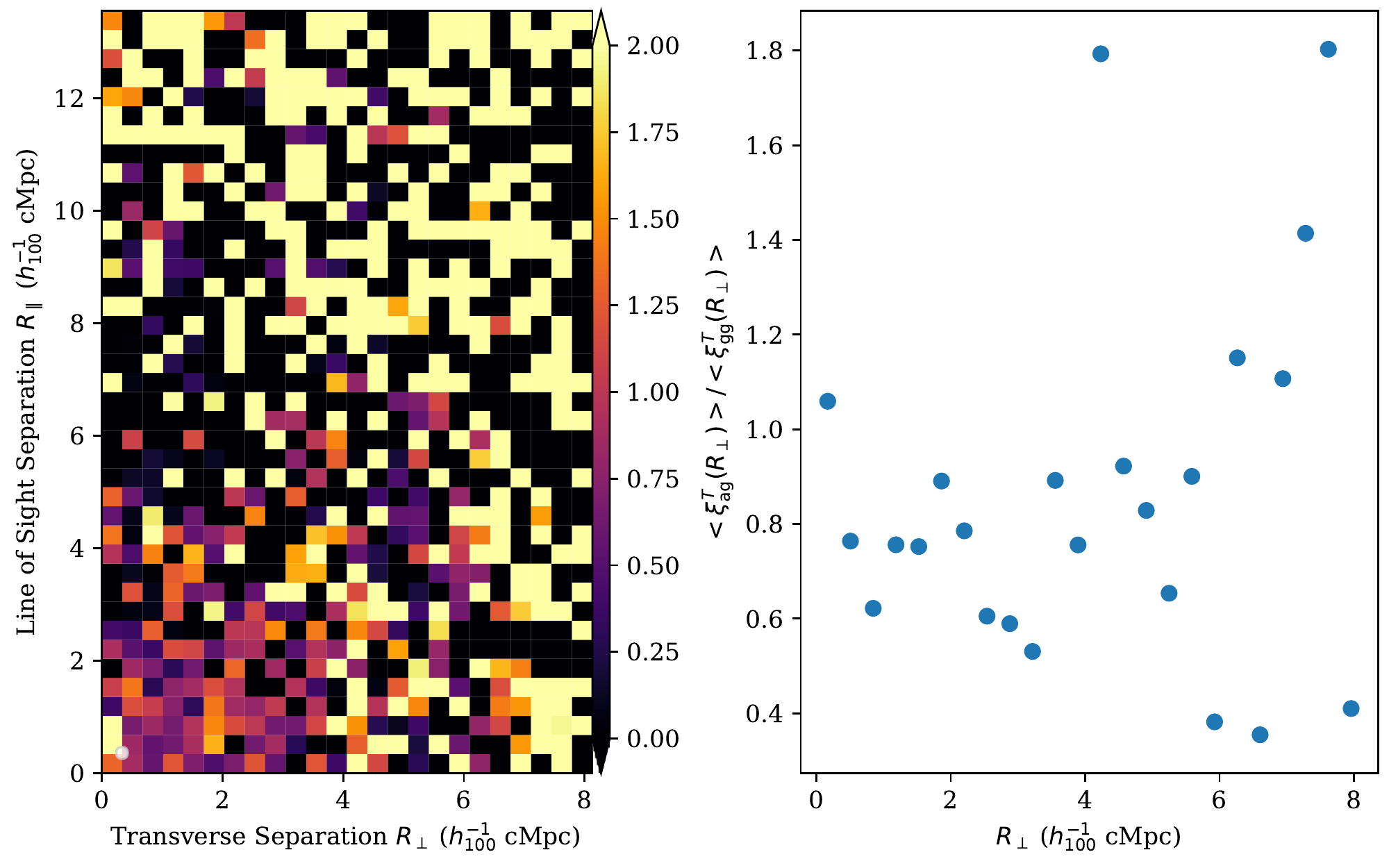}
\caption{
(left) Binned evaluation of the ratio $\mxiag/\mxigg$ which leads
to the evaluation of the bias factor for \ion{O}{6} as given by
Equation~\ref{eqn:bias}.
(right) The same ratio averaged along the line-of-sight to $4.4 \, \hmpc$.
From these values we estimate $\mxiag/\mxigg = 0.76 \pm 0.1$.
}
\label{fig:bias}
\end{center}
\end{figure}

Figure~\ref{fig:bias} compares \xiag/\xigg\ for the binned
evaluations (left) and integrated to $4.4 \hmpc$
and we estimate $\mxiag/\mxigg = 0.76 \pm 0.1$.
This implies that \ovi\ systems are primarily hosted by galaxies
in halos with $M_h^{\rm OVI} \approx \Maghalo M_\odot$,
i.e.,\ sub-$L*$ galaxies.
These measurements strengthen previous assertions that
\ovi\ gas arises primarily in the surroundings of 
sub-$L*$ galaxies based on CGM statistics \citep{pwc+11},
galaxy-absorber clustering on predominantly
smaller scales \citep{cm09}, and linking individual galaxies
to \ovi\ absorbers \citep{stockeetal06,pratt18}.
Future work will synthesize these cross-correlation measurements
with the statistical incidence of \ovi\ to further assess
the physical association of \ovi\ to dark matter
halos \citep[e.g.,][]{ct08}.

\section{Summary}


In this paper we have reviewed the design of a photometric and spectroscopic galaxy redshift survey to support studies of the relationships between QSO absorption-line systems and galaxies, large-scale structures, and other environmental factors. We have reviewed our data handling and measurement methods as well as the content of an on-line public database released with this paper.  Combined with absorption-line measurements from high-resolution ultraviolet CASBaH spectroscopy from HST \citep{casbah}, this redshift survey can be used to investigate the role of circumgalactic and intergalactic gases in galaxy evolution, and subsequent papers will exploit the data for various purposes.  Importantly, both the galaxies and the absorption systems in the CASBaH database are {\it blindly} selected; no explicit preference for galaxies or absorbers of any particular type was imposed on this survey. 

As an initial step in our long-term goal of investigating absorber-galaxy-environment connections, we have analyzed the clustering of \ion{O}{6} absorbers with galaxies in the CASBaH database.  At small impact parameters, the CASBaH \ion{O}{6} systems with $\log \N{O^{+5}} > 13.5$ 
and $z <$ 0.75 have high covering fractions that are consistent with earlier studies with different selection criteria \citep[e.g.,][]{pwc+11,ttw+11,johnson+15}. 
This sample also exhibits a covering fraction that is larger than expected from random realizations out to very large projected distances ($\approx$ 8 pMpc), which indicates strong \ovi-galaxy clustering, i.e.,\
the gas traces the large-scale structures 
that comprise the cosmic web. 

The clustering of \ion{O}{6} with galaxies is reasonably well described by a power law cross-correlation function of the form 
$\xi(r) = (r/r_0)^{-\gamma}$ with $r_0$ = $\gorval$ and $\gamma$ = $\gogval$, and the bias implied by our cross-correlation analysis suggests that \ion{O}{6} absorbers are typically affiliated with dark-matter halos having 
masses $\approx 10^{11} M_{\odot}$ at $z \sim 0.3$. 

All of the spectra and photometry derived from our
efforts are publicly available in a \specdb\
database file that can be downloaded with 
that package\footnote{https://github.com/specdb/specdb}.
The code used to generate the figures and measurements
reported here will be released on GitHub
with the first set of CASBaH science papers.

\acknowledgments

JXP, JKW, JCH, NL, SL, JB, CNAW, and TMT acknowledge financial support for programs HST-GO-11741 and HST-GO-13846 from NASA through grants from the Space Telescope Science Institute, which is operated by the Association of Universities for Research in Astronomy, Inc., under NASA Contract NAS5-26555. JCH also recognizes support from NSF grant AST-1517353, and JKW appreciates support from a 2018 Alfred P. Sloan Research Fellowship. The authors would like to especially thank Michael Cooper for helpful tips in reducing the DEIMOS spectra. 
We also acknowledge J. Tinker for helpful discussions.

This study is also partly based on data acquired using the Large Binocular Telescope (LBT). The LBT is an international collaboration among institutions in the US, Italy, and Germany. LBT Corporation partners are the University of Arizona, on behalf of the Arizona university system; Istituto Nazionale do Astrofisica, Italy; LBT Beteiligungsgesellschaft, Germany, representing the Max Planck Society, the Astrophysical Institute of Postdam, and Heidelberg University; Ohio State University, and the Research Corporation, on behalf of the University of Notre Dame, the University of Minnesota, and the University of Virginia. 

The conclusions of this work are based on data collected
from observatories at the summit of Mauna Kea. The authors
wish to recognize and acknowledge the very significant cultural
role and reverence that the summit of Mauna Kea has always
had within the indigenous Hawaiian community. We are most
fortunate to have the opportunity to conduct observations from
this mountain.

This work has made use of data from the European Space Agency (ESA)
mission {\it Gaia} (\url{https://www.cosmos.esa.int/gaia}), processed by
the {\it Gaia} Data Processing and Analysis Consortium (DPAC,
\url{https://www.cosmos.esa.int/web/gaia/dpac/consortium}). Funding
for the DPAC has been provided by national institutions, in particular
the institutions participating in the {\it Gaia} Multilateral Agreement. We acknowledge use of the SDSS {\url{www.sdss.org}}, which is funded by the Alfred P. Sloan Foundation, the U.S. Department of Energy Office of Science, the National Science Foundation, the US Department of Energy, the National Aeronautics and Space Administration, the Japanese Monbukagakusho and Participating Institutions.

The following python packages were used in our analysis: 
{\sc astropy, linetools, pyigm}
and the authors thank their developers. 

\bibliographystyle{apj}
\bibliography{allrefs_casbah.bib}



\end{document}

%% file: tab_fields.tex
\begin{deluxetable*}{lccccc}
\tablewidth{0pc}
\tablecaption{Survey Overview\label{tab:fields}}
\tabletypesize{\small}
\tablehead{\colhead{Field} & \colhead{RA} & \colhead{DEC} 
& \colhead{\zem} & \colhead{Imaging} 
& \colhead{Spectra} 
} 
\startdata 
PHL1377&38.78077 & -4.03491 &1.437 
&SDSS/{\it ugriz}&BOSS-DR12\\ 
&&&&LBT+LBC&Keck/DEIMOS\\ 
FBQS0751+2919&117.80128 & 29.32730 &0.916 
&SDSS/{\it ugriz}&BOSS-DR12\\ 
&&&&LBT+LBC&MMT/Hectospec\\ 
&&&&&Keck/DEIMOS\\ 
PG1148+549&177.83526 & 54.62586 &0.976 
&SDSS/{\it ugriz}&BOSS-DR12\\ 
&&&&LBT+LBC&MMT/Hectospec\\ 
&&&&&\\ 
PG1206+459&182.24172 & 45.67652 &1.165 
&SDSS/{\it ugriz}&BOSS-DR12\\ 
&&&&LBT+LBC&MMT/Hectospec\\ 
&&&&&Keck/DEIMOS\\ 
PG1338+416&205.25326 & 41.38724 &1.214 
&SDSS/{\it ugriz}&BOSS-DR12\\ 
&&&&LBT+LBC&\\ 
PG1407+265&212.34963 & 26.30587 &0.94\tablenotemark{a} 
&SDSS/{\it ugriz}&BOSS-DR12\\ 
&&&&LBT+LBC&MMT/Hectospec\\ 
&&&&&Keck/DEIMOS\\ 
LBQS1435-0134&219.45118 & -1.78633 &1.311 
&SDSS/{\it ugriz}&BOSS-DR12\\ 
PG1522+101&231.10231 & 9.97494 &1.328 
&SDSS/{\it ugriz}&BOSS-DR12\\ 
&&&&&MMT/Hectospec\\ 
PG1630+377&248.00462 & 37.63055 &1.479
&SDSS/{\it ugriz}&BOSS-DR12\\ 
&&&&LBT+LBC&MMT/Hectospec\\ 
&&&&&Keck/DEIMOS\\ 
\enddata 
\tablenotetext{a}{This QSO lacks typical emission lines (e.g., the Ly$\alpha$ emission line is almost undetectable), and the optical emission lines that are detected span a redshift range of $\approx$ 10000 km s$^{-1}$ \citep{mcdowell95}.  Consequently, the redshift of this QSO is more uncertain than the redshifts of the other targets; \citet{mcdowell95} report a redshift uncertainty of $\pm$0.02.} 
\end{deluxetable*}

%% file: tab_lbt_obs.tex
\begin{deluxetable*}{lcccl}
\tablewidth{0pc}
\tablecaption{LBT/LBC Observing \label{tab:lbt-obs}}
\tabletypesize{\small}
\tablehead{\colhead{Field} & \colhead{Date} & 
\colhead{Seeing} & \colhead{Filters} 
} 
\startdata 
PKS0232-042   & 2010/Oct, 2010/Dec & $1\farcs0$ & $UBVI$ \\
FBQS0751+2919 & 2010/Oct, 2010/Dec & $1\farcs0$ & $UBVI$ \\
PG1148+549    & 2011/May & $0\farcs8$ & $UBVI$ \\
PG1206+459    & 2010/May, 2010/Dec, 2011/May & $1\farcs0$ & $UBVI$ \\
PG1338+416    & 2011/Apr & $1\farcs0$ & $UBVI$ \\
PG1407+265    & 2010/Mar & $1\farcs0$ & $UBVI$ \\
PG1522+101    & 2011/May & $0\farcs8$ & $UBVI$ \\
PG1630+377    & 2009/Jun & $1\farcs0$ & $griz$ &  \\
\enddata 
\end{deluxetable*}

%% file: tab_lbc_depth.tex
\begin{deluxetable}{cccc}
\tablewidth{0pc}
\tablecaption{LBT/LBC Imaging \label{tab:lbc_depth}}
\tabletypesize{\small}
\tablehead{\colhead{Field} & \colhead{Filter} & \colhead{AM$^a$} & 
   \colhead{Mag. Limit$^b$} 
} 
\startdata 
PG1206+459 & U & 1.3  & 27.42  \\
PG1206+459 & B & 1.0  & 26.51 \\
PG1206+459 & V & 0.9  & 26.20 \\
PG1206+459 & I & 0.9  & 25.29 \\
PHL1377 & U & 1.0  & 27.46 \\
PHL1377 & B & 1.0  & 27.12 \\
PHL1377 & V & 0.9  & 26.79 \\
PHL1377 & I & 1.0  & 26.67 \\
PG1407+265 & U & 1.0  & 27.62 \\
PG1407+265 & B & 1.3  & 27.49 \\
PG1407+265 & V & 1.4  & 27.10 \\
PG1407+265 & I & 1.3  & 26.77 \\
PG1448+549 & B & 1.0  & 26.50 \\
PG1448+549 & U & 1.1  & 27.53 \\
PG1630+377 & g & 1.1  & 28.08 \\
PG1630+377 & r & 1.0  & 27.20 \\
\enddata 
\tablenotetext{a}{Median airmass of the observations.}
\tablenotetext{b}{The $5 \sigma$ limiting magnitude for a point source is computed 
by finding the mean value across all images in a given filter of the faintest sources whose magnitude error is 0.198 magnitudes or less using the sextractor $MAG\_AUTO$ parameter. Sextractor is run with a detection threshold of a $2.5 \sigma$ point source with at least 3 pixels above the threshold. }
\end{deluxetable} 

%% file: tab_deimos_criteria.tex
\begin{deluxetable*}{lccccc}
\tablewidth{0pc}
\tablecaption{DEIMOS Targeting Criteria\label{tab:deimos_criteria}}
\tabletypesize{\small}
\tablehead{\colhead{Field} & \colhead{Filter} 
& \colhead{S/G$^a_{\rm max}$} & \colhead{$m_{\rm max}^b$} 
& \colhead{$\theta_{\rm max}^c$} 
} 
\startdata 
PG1630+377&R &0.9 &25.0 &15.0 &\\ 
PG1206+459&V &0.9 &25.0 &15.0 &\\ 
FBQS0751+2919&V &1.0 &24.5 &10.0 &\\ 
PG1407+265&V &0.9 &24.5 &15.0 &\\ 
PHL1377&V &1.0 &24.5 &15.0 &\\ 
\enddata 
\tablenotetext{a}{Star/galaxy classifier from SExtractor with S/G=1 a star-like object.  Ignored for $\theta < 20$.}
\tablenotetext{b}{Faintest magnitude source for targeting.}
\tablenotetext{c}{Largest angular offset for targeting.}
\end{deluxetable*}

%% file: tab_hectolog.tex
\begin{deluxetable*}{lccc}
\tablecaption{Hectospec Observing\label{tab:hectolog}}
\tablehead{
\colhead{Field} & \colhead{Configuration} &\colhead{Date} & \colhead{Total Exposure} \\
 \colhead{} &  \colhead{} &  \colhead{} &  \colhead{sec}
}
\startdata
FBQS0751+2919 & 1  & 2011 Jan 25  &  5400\\
FBQS0751+2919 & 2  & 2011 Jan 26  &  5400\\
FBQS0751+2919 & 3  & 2011 Feb 10  &  5400\\
FBQS0751+2919 & 4  & 2011 Mar 25  &  5400\\
     {}       & {} &   {}       &  {}\\
PG1148+549    & 1  &  2011 Feb 10 &  5400\\
PG1148+549    & 2  &  2011 Feb 11 &  5400\\
PG1148+549    & 3  &  2011 Feb 11 &  5400\\
PG1148+549    & 4  &  2011 Mar 30 &  7187\\
     {}       & {} &     {}     & {}\\
PG1206+459    &  1 &  2012 Jan 21 &  3600\\
PG1206+459    &  2 &  2012 Mar 14 &  3600\\
PG1206+459    &  3 &  2012 Feb 22 &  3600\\
     {}       & {} &   {}       &   {}\\
PG1407+265    &  1 &  2011 Jun 03 &  3600\\
PG1407+265    &  2 &  2011 Jun 03 &  3600\\
PG1407+265    &  3 &  2011 Jun 05 &  3600\\
PG1407+265    &  4 &  2011 Jun 06 &  3600\\
     {}       & {} &        {}  &  {}\\
PG1522+101    &  1 &  2011 Jun 05 &  3600\\
PG1522+101    &  2 &  2012 Feb 22 &  3600\\
     {}       & {} &  {}     & {}\\          
PG1630+377    &  1 &  2011 Jun 03 &  3600\\
PG1630+377    &  2 &  2011 Jun 04 &  2700\\
PG1630+377    &  3 &  2011 Jun 05 &  3120\\
PG1630+377    &  4 &  2011 Jun 06 &  3600\\
\enddata
\end{deluxetable*}

%% file: tab_deimos_spectraobslog.tex
\begin{deluxetable*}{lccc}
\tablewidth{0pc}
\tablecaption{DEIMOS Spectroscopic Observations\label{tab:deimos_obs}}
\tabletypesize{\small}
\tablehead{\colhead{Field} & \colhead{Mask ID} 
& \colhead{N$_{\rm slitlets}$} & \colhead{Date Obs$^{a}$} } 
\startdata 
PHL1377&M1-8005&88 & 2012 Nov 14\\ 
PHL1377&M2-8006&87 & 2012 Nov 14, Dec13\\ 
PHL1377&M3-8007&86 & 2012 Nov 14\\ 
FBQS0751+2919&M4-8008 &87 & 2012 Nov 14, Dec 13\\ 
FBQS0751+2919&M5-8009 &93& 2012 Dec 13\\ 
PG1206+459& M1-8377&72& 2013 May 07, 08\\ 
PG1407+265&M1-8380 &75& 2013 May 07, 08\\ 
PG1407+265&M2-8381 &74& 2013 May 08\\ 
PG1407+265&M3-8382 &74& 2013 May 08\\ 
PG1407+265&M4-8383 &76& 2013 May 07, 08\\ 
PG1630+377&M3-8387 &69 & 2013 May 07\\ 
PG1630+377&M4-8388 &75 & 2013 May 08\\ 
\enddata 
\tablenotetext{a}{Exposure times achieved on each mask were 4 $\times$ 1800s exposures, sometimes spanning two nights. }
\end{deluxetable*}

%% file: tab_sub_redshifts.tex
\begin{deluxetable}{lcccccl}
\tablewidth{0pc}
\tablecaption{CASBaH Redshift Survey$^{a}$ \label{tab:z}}
\tabletypesize{\small}
\tablehead{\colhead{CAS\_ID} & \colhead{RA} 
& \colhead{DEC} & \colhead{$z_{\rm em}$}  & \colhead{$\sigma(z)^b$} & \colhead{ZQ} 
& \colhead{Instr.} 
} 
\startdata 
4494 &212.37009 &26.32546 &0.3273 &0.0001 &4 &SDSS\\ 
4495 &212.38848 &26.47184 &0.2964 &0.0001 &4 &SDSS\\ 
4496 &212.31438 &26.36133 &0.0000 &0.0001 &3 &Hectospec\\ 
4497 &212.45230 &26.21947 &0.5490 &0.0001 &3 &Hectospec\\ 
4498 &212.39738 &26.31966 &0.8081 &0.0001 &3 &Hectospec\\ 
4499 &212.38745 &26.31115 &0.3975 &0.0001 &3 &Hectospec\\ 
4500 &212.46825 &26.34760 &0.4230 &0.0001 &4 &Hectospec\\ 
4501 &212.25992 &26.31462 &0.3698 &0.0001 &3 &Hectospec\\ 
4502 &212.28271 &26.31439 &0.0000 &0.0001 &3 &DEIMOS\\ 
4503 &212.45042 &26.33128 &0.0000 &0.0001 &3 &DEIMOS\\ 
4504 &212.46587 &26.30656 &0.0000 &0.0001 &3 &DEIMOS\\ 
4505 &212.47700 &26.31067 &0.0000 &0.0001 &3 &DEIMOS\\ 
4506 &212.35095 &26.31103 &-1.0000 &0.0001 &0 &DEIMOS\\ 
4507 &212.28904 &26.34083 &0.4264 &0.0001 &3 &DEIMOS\\ 
4508 &231.20121 &9.91829 &0.6041 &0.0002 &3 &SDSS\\ 
4509 &231.19939 &9.93566 &0.1518 &0.0001 &3 &SDSS\\ 
\enddata 
\tablenotetext{a}{The full table is provided in the on-line journal and the database; \\ this small portion of the data is presented to show the table content \\ and format.}
\tablenotetext{b}{We limit the error to a minimum of $10^{-4}$ and advise readers to adopt \\ a minimum uncertainty of 35 \kms.}
\end{deluxetable}

%% file: casbah_galaxies_astroph.bbl
\begin{thebibliography}{104}
\expandafter\ifx\csname natexlab\endcsname\relax\def\natexlab#1{#1}\fi

\bibitem[{Alam {et~al.}(2015)Alam, Albareti, Prieto, Anders, Anderson, Andrews,
  Armengaud, Aubourg, Bailey, Bautista, Beaton, Beers, Bender, Berlind,
  Beutler, Bhardwaj, Bird, Bizyaev, Blake, Blanton, Blomqvist, Bochanski,
  Bolton, Bovy, Bradley, Brandt, Brauer, Brinkmann, Brown, Brownstein, Burden,
  Burtin, Busca, Cai, Capozzi, Rosell, Carrera, Chen, Chiappini, Chojnowski,
  Chuang, Clerc, Comparat, Covey, Croft, Cuesta, Cunha, da~Costa, Da~Rio,
  Davenport, Dawson, De~Lee, Delubac, Deshpande, Dutra-Ferreira, Dwelly, Ealet,
  Ebelke, Edmondson, Eisenstein, Escoffier, Esposito, Fan, Fern{\'a}ndez-Alvar,
  Feuillet, Ak, Finley, Finoguenov, Flaherty, Fleming, Font-Ribera, Foster,
  Frinchaboy, Galbraith-Frew, Garc{\'\i}a-Hern{\'a}ndez, P{\'e}rez, Gaulme, Ge,
  G{\'e}nova-Santos, Ghezzi, Gillespie, Girardi, Goddard, Gontcho,
  Hern{\'a}ndez, Grebel, Grieb, Grieves, Gunn, Guo, Harding, Hasselquist,
  Hawley, Hayden, Hearty, Ho, Hogg, Holley-Bockelmann, Holtzman, Honscheid,
  Huehnerhoff, Jiang, Johnson, Kinemuchi, Kirkby, Kitaura, Klaene, Kneib,
  Koenig, Lam, Lan, Lang, Laurent, Goff, Leauthaud, Lee, Lee, Licquia, Liu,
  Long, L{\'o}pez-Corredoira, Lorenzo-Oliveira, Lucatello, Lundgren, Lupton,
  Mack~III, Mahadevan, Maia, Majewski, Malanushenko, Malanushenko, Manchado,
  Manera, Mao, Maraston, Marchwinski, Margala, Martell, Martig, Masters,
  McBride, McGehee, McGreer, McMahon, M{\'e}nard, Menzel, Merloni,
  M{\'e}sz{\'a}ros, Miller, Miralda-Escud{\'e}, Miyatake, Montero-Dorta, More,
  Morice-Atkinson, Morrison, Muna, Myers, Newman, Neyrinck, Nguyen, Nichol,
  Nidever, Noterdaeme, Nuza, O'Connell, O'Connell, O'Connell, Ogando, Olmstead,
  Oravetz, Oravetz, Osumi, Owen, Padgett, Padmanabhan, Paegert,
  Palanque-Delabrouille, Pan, Parejko, Park, P{\^a}ris, Pattarakijwanich,
  Pellejero-Ibanez, Pepper, Percival, P{\'e}rez-Fournon, P{\'e}rez-R{\`a}fols,
  Petitjean, Pieri, Pinsonneault, de~Mello, Prada, Prakash, Price-Whelan,
  Raddick, Rahman, Reid, Rich, Rix, Robin, Rockosi, Rodrigues,
  Rodr{\'\i}guez-Rottes, Roe, Ross, Ross, Rossi, Ruan,
  Rubi{\~n}o-Mart{\textbackslash}'{\textbackslash}in, Rykoff, Salazar-Albornoz,
  Salvato, Samushia, S{\'a}nchez, Santiago, Sayres, Schiavon, Schlegel,
  Schmidt, Schneider, Schultheis, Schwope, Sc{\'o}ccola, Sellgren, Seo, Shane,
  Shen, Shetrone, Shu, Sivarani, Skrutskie, Slosar, Smith, Sobreira, Stassun,
  Steinmetz, Strauss, Streblyanska, Swanson, Tan, Tayar, Terrien, Thakar,
  Thomas, Thompson, Tinker, Tojeiro, Troup, Vargas-Maga{\~n}a, Vazquez, Verde,
  Viel, Vogt, Wake, Wang, Weaver, Weinberg, Weiner, White, Wilson, Wisniewski,
  Wood-Vasey, Y{\`e}che, York, Zakamska, Zamora, Zasowski, Zehavi, Zhao, Zheng,
  Zhou, Zhou, Zhu, \& Zou}]{Alam:2015kq}
Alam, S., {et~al.} 2015, ApJS, 219, 12, arXiv: 1501.00963

\bibitem[{{Aracil} {et~al.}(2006){Aracil}, {Tripp}, {Bowen}, {Prochaska},
  {Chen}, \& {Frye}}]{araciletal06}
{Aracil}, B., {Tripp}, T.~M., {Bowen}, D.~V., {Prochaska}, J.~X., {Chen},
  H.-W., \& {Frye}, B.~L. 2006, \mnras, 367, 139

\bibitem[{{Bahcall} \& {Peebles}(1969)}]{bp69}
{Bahcall}, J.~N., \& {Peebles}, P.~J.~E. 1969, \apjl, 156, L7+

\bibitem[{{Battisti} {et~al.}(2016){Battisti}, {Calzetti}, \&
  {Chary}}]{Battisti:2016aa}
{Battisti}, A.~J., {Calzetti}, D., \& {Chary}, R.-R. 2016, \apj, 818, 13

\bibitem[{{Battisti} {et~al.}(2017){Battisti}, {Calzetti}, \&
  {Chary}}]{Battisti:2017aa}
---. 2017, \apj, 851, 90

\bibitem[{{Bertin} \& {Arnouts}(1996)}]{bertin1996}
{Bertin}, E., \& {Arnouts}, S. 1996, \aaps, 117, 393

\bibitem[{{Bertin} {et~al.}(2002){Bertin}, {Mellier}, {Radovich}, {Missonnier},
  {Didelon}, \& {Morin}}]{bertin2002}
{Bertin}, E., {Mellier}, Y., {Radovich}, M., {Missonnier}, G., {Didelon}, P.,
  \& {Morin}, B. 2002, in Astronomical Society of the Pacific Conference
  Series, Vol. 281, Astronomical Data Analysis Software and Systems XI, ed.
  D.~A. {Bohlender}, D.~{Durand}, \& T.~H. {Handley}, 228

\bibitem[{{Bielby} \& {et al.}(2019)}]{qsages}
{Bielby}, R., \& {et al.} 2019, in prep.

\bibitem[{{Bond} {et~al.}(1996){Bond}, {Kofman}, \& {Pogosyan}}]{bond+96}
{Bond}, J.~R., {Kofman}, L., \& {Pogosyan}, D. 1996, \nat, 380, 603

\bibitem[{{Booth} {et~al.}(2012){Booth}, {Schaye}, {Delgado}, \& {Dalla
  Vecchia}}]{booth+12}
{Booth}, C.~M., {Schaye}, J., {Delgado}, J.~D., \& {Dalla Vecchia}, C. 2012,
  \mnras, 420, 1053

\bibitem[{{Bowen} {et~al.}(2002){Bowen}, {Pettini}, \& {Blades}}]{bowen+02}
{Bowen}, D.~V., {Pettini}, M., \& {Blades}, J.~C. 2002, \apj, 580, 169

\bibitem[{{Bruzual} \& {Charlot}(2003)}]{Bruzual:2003aa}
{Bruzual}, G., \& {Charlot}, S. 2003, \mnras, 344, 1000

\bibitem[{{Buat} {et~al.}(2011){Buat}, {Giovannoli}, {Heinis}, {Charmandaris},
  {Coia}, {Daddi}, {Dickinson}, {Elbaz}, {Hwang}, {Morrison}, {Dasyra},
  {Aussel}, {Altieri}, {Dannerbauer}, {Kartaltepe}, {Leiton}, {Magdis},
  {Magnelli}, \& {Popesso}}]{Buat:2011aa}
{Buat}, V., {et~al.} 2011, \aap, 533, A93

\bibitem[{{Burchett} {et~al.}(2019){Burchett}, {et al.}, {et al.}, \& {et
  al.}}]{casbah_ovi}
{Burchett}, J., {et al.}, {et al.}, \& {et al.} 2019, in prep.

\bibitem[{{Burchett} {et~al.}(2018){Burchett}, {Tripp}, {Prochaska}, {Werk},
  {Tumlinson}, {Howk}, {Willmer}, {Lehner}, {Meiring}, {Bowen}, {Bordoloi},
  {Peeples}, {Jenkins}, {O'Meara}, {Tejos}, \& {Katz}}]{burchett19a}
{Burchett}, J.~N., {et~al.} 2018, arXiv e-prints

\bibitem[{{Burchett} {et~al.}(2015){Burchett}, {Tripp}, {Prochaska}, {Werk},
  {Tumlinson}, {O'Meara}, {Bordoloi}, {Katz}, \& {Willmer}}]{burchett+15}
---. 2015, \apj, 815, 91

\bibitem[{Burchett {et~al.}(2013)Burchett, Tripp, Werk, Howk, Prochaska, Ford,
  \& Dav{\'e}}]{Burchett:2013qy}
Burchett, J.~N., Tripp, T.~M., Werk, J.~K., Howk, J.~C., Prochaska, J.~X.,
  Ford, A.~B., \& Dav{\'e}, R. 2013, ApJL, 779, L17

\bibitem[{Calzetti {et~al.}(2000)Calzetti, Armus, Bohlin, Kinney, Koornneef, \&
  Storchi-Bergmann}]{Calzetti:2000lr}
Calzetti, D., Armus, L., Bohlin, R.~C., Kinney, A.~L., Koornneef, J., \&
  Storchi-Bergmann, T. 2000, ApJ, 533, 682

\bibitem[{{Cantalupo} {et~al.}(2014){Cantalupo}, {Arrigoni-Battaia},
  {Prochaska}, {Hennawi}, \& {Madau}}]{cantalupo14}
{Cantalupo}, S., {Arrigoni-Battaia}, F., {Prochaska}, J.~X., {Hennawi}, J.~F.,
  \& {Madau}, P. 2014, \nat, 506, 63

\bibitem[{{Cautun} {et~al.}(2013){Cautun}, {van de Weygaert}, \&
  {Jones}}]{cautun+13}
{Cautun}, M., {van de Weygaert}, R., \& {Jones}, B.~J.~T. 2013, \mnras, 429,
  1286

\bibitem[{Chabrier(2003)}]{Chabrier:2003pd}
Chabrier, G. 2003, PASP, 115, 763

\bibitem[{{Chambers} {et~al.}(2016){Chambers}, {Magnier}, {Metcalfe},
  {Flewelling}, {Huber}, {Waters}, {Denneau}, {Draper}, {Farrow}, {Finkbeiner},
  {Holmberg}, {Koppenhoefer}, {Price}, {Saglia}, {Schlafly}, {Smartt},
  {Sweeney}, {Wainscoat}, {Burgett}, {Grav}, {Heasley}, {Hodapp}, {Jedicke},
  {Kaiser}, {Kudritzki}, {Luppino}, {Lupton}, {Monet}, {Morgan}, {Onaka},
  {Stubbs}, {Tonry}, {Banados}, {Bell}, {Bender}, {Bernard}, {Botticella},
  {Casertano}, {Chastel}, {Chen}, {Chen}, {Cole}, {Deacon}, {Frenk},
  {Fitzsimmons}, {Gezari}, {Goessl}, {Goggia}, {Goldman}, {Grebel}, {Hambly},
  {Hasinger}, {Heavens}, {Heckman}, {Henderson}, {Henning}, {Holman}, {Hopp},
  {Ip}, {Isani}, {Keyes}, {Koekemoer}, {Kotak}, {Long}, {Lucey}, {Liu},
  {Martin}, {McLean}, {Morganson}, {Murphy}, {Nieto-Santisteban}, {Norberg},
  {Peacock}, {Pier}, {Postman}, {Primak}, {Rae}, {Rest}, {Riess}, {Riffeser},
  {Rix}, {Roser}, {Schilbach}, {Schultz}, {Scolnic}, {Szalay}, {Seitz},
  {Shiao}, {Small}, {Smith}, {Soderblom}, {Taylor}, {Thakar}, {Thiel},
  {Thilker}, {Urata}, {Valenti}, {Walter}, {Watters}, {Werner}, {White},
  {Wood-Vasey}, \& {Wyse}}]{Chambers:2016aa}
{Chambers}, K.~C., {et~al.} 2016, ArXiv e-prints

\bibitem[{{Chen} \& {Mulchaey}(2009)}]{cm09}
{Chen}, H., \& {Mulchaey}, J.~S. 2009, \apj, 701, 1219

\bibitem[{{Chen} {et~al.}(2005){Chen}, {Prochaska}, {Weiner}, {Mulchaey}, \&
  {Williger}}]{cpw+05}
{Chen}, H.-W., {Prochaska}, J.~X., {Weiner}, B.~J., {Mulchaey}, J.~S., \&
  {Williger}, G.~M. 2005, \apjl, 629, L25

\bibitem[{{Chen} \& {Tinker}(2008)}]{ct08}
{Chen}, H.-W., \& {Tinker}, J.~L. 2008, \apj, 687, 745

\bibitem[{{Coil} {et~al.}(2017){Coil}, {Mendez}, {Eisenstein}, \&
  {Moustakas}}]{coil+17}
{Coil}, A.~L., {Mendez}, A.~J., {Eisenstein}, D.~J., \& {Moustakas}, J. 2017,
  \apj, 838, 87

\bibitem[{{Croft} {et~al.}(2002){Croft}, {Weinberg}, {Bolte}, {Burles},
  {Hernquist}, {Katz}, {Kirkman}, \& {Tytler}}]{cwb+03}
{Croft}, R.~A.~C., {Weinberg}, D.~H., {Bolte}, M., {Burles}, S., {Hernquist},
  L., {Katz}, N., {Kirkman}, D., \& {Tytler}, D. 2002, \apj, 581, 20

\bibitem[{{Cutri} \& {et al.}(2013)}]{Cutri:2013aa}
{Cutri}, R.~M., \& {et al.} 2013, VizieR Online Data Catalog, 2328

\bibitem[{{Dale} {et~al.}(2014){Dale}, {Helou}, {Magdis}, {Armus},
  {D{\'{\i}}az-Santos}, \& {Shi}}]{Dale:2014aa}
{Dale}, D.~A., {Helou}, G., {Magdis}, G.~E., {Armus}, L., {D{\'{\i}}az-Santos},
  T., \& {Shi}, Y. 2014, \apj, 784, 83

\bibitem[{{Danforth} {et~al.}(2016){Danforth}, {Keeney}, {Tilton}, {Shull},
  {Stocke}, {Stevans}, {Pieri}, {Savage}, {France}, {Syphers}, {Smith},
  {Green}, {Froning}, {Penton}, \& {Osterman}}]{danforth16}
{Danforth}, C.~W., {et~al.} 2016, \apj, 817, 111

\bibitem[{{Davis} {et~al.}(1985){Davis}, {Efstathiou}, {Frenk}, \&
  {White}}]{davis+85}
{Davis}, M., {Efstathiou}, G., {Frenk}, C.~S., \& {White}, S.~D.~M. 1985, \apj,
  292, 371

\bibitem[{{Dawson} {et~al.}(2013){Dawson}, {Schlegel}, {Ahn}, {Anderson},
  {Aubourg}, {Bailey}, {Barkhouser}, {Bautista}, {Beifiori}, {Berlind},
  {Bhardwaj}, {Bizyaev}, {Blake}, {Blanton}, {Blomqvist}, {Bolton}, {Borde},
  {Bovy}, {Brandt}, {Brewington}, {Brinkmann}, {Brown}, {Brownstein}, {Bundy},
  {Busca}, {Carithers}, {Carnero}, {Carr}, {Chen}, {Comparat}, {Connolly},
  {Cope}, {Croft}, {Cuesta}, {da Costa}, {Davenport}, {Delubac}, {de Putter},
  {Dhital}, {Ealet}, {Ebelke}, {Eisenstein}, {Escoffier}, {Fan}, {Filiz Ak},
  {Finley}, {Font-Ribera}, {G{\'e}nova-Santos}, {Gunn}, {Guo}, {Haggard},
  {Hall}, {Hamilton}, {Harris}, {Harris}, {Ho}, {Hogg}, {Holder}, {Honscheid},
  {Huehnerhoff}, {Jordan}, {Jordan}, {Kauffmann}, {Kazin}, {Kirkby}, {Klaene},
  {Kneib}, {Le Goff}, {Lee}, {Long}, {Loomis}, {Lundgren}, {Lupton}, {Maia},
  {Makler}, {Malanushenko}, {Malanushenko}, {Mandelbaum}, {Manera}, {Maraston},
  {Margala}, {Masters}, {McBride}, {McDonald}, {McGreer}, {McMahon}, {Mena},
  {Miralda-Escud{\'e}}, {Montero-Dorta}, {Montesano}, {Muna}, {Myers},
  {Naugle}, {Nichol}, {Noterdaeme}, {Nuza}, {Olmstead}, {Oravetz}, {Oravetz},
  {Owen}, {Padmanabhan}, {Palanque-Delabrouille}, {Pan}, {Parejko},
  {P{\^a}ris}, {Percival}, {P{\'e}rez-Fournon}, {P{\'e}rez-R{\`a}fols},
  {Petitjean}, {Pfaffenberger}, {Pforr}, {Pieri}, {Prada}, {Price-Whelan},
  {Raddick}, {Rebolo}, {Rich}, {Richards}, {Rockosi}, {Roe}, {Ross}, {Ross},
  {Rossi}, {Rubi{\~n}o-Martin}, {Samushia}, {S{\'a}nchez}, {Sayres}, {Schmidt},
  {Schneider}, {Sc{\'o}ccola}, {Seo}, {Shelden}, {Sheldon}, {Shen}, {Shu},
  {Slosar}, {Smee}, {Snedden}, {Stauffer}, {Steele}, {Strauss}, {Streblyanska},
  {Suzuki}, {Swanson}, {Tal}, {Tanaka}, {Thomas}, {Tinker}, {Tojeiro},
  {Tremonti}, {Vargas Maga{\~n}a}, {Verde}, {Viel}, {Wake}, {Watson}, {Weaver},
  {Weinberg}, {Weiner}, {West}, {White}, {Wood-Vasey}, {Yeche}, {Zehavi},
  {Zhao}, \& {Zheng}}]{eBOSS}
{Dawson}, K.~S., {et~al.} 2013, \aj, 145, 10

\bibitem[{{de Jong} {et~al.}(2017){de Jong}, {Verdois Kleijn}, {Erben},
  {Hildebrandt}, {Kuijken}, {Sikkema}, {Brescia}, {Bilicki}, {Napolitano},
  {Amaro}, {Begeman}, {Boxhoorn}, {Buddelmeijer}, {Cavuoti}, {Getman}, {Grado},
  {Helmich}, {Huang}, {Irisarri}, {La Barbera}, {Longo}, {McFarland},
  {Nakajima}, {Paolillo}, {Puddu}, {Radovich}, {Rifatto}, {Tortora},
  {Valentijn}, {Vellucci}, {Vriend}, {Amon}, {Blake}, {Choi}, {Conti}, {Gwyn},
  {Herbonnet}, {Heymans}, {Hoekstra}, {Klaes}, {Merten}, {Miller}, {Schneider},
  \& {Viola}}]{de-Jong:2017aa}
{de Jong}, J.~T.~A., {et~al.} 2017, \aap, 604, A134

\bibitem[{{Dey} {et~al.}(2018){Dey}, {Schlegel}, {Lang}, {Blum}, {Burleigh},
  {Fan}, {Findlay}, {Finkbeiner}, {Herrera}, {Juneau}, {Landriau}, {Levi},
  {McGreer}, {Meisner}, {Myers}, {Moustakas}, {Nugent}, {Patej}, {Schlafly},
  {Walker}, {Valdes}, {Weaver}, {Yeche}, {Zou}, {Zhou}, {Abareshi}, {Abbott},
  {Abolfathi}, {Aguilera}, {Allen}, {Alvarez}, {Annis}, {Aubert}, {Bell},
  {BenZvi}, {Bielby}, {Bolton}, {Briceno}, {Buckley-Geer}, {Butler},
  {Calamida}, {Carlberg}, {Carter}, {Casas}, {Castander}, {Choi}, {Comparat},
  {Cukanovaite}, {Delubac}, {DeVries}, {Dey}, {Dhungana}, {Dickinson}, {Ding},
  {Donaldson}, {Duan}, {Duckworth}, {Eftekharzadeh}, {Eisenstein}, {Etourneau},
  {Fagrelius}, {Farihi}, {Fitzpatrick}, {Font-Ribera}, {Fulmer}, {Gansicke},
  {Gaztanaga}, {George}, {Gerdes}, {Gontcho}, {Green}, {Guy}, {Harmer},
  {Hernandez}, {Honscheid}, {Lijuan}, {Huang}, {James}, {Jannuzi}, {Jiang},
  {Joyce}, {Karcher}, {Karkar}, {Kehoe}, {Kneib}, {Kueter-Young}, {Lan},
  {Lauer}, {Le Guillou}, {Le Van Suu}, {Lee}, {Lesser}, {Li}, {Mann},
  {Marshall}, {Mart{\'{\i}}nez-V{\'a}zquez}, {Martini}, {du Mas des Bourboux},
  {McManus}, {Menard}, {Metcalfe}, {Mu{\~n}oz-Guti{\'e}rrez}, {Najita},
  {Napier}, {Narayan}, {Newman}, {Nie}, {Nord}, {Norman}, {Olsen}, {Paat},
  {Palanque-Delabrouille}, {Peng}, {Poppett}, {Poremba}, {Prakash},
  {Rabinowitz}, {Raichoor}, {Rezaie}, {Robertson}, {Roe}, {Ross}, {Ross},
  {Rudnick}, {Safonova}, {Saha}, {Sanchez}, {Schweiker}, {Scott}, {Seo},
  {Shan}, {Silva}, {Soto}, {Sprayberry}, {Staten}, {Stillman}, {Stupak},
  {Summers}, {Sien Tie}, {Tirado}, {Vargas-Magana}, {Vivas}, {Wechsler},
  {Williams}, {Yang}, {Yang}, {Yapici}, {Zaritsky}, {Zenteno}, {Zhang},
  {Zhang}, {Zhou}, \& {Zhou}}]{Dey:2018aa}
{Dey}, A., {et~al.} 2018, ArXiv e-prints

\bibitem[{{Eisenstein} {et~al.}(2001){Eisenstein}, {Annis}, {Gunn}, {Szalay},
  {Connolly}, {Nichol}, {Bahcall}, {Bernardi}, {Burles}, {Castander},
  {Fukugita}, {Hogg}, {Ivezi{\'c}}, {Knapp}, {Lupton}, {Narayanan}, {Postman},
  {Reichart}, {Richmond}, {Schneider}, {Schlegel}, {Strauss}, {SubbaRao},
  {Tucker}, {Vanden Berk}, {Vogeley}, {Weinberg}, \& {Yanny}}]{eisenstein+01}
{Eisenstein}, D.~J., {et~al.} 2001, \aj, 122, 2267

\bibitem[{{Faber} {et~al.}(2003){Faber}, {Phillips}, {Kibrick}, {Alcott},
  {Allen}, {Burrous}, {Cantrall}, {Clarke}, {Coil}, {Cowley}, {Davis}, {Deich},
  {Dietsch}, {Gilmore}, {Harper}, {Hilyard}, {Lewis}, {McVeigh}, {Newman},
  {Osborne}, {Schiavon}, {Stover}, {Tucker}, {Wallace}, {Wei}, {Wirth}, \&
  {Wright}}]{fpk+03}
{Faber}, S.~M., {et~al.} 2003, in Proceedings of the SPIE, ed. M.~Iye \&
  A.~F.~M. Moorwood, 1657--1669

\bibitem[{{Fabricant} {et~al.}(2005){Fabricant}, {Fata}, {Roll}, {Hertz},
  {Caldwell}, {Gauron}, {Geary}, {McLeod}, {Szentgyorgyi}, {Zajac}, {Kurtz},
  {Barberis}, {Bergner}, {Brown}, {Conroy}, {Eng}, {Geller}, {Goddard},
  {Honsa}, {Mueller}, {Mink}, {Ordway}, {Tokarz}, {Woods}, {Wyatt}, {Epps}, \&
  {Dell'Antonio}}]{hectospec}
{Fabricant}, D., {et~al.} 2005, \pasp, 117, 1411

\bibitem[{{Finn} {et~al.}(2016){Finn}, {Morris}, {Tejos}, {Crighton}, {Perry},
  {Fumagalli}, {Bielby}, {Theuns}, {Schaye}, {Shanks}, {Liske}, {Gunawardhana},
  \& {Bartle}}]{finn+16}
{Finn}, C.~W., {et~al.} 2016, \mnras, 460, 590

\bibitem[{{Flewelling} {et~al.}(2016){Flewelling}, {Magnier}, {Chambers},
  {Heasley}, {Holmberg}, {Huber}, {Sweeney}, {Waters}, {Chen}, {Farrow},
  {Hasinger}, {Henderson}, {Long}, {Metcalfe}, {Nieto-Santisteban}, {Norberg},
  {Saglia}, {Szalay}, {Rest}, {Thakar}, {Tonry}, {Valenti}, {Werner}, {White},
  {Denneau}, {Draper}, {Hodapp}, {Jedicke}, {Kaiser}, {Kudritzki}, {Price},
  {Wainscoat}, {Chastel}, {McClean}, {Postman}, \& {Shiao}}]{Flewelling:2016aa}
{Flewelling}, H.~A., {et~al.} 2016, ArXiv e-prints

\bibitem[{{Ford} {et~al.}(2016){Ford}, {Werk}, {Dav{\'e}}, {Tumlinson},
  {Bordoloi}, {Katz}, {Kollmeier}, {Oppenheimer}, {Peeples}, {Prochaska}, \&
  {Weinberg}}]{ford+16}
{Ford}, A.~B., {et~al.} 2016, \mnras, 459, 1745

\bibitem[{{Gaia Collaboration} {et~al.}(2016{\natexlab{a}}){Gaia
  Collaboration}, {Brown}, {Vallenari}, {Prusti}, {de Bruijne}, {Mignard},
  {Drimmel}, {Babusiaux}, {Bailer-Jones}, {Bastian}, \&
  et~al.}]{gaia-collaboration2016b}
{Gaia Collaboration} {et~al.} 2016{\natexlab{a}}, \aap, 595, A2

\bibitem[{{Gaia Collaboration} {et~al.}(2016{\natexlab{b}}){Gaia
  Collaboration}, {Prusti}, {de Bruijne}, {Brown}, {Vallenari}, {Babusiaux},
  {Bailer-Jones}, {Bastian}, {Biermann}, {Evans}, \&
  et~al.}]{gaia-collaboration2016a}
---. 2016{\natexlab{b}}, \aap, 595, A1

\bibitem[{{Ganguly} {et~al.}(2013){Ganguly}, {Lynch}, {Charlton}, {Eracleous},
  {Tripp}, {Palma}, {Sembach}, {Misawa}, {Masiero}, {Milutinovic}, {Lackey}, \&
  {Jones}}]{ganguly+13}
{Ganguly}, R., {et~al.} 2013, \mnras, 435, 1233

\bibitem[{{Giallongo} {et~al.}(2008){Giallongo}, {Ragazzoni}, {Grazian},
  {Baruffolo}, {Beccari}, {de Santis}, {Diolaiti}, {di Paola}, {Farinato},
  {Fontana}, {Gallozzi}, {Gasparo}, {Gentile}, {Green}, {Hill}, {Kuhn},
  {Pasian}, {Pedichini}, {Radovich}, {Salinari}, {Smareglia}, {Speziali},
  {Testa}, {Thompson}, {Vernet}, \& {Wagner}}]{lbt-lbc}
{Giallongo}, E., {et~al.} 2008, \aap, 482, 349

\bibitem[{{Gould} \& {Weinberg}(1996)}]{GW96}
{Gould}, A., \& {Weinberg}, D.~H. 1996, \apj, 468, 462

\bibitem[{{Green} {et~al.}(2012){Green}, {Froning}, {Osterman}, {Ebbets},
  {Heap}, {Leitherer}, {Linsky}, {Savage}, {Sembach}, {Shull}, {Siegmund},
  {Snow}, {Spencer}, {Stern}, {Stocke}, {Welsh}, {B{\'e}land}, {Burgh},
  {Danforth}, {France}, {Keeney}, {McPhate}, {Penton}, {Andrews},
  {Brownsberger}, {Morse}, \& {Wilkinson}}]{cos}
{Green}, J.~C., {et~al.} 2012, \apj, 744, 60

\bibitem[{{Hennawi} \& {Prochaska}(2007)}]{QPQ2}
{Hennawi}, J.~F., \& {Prochaska}, J.~X. 2007, \apj, 655, 735

\bibitem[{{Hewett} \& {Wild}(2010)}]{hw11}
{Hewett}, P.~C., \& {Wild}, V. 2010, \mnras, 405, 2302

\bibitem[{{Howk}(2019)}]{howk2019}
{Howk}, J.~C. e.~a. 2019, in prep.

\bibitem[{{Hudelot} {et~al.}(2012){Hudelot}, {Cuillandre}, {Withington},
  {Goranova}, {McCracken}, {Magnard}, {Mellier}, {Regnault}, {Betoule},
  {Aussel}, {Kavelaars}, {Fernique}, {Bonnarel}, {Ochsenbein}, \&
  {Ilbert}}]{Hudelot:2012aa}
{Hudelot}, P., {et~al.} 2012, VizieR Online Data Catalog, 2317

\bibitem[{{Johnson} {et~al.}(2015){Johnson}, {Chen}, \&
  {Mulchaey}}]{johnson+15}
{Johnson}, S.~D., {Chen}, H.-W., \& {Mulchaey}, J.~S. 2015, \mnras, 449, 3263

\bibitem[{{Keeney} {et~al.}(2018){Keeney}, {Stocke}, {Pratt}, {Davis},
  {Syphers}, {Danforth}, {Shull}, {Froning}, {Green}, {Penton}, \&
  {Savage}}]{keeney+18}
{Keeney}, B.~A., {et~al.} 2018, \apjs, 237, 11

\bibitem[{{Lan} \& {Mo}(2018)}]{lm18}
{Lan}, T.-W., \& {Mo}, H. 2018, ArXiv e-prints, arXiv:1806.05786

\bibitem[{{Landy} \& {Szalay}(1993)}]{landy93}
{Landy}, S.~D., \& {Szalay}, A.~S. 1993, \apj, 412, 64

\bibitem[{{Lawrence} {et~al.}(2007){Lawrence}, {Warren}, {Almaini}, {Edge},
  {Hambly}, {Jameson}, {Lucas}, {Casali}, {Adamson}, {Dye}, {Emerson},
  {Foucaud}, {Hewett}, {Hirst}, {Hodgkin}, {Irwin}, {Lodieu}, {McMahon},
  {Simpson}, {Smail}, {Mortlock}, \& {Folger}}]{Lawrence:2007aa}
{Lawrence}, A., {et~al.} 2007, \mnras, 379, 1599

\bibitem[{{Lehner} {et~al.}(2013){Lehner}, {Howk}, {Tripp}, {Tumlinson},
  {Prochaska}, {O'Meara}, {Thom}, {Werk}, {Fox}, \& {Ribaudo}}]{lht+13}
{Lehner}, N., {et~al.} 2013, \apj, 770, 138

\bibitem[{{Lo Faro} {et~al.}(2017){Lo Faro}, {Buat}, {Roehlly},
  {Alvarez-Marquez}, {Burgarella}, {Silva}, \& {Efstathiou}}]{Lo-Faro:2017aa}
{Lo Faro}, B., {Buat}, V., {Roehlly}, Y., {Alvarez-Marquez}, J., {Burgarella},
  D., {Silva}, L., \& {Efstathiou}, A. 2017, \mnras, 472, 1372

\bibitem[{{Luki{\'c}} {et~al.}(2015){Luki{\'c}}, {Stark}, {Nugent}, {White},
  {Meiksin}, \& {Almgren}}]{lukic+15}
{Luki{\'c}}, Z., {Stark}, C.~W., {Nugent}, P., {White}, M., {Meiksin}, A.~A.,
  \& {Almgren}, A. 2015, \mnras, 446, 3697

\bibitem[{{McDowell} {et~al.}(1995){McDowell}, {Canizares}, {Elvis},
  {Lawrence}, {Markoff}, {Mathur}, \& {Wilkes}}]{mcdowell95}
{McDowell}, J.~C., {Canizares}, C., {Elvis}, M., {Lawrence}, A., {Markoff}, S.,
  {Mathur}, S., \& {Wilkes}, B.~J. 1995, \apj, 450, 585

\bibitem[{{Meiring} {et~al.}(2013){Meiring}, {Tripp}, {Werk}, {Howk},
  {Jenkins}, {Prochaska}, {Lehner}, \& {Sembach}}]{mtw+13}
{Meiring}, J.~D., {Tripp}, T.~M., {Werk}, J.~K., {Howk}, J.~C., {Jenkins},
  E.~B., {Prochaska}, J.~X., {Lehner}, N., \& {Sembach}, K.~R. 2013, \apj, 767,
  49

\bibitem[{{Miralda-Escud{\'e}} {et~al.}(1996){Miralda-Escud{\'e}}, {Cen},
  {Ostriker}, \& {Rauch}}]{mco+96}
{Miralda-Escud{\'e}}, J., {Cen}, R., {Ostriker}, J.~P., \& {Rauch}, M. 1996,
  \apj, 471, 582

\bibitem[{{Misawa} {et~al.}(2007){Misawa}, {Tytler}, {Iye}, {Kirkman},
  {Suzuki}, {Lubin}, \& {Kashikawa}}]{misawa+07}
{Misawa}, T., {Tytler}, D., {Iye}, M., {Kirkman}, D., {Suzuki}, N., {Lubin},
  D., \& {Kashikawa}, N. 2007, \aj, 134, 1634

\bibitem[{{Morris} {et~al.}(1993){Morris}, {Weymann}, {Dressler}, {McCarthy},
  {Smith}, {Terrile}, {Giovanelli}, \& {Irwin}}]{mwd+93}
{Morris}, S.~L., {Weymann}, R.~J., {Dressler}, A., {McCarthy}, P.~J., {Smith},
  B.~A., {Terrile}, R.~J., {Giovanelli}, R., \& {Irwin}, M. 1993, \apj, 419,
  524

\bibitem[{{Moustakas} {et~al.}(2013){Moustakas}, {Coil}, {Aird}, {Blanton},
  {Cool}, {Eisenstein}, {Mendez}, {Wong}, {Zhu}, \& {Arnouts}}]{moustakas+13}
{Moustakas}, J., {et~al.} 2013, \apj, 767, 50

\bibitem[{{Muzahid} {et~al.}(2013){Muzahid}, {Srianand}, {Arav}, {Savage}, \&
  {Narayanan}}]{muzahid+13}
{Muzahid}, S., {Srianand}, R., {Arav}, N., {Savage}, B.~D., \& {Narayanan}, A.
  2013, \mnras, 431, 2885

\bibitem[{{Newman} {et~al.}(2013){Newman}, {Cooper}, {Davis}, {Faber}, {Coil},
  {Guhathakurta}, {Koo}, {Phillips}, {Conroy}, {Dutton}, {Finkbeiner}, {Gerke},
  {Rosario}, {Weiner}, {Willmer}, {Yan}, {Harker}, {Kassin}, {Konidaris},
  {Lai}, {Madgwick}, {Noeske}, {Wirth}, {Connolly}, {Kaiser}, {Kirby},
  {Lemaux}, {Lin}, {Lotz}, {Luppino}, {Marinoni}, {Matthews}, {Metevier}, \&
  {Schiavon}}]{Newman2013}
{Newman}, J.~A., {et~al.} 2013, \apjs, 208, 5

\bibitem[{{Noll} {et~al.}(2009){Noll}, {Burgarella}, {Giovannoli}, {Buat},
  {Marcillac}, \& {Mu{\~n}oz-Mateos}}]{Noll:2009aa}
{Noll}, S., {Burgarella}, D., {Giovannoli}, E., {Buat}, V., {Marcillac}, D., \&
  {Mu{\~n}oz-Mateos}, J.~C. 2009, \aap, 507, 1793

\bibitem[{{Nuza} {et~al.}(2013){Nuza}, {S{\'a}nchez}, {Prada}, {Klypin},
  {Schlegel}, {Gottl{\"o}ber}, {Montero-Dorta}, {Manera}, {McBride}, {Ross},
  {Angulo}, {Blanton}, {Bolton}, {Favole}, {Samushia}, {Montesano}, {Percival},
  {Padmanabhan}, {Steinmetz}, {Tinker}, {Skibba}, {Schneider}, {Guo}, {Zehavi},
  {Zheng}, {Bizyaev}, {Malanushenko}, {Malanushenko}, {Oravetz}, {Oravetz}, \&
  {Shelden}}]{nuza13}
{Nuza}, S.~E., {et~al.} 2013, \mnras, 432, 743

\bibitem[{{Palanque-Delabrouille} {et~al.}(2013){Palanque-Delabrouille},
  {Y{\`e}che}, {Borde}, {Le Goff}, {Rossi}, {Viel}, {Aubourg}, {Bailey},
  {Bautista}, {Blomqvist}, {Bolton}, {Bolton}, {Busca}, {Carithers}, {Croft},
  {Dawson}, {Delubac}, {Font-Ribera}, {Ho}, {Kirkby}, {Lee}, {Margala},
  {Miralda-Escud{\'e}}, {Muna}, {Myers}, {Noterdaeme}, {P{\^a}ris},
  {Petitjean}, {Pieri}, {Rich}, {Rollinde}, {Ross}, {Schlegel}, {Schneider},
  {Slosar}, \& {Weinberg}}]{palanque+13}
{Palanque-Delabrouille}, N., {et~al.} 2013, \aap, 559, A85

\bibitem[{Pedregosa {et~al.}(2011)Pedregosa, Varoquaux, Gramfort, Michel,
  Thirion, Grisel, Blondel, Prettenhofer, Weiss, Dubourg, Vanderplas, Passos,
  Cournapeau, Brucher, Perrot, \& Duchesnay}]{scikit-learn}
Pedregosa, F., {et~al.} 2011, Journal of Machine Learning Research, 12, 2825

\bibitem[{{Penton} {et~al.}(2002){Penton}, {Stocke}, \& {Shull}}]{pss02}
{Penton}, S.~V., {Stocke}, J.~T., \& {Shull}, J.~M. 2002, \apj, 565, 720

\bibitem[{{Pratt} {et~al.}(2018){Pratt}, {Stocke}, {Keeney}, \&
  {Danforth}}]{pratt18}
{Pratt}, C.~T., {Stocke}, J.~T., {Keeney}, B.~A., \& {Danforth}, C.~W. 2018,
  \apj, 855, 18

\bibitem[{{Prochaska} {et~al.}(2013){Prochaska}, {Hennawi}, {Lee}, {Cantalupo},
  {Bovy}, {Djorgovski}, {Ellison}, {Wingyee Lau}, {Martin}, {Myers}, {Rubin},
  \& {Simcoe}}]{qpq6}
{Prochaska}, J.~X., {et~al.} 2013, \apj, 776, 136

\bibitem[{{Prochaska} {et~al.}(2011{\natexlab{a}}){Prochaska}, {Weiner},
  {Chen}, {Cooksey}, \& {Mulchaey}}]{ovi_paper4}
{Prochaska}, J.~X., {Weiner}, B., {Chen}, H.-W., {Cooksey}, K.~L., \&
  {Mulchaey}, J.~S. 2011{\natexlab{a}}, \apjs, 193, 28

\bibitem[{{Prochaska} {et~al.}(2011{\natexlab{b}}){Prochaska}, {Weiner},
  {Chen}, {Mulchaey}, \& {Cooksey}}]{pwc+11}
{Prochaska}, J.~X., {Weiner}, B., {Chen}, H.-W., {Mulchaey}, J., \& {Cooksey},
  K. 2011{\natexlab{b}}, \apj, 740, 91

\bibitem[{{Ribaudo} {et~al.}(2011{\natexlab{a}}){Ribaudo}, {Lehner}, \&
  {Howk}}]{ribaudo11}
{Ribaudo}, J., {Lehner}, N., \& {Howk}, J.~C. 2011{\natexlab{a}}, \apj, 736, 42

\bibitem[{{Ribaudo} {et~al.}(2011{\natexlab{b}}){Ribaudo}, {Lehner}, {Howk},
  {Werk}, {Tripp}, {Prochaska}, {Meiring}, \& {Tumlinson}}]{rlh+11}
{Ribaudo}, J., {Lehner}, N., {Howk}, J.~C., {Werk}, J.~K., {Tripp}, T.~M.,
  {Prochaska}, J.~X., {Meiring}, J.~D., \& {Tumlinson}, J. 2011{\natexlab{b}},
  \apj, 743, 207

\bibitem[{Rodrigo {et~al.}(2012)Rodrigo, Solano, \& Bayo}]{svoFilterService}
Rodrigo, C., Solano, E., \& Bayo, A. 2012, The SVO Filter Profile Service

\bibitem[{{Sand} {et~al.}(2009){Sand}, {Olszewski}, {Willman}, {Zaritsky},
  {Seth}, {Harris}, {Piatek}, \& {Saha}}]{sand2009}
{Sand}, D.~J., {Olszewski}, E.~W., {Willman}, B., {Zaritsky}, D., {Seth}, A.,
  {Harris}, J., {Piatek}, S., \& {Saha}, A. 2009, \apj, 704, 898

\bibitem[{{Savage} {et~al.}(2010){Savage}, {Narayanan}, {Wakker}, {Stocke},
  {Keeney}, {Shull}, {Sembach}, {Yao}, \& {Green}}]{savage10}
{Savage}, B.~D., {et~al.} 2010, \apj, 719, 1526

\bibitem[{{Schlafly} \& {Finkbeiner}(2011)}]{Schlafly:2011aa}
{Schlafly}, E.~F., \& {Finkbeiner}, D.~P. 2011, \apj, 737, 103

\bibitem[{{Simcoe} {et~al.}(2004){Simcoe}, {Sargent}, \& {Rauch}}]{simcoe04}
{Simcoe}, R.~A., {Sargent}, W.~L.~W., \& {Rauch}, M. 2004, \apj, 606, 92

\bibitem[{{Slosar} {et~al.}(2013){Slosar}, {Ir{\v s}i{\v c}}, {Kirkby},
  {Bailey}, {Busca}, {Delubac}, {Rich}, {Aubourg}, {Bautista}, {Bhardwaj},
  {Blomqvist}, {Bolton}, {Bovy}, {Brownstein}, {Carithers}, {Croft}, {Dawson},
  {Font-Ribera}, {Le Goff}, {Ho}, {Honscheid}, {Lee}, {Margala}, {McDonald},
  {Medolin}, {Miralda-Escud{\'e}}, {Myers}, {Nichol}, {Noterdaeme},
  {Palanque-Delabrouille}, {P{\^a}ris}, {Petitjean}, {Pieri}, {Pi{\v s}kur},
  {Roe}, {Ross}, {Rossi}, {Schlegel}, {Schneider}, {Suzuki}, {Sheldon},
  {Seljak}, {Viel}, {Weinberg}, \& {Y{\`e}che}}]{sik+13}
{Slosar}, A., {et~al.} 2013, \jcap, 4, 26

\bibitem[{{Smith} {et~al.}(2003){Smith}, {Peacock}, {Jenkins}, {White},
  {Frenk}, {Pearce}, {Thomas}, {Efstathiou}, \& {Couchman}}]{smith+03}
{Smith}, R.~E., {et~al.} 2003, \mnras, 341, 1311

\bibitem[{{Stocke} {et~al.}(2006){Stocke}, {Penton}, {Danforth}, {Shull},
  {Tumlinson}, \& {McLin}}]{stockeetal06}
{Stocke}, J.~T., {Penton}, S.~V., {Danforth}, C.~W., {Shull}, J.~M.,
  {Tumlinson}, J., \& {McLin}, K.~M. 2006, \apj, 641, 217

\bibitem[{{Tejos} {et~al.}(2012){Tejos}, {Morris}, {Crighton}, {Theuns},
  {Altay}, \& {Finn}}]{tejos+12}
{Tejos}, N., {Morris}, S.~L., {Crighton}, N.~H.~M., {Theuns}, T., {Altay}, G.,
  \& {Finn}, C.~W. 2012, \mnras, 425, 245

\bibitem[{{Tejos} {et~al.}(2014){Tejos}, {Morris}, {Finn}, {Crighton},
  {Bechtold}, {Jannuzi}, {Schaye}, {Theuns}, {Altay}, {Le F{\`e}vre},
  {Ryan-Weber}, \& {Dav{\'e}}}]{tejos+14}
{Tejos}, N., {et~al.} 2014, \mnras, 437, 2017

\bibitem[{{Tempel} {et~al.}(2014){Tempel}, {Stoica}, {Mart{\'{\i}}nez},
  {Liivam{\"a}gi}, {Castellan}, \& {Saar}}]{tempel+14}
{Tempel}, E., {Stoica}, R.~S., {Mart{\'{\i}}nez}, V.~J., {Liivam{\"a}gi},
  L.~J., {Castellan}, G., \& {Saar}, E. 2014, \mnras, 438, 3465

\bibitem[{{Tripp}(2013)}]{Tripp13}
{Tripp}, T. 2013, Science white paper submitted to NASA Solicitation
  NNH12ZDA008L: Science Objectives and Requirements for the Next NASA
  UV/Visible Astrophysics Mission Concepts, arXiv:1303.0043

\bibitem[{{Tripp} {et~al.}(2019){Tripp}, {et al.}, {et al.}, \& {et
  al.}}]{casbah}
{Tripp}, T., {et al.}, {et al.}, \& {et al.} 2019, in prep.

\bibitem[{{Tripp} {et~al.}(1998){Tripp}, {Lu}, \& {Savage}}]{tripp+98}
{Tripp}, T.~M., {Lu}, L., \& {Savage}, B.~D. 1998, \apj, 508, 200

\bibitem[{{Tripp} {et~al.}(2011){Tripp}, {Meiring}, {Prochaska}, {Willmer},
  {Howk}, {Werk}, {Jenkins}, {Bowen}, {Lehner}, {Sembach}, {Thom}, \&
  {Tumlinson}}]{tmp+11}
{Tripp}, T.~M., {et~al.} 2011, Science, 334, 952

\bibitem[{{Tripp} {et~al.}(2008){Tripp}, {Sembach}, {Bowen}, {Savage},
  {Jenkins}, {Lehner}, \& {Richter}}]{tripp08}
{Tripp}, T.~M., {Sembach}, K.~R., {Bowen}, D.~V., {Savage}, B.~D., {Jenkins},
  E.~B., {Lehner}, N., \& {Richter}, P. 2008, \apjs, 177, 39

\bibitem[{{Tumlinson} {et~al.}(2011){Tumlinson}, {Thom}, {Werk}, {Prochaska},
  {Tripp}, {Weinberg}, {Peeples}, {O'Meara}, {Oppenheimer}, {Meiring}, {Katz},
  {Dav{\'e}}, {Ford}, \& {Sembach}}]{ttw+11}
{Tumlinson}, J., {et~al.} 2011, Science, 334, 948

\bibitem[{{Verner} {et~al.}(1994){Verner}, {Barthel}, \& {Tytler}}]{Verner94}
{Verner}, D.~A., {Barthel}, P.~D., \& {Tytler}, D. 1994, \aaps, 108, 287

\bibitem[{{Wakker} {et~al.}(2015){Wakker}, {Hernandez}, {French}, {Kim},
  {Oppenheimer}, \& {Savage}}]{wakker+15}
{Wakker}, B.~P., {Hernandez}, A.~K., {French}, D.~M., {Kim}, T.-S.,
  {Oppenheimer}, B.~D., \& {Savage}, B.~D. 2015, \apj, 814, 40

\bibitem[{{Wakker} \& {Savage}(2009)}]{wakker09}
{Wakker}, B.~P., \& {Savage}, B.~D. 2009, \apjs, 182, 378

\bibitem[{{Warren} {et~al.}(2007){Warren}, {Cross}, {Dye}, {Hambly}, {Almaini},
  {Edge}, {Hewett}, {Hodgkin}, {Irwin}, {Jameson}, {Lawrence}, {Lucas},
  {Mortlock}, {Adamson}, {Bryant}, {Collins}, {Davis}, {Emerson}, {Evans},
  {Gonzales-Solares}, {Hirst}, {Kerr}, {Lewis}, {Mann}, {Rawlings}, {Read},
  {Riello}, {Sutorius}, \& {Varricatt}}]{Warren:2007aa}
{Warren}, S.~J., {et~al.} 2007, ArXiv Astrophysics e-prints

\bibitem[{{Werk} {et~al.}(2013){Werk}, {Prochaska}, {Thom}, {Tumlinson},
  {Tripp}, {O'Meara}, \& {Peeples}}]{werk+13}
{Werk}, J.~K., {Prochaska}, J.~X., {Thom}, C., {Tumlinson}, J., {Tripp}, T.~M.,
  {O'Meara}, J.~M., \& {Peeples}, M.~S. 2013, \apjs, 204, 17

\bibitem[{{Woodgate} {et~al.}(1998){Woodgate}, {Kimble}, {Bowers}, {Kraemer},
  {Kaiser}, {Danks}, {Grady}, {Loiacono}, {Brumfield}, {Feinberg}, {Gull},
  {Heap}, {Maran}, {Lindler}, {Hood}, {Meyer}, {Vanhouten}, {Argabright},
  {Franka}, {Bybee}, {Dorn}, {Bottema}, {Woodruff}, {Michika}, {Sullivan},
  {Hetlinger}, {Ludtke}, {Stocker}, {Delamere}, {Rose}, {Becker}, {Garner},
  {Timothy}, {Blouke}, {Joseph}, {Hartig}, {Green}, {Jenkins}, {Linsky},
  {Hutchings}, {Moos}, {Boggess}, {Roesler}, \& {Weistrop}}]{STIS}
{Woodgate}, B.~E., {et~al.} 1998, \pasp, 110, 1183

\bibitem[{{Zehavi} {et~al.}(2011){Zehavi}, {Zheng}, {Weinberg}, {Blanton},
  {Bahcall}, {Berlind}, {Brinkmann}, {Frieman}, {Gunn}, {Lupton}, {Nichol},
  {Percival}, {Schneider}, {Skibba}, {Strauss}, {Tegmark}, \& {York}}]{zzw+11}
{Zehavi}, I., {et~al.} 2011, \apj, 736, 59

\bibitem[{{Zhu} {et~al.}(2014){Zhu}, {M{\'e}nard}, {Bizyaev}, {Brewington},
  {Ebelke}, {Ho}, {Kinemuchi}, {Malanushenko}, {Malanushenko}, {Marchante},
  {More}, {Oravetz}, {Pan}, {Petitjean}, \& {Simmons}}]{zhu+14}
{Zhu}, G., {et~al.} 2014, \mnras, 439, 3139

\bibitem[{{Zou} {et~al.}(2017{\natexlab{a}}){Zou}, {Zhang}, {Zhou}, {Nie},
  {Peng}, {Zhou}, {Jiang}, {Cai}, {Dey}, {Fan}, {Fan}, {Guo}, {He}, {Jiang},
  {Lang}, {Lesser}, {Li}, {Ma}, {Mao}, {McGreer}, {Schlegel}, {Shao}, {Wang},
  {Wang}, {Wu}, {Wu}, {Yang}, \& {Yue}}]{Zou:2017ab}
{Zou}, H., {et~al.} 2017{\natexlab{a}}, \aj, 153, 276

\bibitem[{{Zou} {et~al.}(2017{\natexlab{b}}){Zou}, {Zhou}, {Fan}, {Zhang},
  {Zhou}, {Nie}, {Peng}, {McGreer}, {Jiang}, {Dey}, {Fan}, {He}, {Jiang},
  {Lang}, {Lesser}, {Ma}, {Mao}, {Schlegel}, \& {Wang}}]{Zou:2017aa}
---. 2017{\natexlab{b}}, \pasp, 129, 064101

\end{thebibliography}
